\documentclass[twocolumn]{aastex63}

\usepackage{placeins}

\let\tablenum\relax
\usepackage{siunitx}
\DeclareSIUnit\parsec{pc}
\DeclareSIUnit\year{yr}

\defcitealias{palla1999star}{PS99}
\defcitealias{siess2000}{SFD00}
\defcitealias{baraffe2015}{BHAC15}
\defcitealias{feiden2016magnetic}{F16}
\defcitealias{bressan2012parsec}{Br12}
\defcitealias{chen2014improvingPARSEC}{C14}

\begin{document}

\title{Dynamical stellar masses of pre-main sequence stars in Lupus and Taurus obtained with ALMA surveys in comparison with stellar evolutionary models}

\author{Teresa A. M. Braun}
\affiliation{Academia Sinica Institute of Astronomy and Astrophysics, 11F of Astro-Math Bldg, 1, Sec. 4, Roosevelt Rd, Taipei 10617, Taiwan}

\author{Hsi-Wei Yen}
\affiliation{Academia Sinica Institute of Astronomy and Astrophysics, 11F of Astro-Math Bldg, 1, Sec. 4, Roosevelt Rd, Taipei 10617, Taiwan}

\author{Patrick M. Koch}
\affiliation{Academia Sinica Institute of Astronomy and Astrophysics, 11F of Astro-Math Bldg, 1, Sec. 4, Roosevelt Rd, Taipei 10617, Taiwan}

\author{Carlo F. Manara}
\affiliation{European Southern Observatory (ESO), Karl-Schwarzschild-Str. 2, D-85748 Garching, Germany}

\author{Anna Miotello}
\affiliation{European Southern Observatory (ESO), Karl-Schwarzschild-Str. 2, D-85748 Garching, Germany}

\author{Leonardo Testi}
\affiliation{European Southern Observatory (ESO), Karl-Schwarzschild-Str. 2, D-85748 Garching, Germany}
\affiliation{INAF/Osservatorio Astrofisico di Arcetri, Largo Enrico Fermi 5, I-50125 Florence, Italy}

\begin{abstract} 
    We analysed archival molecular line data of pre-main sequence (PMS) stars in the Lupus and Taurus star-forming regions obtained with ALMA surveys with an integration time of a few minutes per source. We stacked the data of $^{13}$CO and C$^{18}$O ($J$ = 2--1 \& 3--2) and CN ($N$ = 3--2, $J$ = 7/2--5/2) lines to enhance the signal-to-noise ratios, and measured the stellar masses of 45 out of 67 PMS stars from the Keplerian rotation in their circumstellar disks. The measured dynamical stellar masses were compared to the stellar masses estimated from the spectroscopic measurements with seven different stellar evolutionary models. We found that the magnetic model of \citet{feiden2016magnetic} provides the best estimate of the stellar masses in the mass range of $0.6~M_{\odot}\leq M_{\star} \leq 1.3~M_{\odot}$ with a deviation of $<$0.7$\sigma$ from the dynamical masses, while all the other models underestimate the stellar masses in this mass range by 20\% to 40\%. In the mass range of $<$0.6~$M_{\odot}$, the stellar masses estimated with the magnetic model of \citet{feiden2016magnetic} have a larger deviation ($>2\sigma$) from the dynamical masses, and other, non-magnetic stellar evolutionary models of \citet{siess2000}, \citet{baraffe2015} and \citet{feiden2016magnetic} show better agreements with the dynamical masses with the deviations of 1.4$\sigma$ to 1.6$\sigma$. Our results show the mass dependence of the accuracy of these stellar evolutionary models.

\end{abstract}


\section{Introduction}
\label{sec:introduction}

The mass is an important characteristic of a star. It determines not only the luminosity and temperature of the star, but also its evolution and feedback to the interstellar medium \citep{burkert2004stellar_feedback,ceverino2009stellar_feedback}. Furthermore, mass determination of a large sample of young stars is essential to constrain the initial mass function (IMF). The IMF is a key to understand the physics of star formation and is also important to understand the evolution of stellar clusters and galaxies \citep{jeffries2012measuring_IMF,hopkins2018_IMF}. 

One method widely adopted to determine stellar masses uses the information of the luminosity and effective temperature of a star. By incorporating the physical processes acting on a star, theoretical models were established to describe the relationship between luminosity, effective temperature, mass and age of a star \citep[e.g.,][]{dantona1997,baraffe1998,palla1999star,siess2000,bressan2012parsec,chen2014improvingPARSEC,baraffe2015,feiden2016magnetic}. The location of a star in the Hertzsprung Russel Diagram (HR Diagram) is compared to the theoretical evolutionary tracks from the stellar evolutionary models to constrain the mass and age of the star. The advantage of this approach is the large number of stars that it can be applied to. With existing facilities it is possible to perform surveys on large samples of stars and provide data for the spectroscopic method to determine stellar masses \citep[e.g.,][]{rigliaco2012xShooter,herczeg2014specSurvey,alcala2017xshootersurveyMspec,manara2017XshootersurveyMspec}.

The different stellar evolutionary models are in good agreement for main sequence (MS) stars and have been compared with various precise direct measurements of stellar mass (e.g., from orbital motion of binaries). 
Therefore, this method is considered reliable for MS stars \citep{hillenbrand2004}. However, the spectroscopic method has larger uncertainties for pre-main sequence (PMS) stars. The age and mass of a PMS star estimated with different models can show discrepancies of 10\% to more than 50\% \citep{sheehan2019high}. The theoretical models adopt different assumptions of convection, atmosphere, opacities and the equation of state as well as different treatments of those processes, resulting in significant deviations in the estimated masses for PMS stars. Other influences, like magnetic fields, accretion, and dust in the atmosphere of a star, further complicate the models \citep{siess2000,baraffe2009accretion,cassisi2012uncertainties_EvModels,baraffe2015}. 
Therefore, further calibration of the stellar models of PMS stars is necessary.

One method to calibrate the stellar evolutionary models makes use of the orbital motions of eclipsing or resolved close binaries \citep[e.g.,][]{stassun2014empirical,rizzuto2016dynamical,rizzuto2019dynamical}. 
By monitoring the motions of the stars, the stellar masses can be measured independently from the stellar evolutionary models. Because of the rareness of such binary systems and the time needed to closely monitor the orbital motions, the sample of stars which this method can be applied to is limited. 

Another approach is to observe the rotation of protoplanetary disks around PMS stars \citep{sheehan2019high, simon2000dynamical,simon2017dynamical,simon2019masses}. By measuring the velocity and radius of the material around a star by spectroscopic observations, it is possible to determine the enclosed mass with a minimum of assumptions. Different from the approach with binary systems, only one measurement is needed and the method is applicable to every star surrounded by a disk in Keplerian rotation. Hence, this can be used to obtain measurements of a large sample of PMS stars, suitable to test the theoretical models. 

Dynamical mass measurements have been used in several surveys to test a number of different evolutionary models. Among these, there are smaller studies with three to nine stars \citep{simon2000dynamical, sheehan2019high}, as well as larger studies, including 25 to 32 PMS stars \citep{hillenbrand2004, simon2017dynamical, simon2019masses, stassun2014empirical}. 
For PMS stars, it was found that the spectroscopic and dynamical masses are in good agreement for masses larger than about 1.2~$M_{\odot}$ \citep{hillenbrand2004, stassun2014empirical}. \citet{hillenbrand2004} reported, that the evolutionary models examined in their paper tend to underestimate the stellar masses in the mass range below 1.2~$M_{\odot}$. Whereas 
\citet{stassun2014empirical} found a tendency to overestimate stellar masses in this mass range by examining eclipsing binary systems. This tendency gets less significant when excluding the systems with tertiary components. The discrepancy in the results of \citet{hillenbrand2004} and \citet{stassun2014empirical} might be explained by the different stellar evolutionary models adopted because there are only two common models in these studies. 
\citet{simon2019masses} updated several dynamical mass measurements with distances measured by Gaia \citep{prusti2016gaia, brown2018gaia} and found an underestimation of stellar masses in the mass range of 0.4 to 1.4~$M_{\odot}$ for evolutionary models which do not include magnetic fields \citep{baraffe2015, feiden2016magnetic}. For the stellar evolutionary model of \citet{feiden2016magnetic} with magnetic fields, the discrepancy between dynamical and spectroscopic mass measurements becomes insignificant.

In order to further investigate the agreement of spectroscopic and dynamical mass measurements, wider mass ranges and different star-forming regions need to be examined. Especially the sample size of low mass stars ($<$ 0.5 $M_\odot$) is so far limited. A larger sample of PMS stars is essential to verify the findings in previous studies.

With current facilities like ALMA it is possible to detect protoplanetary disks for large samples of stars in continuum emission at relatively high angular resolutions with short integration times of only a few minutes per source \citep[e.g.,][]{barenfeld2016almaSurvey_disks,Ansdell2016,ansdell2017almasurvey,Ansdell2018,long2018taurus,eisner2018alma_survey,cazzoletti2019almasurvey,williams2019ophiuchus_almasurvey,vanterwisga2020almasurvey_disks}. 
In these observations, several molecular lines are often simultaneously observed. 
The obtained molecular-line data have lower signal-to-noise ratios (S/N) compared to the continuum data, which could be too low to constrain the velocity patterns in disks.
However, by stacking data of different molecular lines and/or aligning line emission from different positions in a disk, it is possible to obtain sufficient signals and measure disk rotation and stellar mass \citep{Yen2016,matra2017aligningSpectra,salinas2017KepMasking,teague2018KepMasking,teague2018stacking,Yen2018}.

In this work, we aim to measure the masses of a large sample of PMS stars by making use of ALMA archival data obtained with shallow, but large surveys of nearby star-forming regions, and to compare the results to the spectroscopic masses determined with stellar evolutionary models. 
The structure of this paper is as follows: After introducing the sample and data in Section \ref{sec:sample selection} and \ref{sec:observation}, the analysis is explained in Section \ref{sec:analysis}. The results of the mass measurements are described in Section \ref{sec:results}, followed by the discussion and comparison with stellar evolutionary models in Section \ref{sec:discussion}. In the end the obtained insights are summarised, and future prospects are discussed in Section \ref{sec:summary}.
\section{Sample Selection}
\label{sec:sample selection}

The sample of this study is selected from ALMA surveys at 0$.\!\!^{\prime\prime}$12 - 0$.\!\!^{\prime\prime}$25 resolutions toward Young Stellar Objects (YSOs) in the young (\SIrange{\sim1}{3}{\mega\year}) and nearby (\SIrange{\sim150}{160}{\parsec}) star-forming regions, Lupus and Taurus \citep{Ansdell2016,Ansdell2018,long2019taurus}.
The sample of the ALMA Lupus survey consists of 93 YSOs with spectroscopic masses of $M_{\star} > 0.1~M_{\odot}$ and Class II or flat IR excess spectra located in the Lupus I, III and IV clouds \citep{Ansdell2016,Ansdell2018}.
The sample of the ALMA Taurus survey consists of 42 YSOs of spectral types earlier than M3, excluding known, close binaries with separations of 0$.\!\!^{\prime\prime}$1--0$.\!\!^{\prime\prime}$5 \citep{long2019taurus}. 
From these Lupus and Taurus samples, 
we selected those YSOs with resolved circumstellar disks in 1.3~mm continuum ALMA surveys, 
so that the inclination angles of the disks and their orientations on the plane of sky could be measured \citep{Ansdell2018, Tazzari2017,long2019taurus, manara2019taurus}.
Thus, the sample in our study consists of 30 YSOs in the Lupus region and 37 YSOs in the Taurus region. 
The basic properties of our sample sources and the inclination angles and orientations of their circumstellar disks are listed in Table \ref{tab:parameter_lupus} and \ref{tab:parameter_taurus}. The distribution of the effective temperatures and luminosities of the stars included in this paper are shown in Appendix \ref{sec:appendix_histograms}.

\begin{deluxetable*}{lccccccccccccc}
\label{tab:parameter_lupus}
\tablecaption{Parameters adopted from the literature for the analysis for stars in the Lupus region}
\tablehead{\colhead{Name} & \colhead{RA} & \colhead{Dec} & \colhead{$i$} & \colhead{$PA$} & \colhead{dist} & \colhead{$T_{\mathrm{eff}}$} & \colhead{$L_{\star}$} & \colhead{Ref.} \\
\colhead{} & \colhead{} & \colhead{} & \colhead{[$^{\circ}$]} & \colhead{[$^{\circ}$]} & \colhead{[pc]} & \colhead{[K]} & \colhead{[L$_{\odot}$]} &  \colhead{}
}
\startdata
Sz65 & 15:39:27.75 & -34:46:17.56 & 61.46$\pm$0.88 & 108.63$\pm$0.37 & 155 & 4060 & 0.89 & 1 \\
J15450887-3417333 & 15:45:08.85 & -34:17:33.81 & 36.30$\pm$5.56 & 2.41$\pm$2.53 & 155 & 3060 & 0.06 & 1 \\
Sz68 & 15:45:12.84 & -34:17:30.98 & 32.89$\pm$3.32 & 175.78$\pm$3.13 & 154 & 4900 & 5.42 & 1 \\
Sz69 & 15:45:17.39 & -34:18:28.66 & 43.53$\pm$8.65 & 124.28$\pm$17.10 & 155 & 3197 & 0.09 & 1 \\
Sz71 & 15:46:44.71 & -34:30:36.05 & 40.82$\pm$0.71 & 37.51$\pm$0.01 & 156 & 3632 & 0.33 & 1 \\
Sz73 & 15:47:56.92 & -35:14:35.15 & 49.76$\pm$3.95 & 94.71$\pm$5.17 & 157 & 4060 & 0.46 & 1 \\
Sz75 & 15:49:12.09 & -35:39:05.46 & 60.2$\pm$5.0 & 169.0$\pm$5.0 & 152 & 4205 & 1.48 & 2 \\
Sz76 & 15:49:30.72 & -35:49:51.83 & -60.0$\pm$5.0 & 65.0$\pm$5.0 & 160 & 3270 & 0.18 & 2 \\
RXJ1556.1-3655 & 15:56:02.08 & -36:55:28.67 & 53.5$\pm$5.0 & 55.6$\pm$5.0 & 158 & 3705 & 0.26 & 2 \\
Sz83 & 15:56:42.29 & -37:49:15.82 & 3.31$\pm$2.90 & 163.76$\pm$5.94 & 160 & 4060 & 1.49 & 1 \\
Sz84 & 15:58:02.50 & -37:36:03.08 & 73.99$\pm$1.56 & 167.31$\pm$0.77 & 153 & 3125 & 0.13 & 1 \\
Sz129 & 15:59:16.45 & -41:57:10.66 & 31.74$\pm$0.75 & 154.94$\pm$0.43 & 162 & 4060 & 0.43 & 1 \\
RYLup & 15:59:28.37 & -40:21:51.56 & 68.0$\pm$5.0 & 109.0$\pm$5.0 & 159 & 4900 & 1.87 & 2 \\
J16000236-4222145 & 16:00:02.34 & -42:22:14.99 & 65.71$\pm$0.36 & 160.45$\pm$0.02 & 164 & 3270 & 0.18 & 1 \\
MYLup & 16:00:44.50 & -41:55:31.27 & 72.98$\pm$0.35 & 58.94$\pm$0.12 & 157 & 5100 & 0.85 & 1 \\
EXLup & 16:03:05.48 & -40:18:25.83 & -30.5$\pm$5.0 & 70.0$\pm$5.0 & 158 & 3850 & 0.76 & 2 \\
Sz133 & 16:03:29.37 & -41:40:02.14 & 78.53$\pm$0.65 & 126.29$\pm$0.09 & 153 & 4350 & 0.07 & 1 \\
Sz90 & 16:07:10.05 & -39:11:03.64 & 61.31$\pm$5.34 & 123.00$\pm$4.86 & 160 & 4060 & 0.42 & 1 \\
Sz98 & 16:08:22.48 & -39:04:46.81 & 47.10$\pm$0.70 & 111.58$\pm$0.06 & 156 & 4060 & 1.53 & 1 \\
Sz100 & 16:08:25.75 & -39:06:01.60 & 45.11$\pm$0.97 & 60.20$\pm$0.06 & 137 & 3057 & 0.08 & 1 \\
J16083070-3828268 & 16:08:30.69 & -38:28:27.28 & 74.0$\pm$5.0 & 107.0$\pm$5.0 & 156 & 4900 & 1.84 & 2 \\
SSTc2dJ160836.2-392302 & 16:08:36.16 & -39:23:02.88 & -55.4$\pm$5.0 & 110.0$\pm$5.0 & 154 & 4205 & 1.15 & 2 \\
Sz108B & 16:08:42.86 & -39:06:15.04 & 49.09$\pm$5.34 & 151.76$\pm$6.05 & 169 & 3125 & 0.11 & 1 \\
J16085324-3914401 & 16:08:53.22 & -39:14:40.53 & 60.72$\pm$4.00 & 100.31$\pm$5.47 & 168 &3415 & 0.21 & 1 \\
Sz111 & 16:08:54.67 & -39:37:43.53 & 53.0$\pm$5.0 & 40.0$\pm$5.0 & 158 & 3705 & 0.21 & 2 \\
Sz113 & 16:08:57.78 & -39:02:23.21 & 10.78$\pm$9.19 & 147.36$\pm$14.20 & 163 & 3197 & 0.04 & 1  \\
Sz114 & 16:09:01.83 & -39:05:12.79 & 15.84$\pm$3.39 & 148.73$\pm$6.87 & 162 & 3175 & 0.21 & 1 \\
J16102955-3922144 & 16:10:29.53 & -39:22:14.83 & 66.54$\pm$9.21 & 118.86$\pm$9.49 & 163 &3200 & 0.11 & 1 \\
Sz123A & 16:10:51.57 & -38:53:14.17 & 43.0$\pm$5.0 & 145.0$\pm$5.0 & 159 & 3705 & 0.13 & 2 \\
J16124373-3815031 & 16:12:43.73 & -38:15:03.40 & 43.69$\pm$7.39 & 22.99$\pm$8.85 & 160 & 3705 & 0.39 & 1 \\
\enddata
\tablecomments{The coordinates were adopted from \citet{Ansdell2016}, the inclination angle $i$ and the position angle $PA$ were adopted from (1) \citet{Tazzari2017} and (2) \citet{Ansdell2018}. The distances were 
obtained by inverting the parallaxes measured by the Gaia mission \citep{prusti2016gaia, brown2018gaia}. The luminosity $L_{\star}$ and effective temperature $T_{\mathrm{eff}}$ of the stars were taken from \citet{alcala2014xshootersurvey, alcala2017xshootersurveyMspec}, according to the updated distances estimates \citep{alcala2019updatedTeffL}. The distance for Sz123A could not be obtained from the Gaia data, therefore the mean distance to the Lupus Clouds was adopted.}
\end{deluxetable*}

\begin{deluxetable*}{lccccccccccccc}
\label{tab:parameter_taurus}
\tablecaption{Parameters adopted from the literature for the analysis for stars in the Taurus region}
\tablehead{\colhead{Name} & \colhead{RA} & \colhead{Dec} & \colhead{$i$} & \colhead{$PA$} & \colhead{dist} & \colhead{$T_{\mathrm{eff}}$} & \colhead{$L_{\star}$} \\
\colhead{} & \colhead{} & \colhead{} & \colhead{[$^{\circ}$]} & \colhead{[$^{\circ}$]} & \colhead{[pc]} & \colhead{[K]} & \colhead{[L$_{\odot}$]} 
}
\startdata
CITau    & 04:33:52.03   & +22:50:29.81  & 50.0$\pm$0.3           & 11.2$\pm$0.4            &  158 & 4277 & 0.81  \\
CIDA9A   &   05:05:22.82 & +25:31:30.50  &   45.6$\pm$0.5         & 102.7$\pm$0.7           &  171 & 3589 & 0.20  \\
DLTau    & 04:33:39.09   & +25:20:37.79  & 45.0$\pm$0.2           & 52.1$\pm$0.4            &  159 & 4277 & 0.65  \\
DNTau    & 04:35:27.39   & +24:14:58.55  & 35.2$^{+0.5}_{-0.6}$   & 79.2$\pm$1.0            &  128 & 3806 & 0.70  \\
DSTau    &   04:47:48.60 & +29:25:10.76  & 65.2$\pm$0.3           & 159.6$\pm$0.4           &  159 & 3792 & 0.25  \\
FTTau    &   04:23:39.20 & +24:56:13.86  & 35.5$\pm$0.4           & 121.8$\pm$0.7           &  127 & 3444 & 0.15  \\
GOTau    & 04:43:03.08   & +25:20:18.35  & 53.9$\pm$0.5           & 20.9$\pm$0.6            &  144 & 3516 & 0.21  \\
IPTau    &   04:24:57.09 & +27:11:56.07  & 45.2$^{+0.8}_{-0.9}$   & 173.0$\pm$1.1           &  130 & 3763 & 0.34  \\
IQTau    & 04:29:51.57   & +26:06:44.45  & 62.1$\pm$0.5           & 42.4$\pm$0.6            &  131 & 3690 & 0.22  \\
MWC480   &   04:58:46.27 & +29:50:36.51  &   36.5$\pm$0.2         & 147.5$\pm$0.3           &  161 & 8400 & 17.38 \\
RYTau    & 04:21:57.42   & +28:26:35.09  & 65.0$\pm$0.1           & 23.1$\pm$0.1            &  128 & 6220 & 12.30 \\
UZTauE   & 04:32:43.08   & +25:52:30.63  &   56.1$\pm$0.4         & 90.4$\pm$0.4            &  131 & 3574 & 0.35  \\
UZTauW   & 04:32:42.8    & +25:52:31.2   & 61.2$^{+1.1}_{-1.0}$   & 91.5$^{+0.8}_{-0.9}$    &  131 &\nodata&\nodata\\
BPTau    &   04:19:15.85 & +29:06:26.48  & 38.2$\pm$0.5           & 151.1$\pm$1.0           &  129 & 3777 & 0.40  \\
DOTau    &   04:38:28.60 & +26:10:49.08  & 27.6$\pm$0.3           & 170.0$\pm$0.9           &  139 & 3806 & 0.23  \\
DQTau    & 04:46:53.06   & +16:59:59.89  & 16.1$\pm$1.2           & 20.3$\pm$4.3            &  197 & 3763 & 1.17  \\
DRTau    &  04:47:06.22  & +16:58:42.55  & 5.4$^{+2.1}_{-2.6}$    &  3.4$^{+8.2}_{-8.0}$    &  195 & 4205 & 0.63  \\
GITau    &   04:33:34.07 & +24:21:16.70  & 43.8$\pm$1.1           & 143.7$^{+1.9}_{-1.6}$   &  130 & 3792 & 0.49  \\
GKTau    &   04:33:34.57 & +24:21:05.49  & 40.2$^{+5.9}_{-6.2}$   & 119.9$^{+8.9}_{-9.1}$   &  129 & 4007 & 0.80  \\
Haro6-13 &   04:32:15.42 & +24:28:59.21  & 41.1$\pm$0.3           & 154.2$\pm$0.3           &  130 & 4277 & 0.79  \\
HOTau    &   04:35:20.22 & +22:32:14.27  & 55.0$\pm$0.8           & 116.3$\pm$1.0           &  161 & 3386 & 0.14  \\
HPTau    & 04:35:52.79   & +22:54:22.93  & 18.3$^{+1.2}_{-1.4}$   & 56.5$^{+4.6}_{-4.3}$    &  177 & 4590 & 1.30  \\
HQTau    &   04:35:47.35 & +22:50:21.36  & 53.8$\pm$3.2           & 179.1$^{+3.2}_{-3.4}$   &  158 & 4900 & 4.34  \\
V409Tau  & 04:18:10.79   & +25:19:56.97  &  69.3$\pm$0.3          & 44.8$\pm$0.5            &  131 & 3763 & 0.66  \\
V836Tau  &   05:03:06.60 & +25:23:19.29  &  43.1$\pm$0.8          & 117.6$\pm$1.3           &  169 & 3734 & 0.44  \\
DHTauA   & 04:29:41.56   & +26:32:57.76  &   16.9$^{+2.0}_{-2.2}$ & 18.8$^{+7.1}_{-7.2}$    &  135 & 3516 & 0.20  \\
DKTauA   & 04:30:44.25   & +26:01:24.35  &   12.8$^{+2.5}_{-2.8}$ & 4.4$^{+10.1}_{-9.4}$    &  128 & 3902 & 0.45  \\
DKTauB   & 04:30:44.4    & +26:01:23.20  & 78.0$^{+6.1}_{-11.0}$  & 28.0$^{+5.2}_{-5.4}$    &  128 &\nodata&\nodata\\
HKTauA   &   04:31:50.58 & +24:24:17.37  &   56.9$\pm$0.5         & 174.9$\pm$0.5           &  133 & 3632 & 0.27  \\
HKTauB   & 04:31:50.6    & +24:24:15.09  & 83.2$\pm$0.2           & 41.2$\pm$0.2            &  133 &\nodata&\nodata\\
HNTauA   & 04:33:39.38   & +17:51:51.98  &   69.8$^{+1.4}_{-1.3}$ & 85.3$^{+0.7}_{-0.6}$    &  136 & 4730 & 0.16  \\
RWAurA   & 05:07:49.57   & +30:24:04.70  &   55.1$^{+0.5}_{-0.4}$ & 41.1$\pm$0.6            &  163 & 5250 & 0.99  \\
RWAurB   & 05:07:49.5    & +30:24:04.29  & 74.6$^{+3.8}_{-8.2}$   & 41.0$^{+3.6}_{-3.7}$    &  163 &\nodata&\nodata\\
TTauN    & 04:21:59.45   & +19:32:06.18  & 28.2$\pm$0.2           & 87.5$\pm$0.5            &  144 & 5250 & 6.82  \\
TTauS    & 04:21:59.4    & +19:32:05.52  & 61.6$^{+8.8}_{-4.8}$   & 7.9$^{+3.7}_{-3.5}$     &  144 &\nodata&\nodata\\
UYAurA   & 04:51:47.40   & +30:47:13.10  &   23.5$^{+7.8}_{-6.6}$ & 125.7$^{+10.3}_{-10.9}$ &  155 & 4060 & 1.05  \\
V710TauA & 04:31:57.81   & +18:21:37.64  & 48.9$\pm$0.3           & 84.3$\pm$0.4            &  142 & 3603 & 0.26  \\
\enddata
\tablecomments{The coordinates, inclination angle $i$ and position angle $PA$ were adopted from \citet{long2019taurus}. The effective temperature T$_{\mathrm{eff}}$ and luminosity L$_{\star}$ were derived from \citet{herczeg2014specSurvey} and were re-scaled for updated distances in \citet{long2019taurus}.
The parameters for secondary stars in binary systems were adopted from \citet{manara2019taurus}. The distances were obtained by inverting the parallaxes measured by the Gaia mission \citep{prusti2016gaia, brown2018gaia}.}
\end{deluxetable*}

\section{Observations}
\label{sec:observation}
\subsection{Lupus Region}
For the YSOs located in the Lupus star forming region, 
we retrieved three data sets obtained with the Lupus survey from the ALMA archive, 2013.1.00220.S, 2015.1.00222.S (PI: J. Williams), and 2016.1.01239.S (PI: S. van Terwisga).
The observations were conducted in Band 7 in Cycle 2 and in Band 6 in Cycle 3 with an integration time of approximately one minute per source. 
Additional observations to complete the sample were conducted in both Band 6 and 7 in Cycle 4 with an integration time of approximately four minutes per source.
Apart from the continuum, several molecular lines were observed, including $^{13}$CO $J$ = 2--1 and 3--2 (\SI{220.399}{\giga\hertz} and \SI{330.588}{\giga\hertz}), C$^{18}$O $J$ = 2--1 and 3--2 (\SI{219.560}{\giga\hertz} and \SI{329.331}{\giga\hertz}) and the CN $N$ = 3--2 transition. For the CN lines, we focus on the brightest hyperfine structure transitions of CN $J$ = 7/2--5/2, with $F$ = 7/2--5/2 and $F$ = 9/2--7/2 at \SI{340.247770}{\giga\hertz}.

For the Band 7 observation in Cycle 2, 37 12-m antennas were used with baseline lengths from \SIrange{21.4}{783.5}{\m}. The native velocity resolution to observe the CO isotopologue and CN lines in Band 7 are \SI{0.11}{\km\per\s} and \SI{0.31}{\km\per\s}, respectively. 37 to 42 12-m antennas were used for the observation in Band 6 in Cycle 3 with baseline lengths from \SIrange{15}{2483}{\m}, and the CO isotopologue lines were observed at a native velocity resolution of \SI{0.16}{\km\per\s}. The observations in Cycle 4 were carried out with 44 12-m antennas and baseline lengths from \SIrange{2600}{16700}{\m} in Band 6 and 41 12-m antennas and baseline lengths from \SIrange{1100}{15100}{\m} in Band 7. The native velocity resolutions in Cycle 4 are \SI{0.11}{\km\per\s} and \SI{0.17}{\km\per\s} for the CO isotopologue lines in Band 7 and Band 6, respectively, and \SI{0.22}{\km\per\s} for CN. Further details of the observations are described in \citet{Ansdell2016, Ansdell2018} and \citet{vanterwisga}.

\subsection{Taurus Region}
For the YSOs located in the Taurus star forming region, 
we retrieved the data set obtained with the project 2016.1.01164.S (PI: G. Herczeg) from the ALMA archive. 
The observations were conducted in Band 6 in Cycle 4 with an integration time of approximately four minutes per source for bright sources and ten minutes per source for the other sources. 
The observations were carried out using 45 to 47 12-m antennas with baselines of \SIrange{21}{3697}{\m}.
In addition to the 1.3 mm continuum emission, 
$^{13}$CO and C$^{18}$O $J$ = 2--1 were observed at a native velocity resolution of \SI{0.16}{\km\per\s}. 
Further information about the observations can be found in \citet{long2018taurus, long2019taurus}.

\subsection{Data Reduction and Imaging}
After obtaining the raw visibility data from the ALMA archive, the data were calibrated with the pipeline of the Common Astronomy Software Applications package CASA. Each data set was calibrated with the CASA version recommended by the ALMA observatory. For the Lupus sample the following CASA versions were used: 2013.1.00220.S with CASA 4.2.2, 2015.1.00222.S with CASA 5.1.1 and 2016.1.01239.S with CASA 4.7.2. The data of the Taurus sample were partly calibrated with CASA 4.7.2 and CASA 5.1.1. More information about the calibration of the Taurus data is given in \citet{long2018taurus, long2019taurus}. 

We extracted the molecular lines by subtracting the continuum emission from the calibrated visibility in the uv-plane. The ranges of line-free velocity channels were inspected manually. Afterwards the images were generated from the visibility data and cleaned using the task tclean in CASA, with briggs weighting with a robust parameter of +0.5. 
The channel widths of \SI{0.17}{\km\per\s} and \SI{0.12}{\km\per\s} were adopted to generate images of the CO isotopologue $J$ = 2--1 and 3--2 lines, respectively. Because of the coarse native spectral resolution of the CN data, the channel width of \SI{0.22}{\km\per\s} was adopted to generate CN images. 
For further analysis, all the images for a given source were convolved to the coarsest resolution among them using the CASA task imsmooth and regridded to have the same spatial axes.
In addition, the intensity of the $J$ = 3--2 emission is higher than that of the $J$ = 2--1 emission at a lower frequency at a given temperature. 
Thus, we converted the intensity in our data from units of Jy~beam$^{-1}$ to brightness temperature in K to correct for this frequency dependence for direct comparison of different molecular lines and for the following analysis.
\section{Analysis}
\label{sec:analysis}

\subsection{Stacking Molecular-Line Data}
The S/Ns of the images of individual molecular lines are limited because of the short integration time of these surveys.
Thus, we first stacked the data of different molecular lines to enhance the S/Ns.
Different transitions of the same molecules are expected to trace a similar region in a disk, so we first stacked the $J$ = 2--1 and $J$ = 3--2 image cubes of $^{13}$CO and of C$^{18}$O for the YSOs in the Lupus star-forming region.
The image cubes were combined by calculating the noise weighted mean of them.  Hereafter, the images of the stacked $J$ = 2--1 and 3--2 data of $^{13}$CO and C$^{18}$O are called the $^{13}$CO and C$^{18}$O images, respectively.
In addition, because different molecules in a disk are expected to follow the same Keplerian rotation, we also stacked the $^{13}$CO, C$^{18}$O and CN data with various combinations, hereafter $^{13}$CO+CN, C$^{18}$O+CN, $^{13}$CO+C$^{18}$O and $^{13}$CO+C$^{18}$O+CN, to further achieve higher S/Ns. For the stars in the Taurus region no $J$ = 3--2 and CN data was available. Therefore, only stacking of $^{13}$CO and C$^{18}$O $J$ = 2--1 was possible.

\begin{figure}
    \centering
    \includegraphics[width=\linewidth]{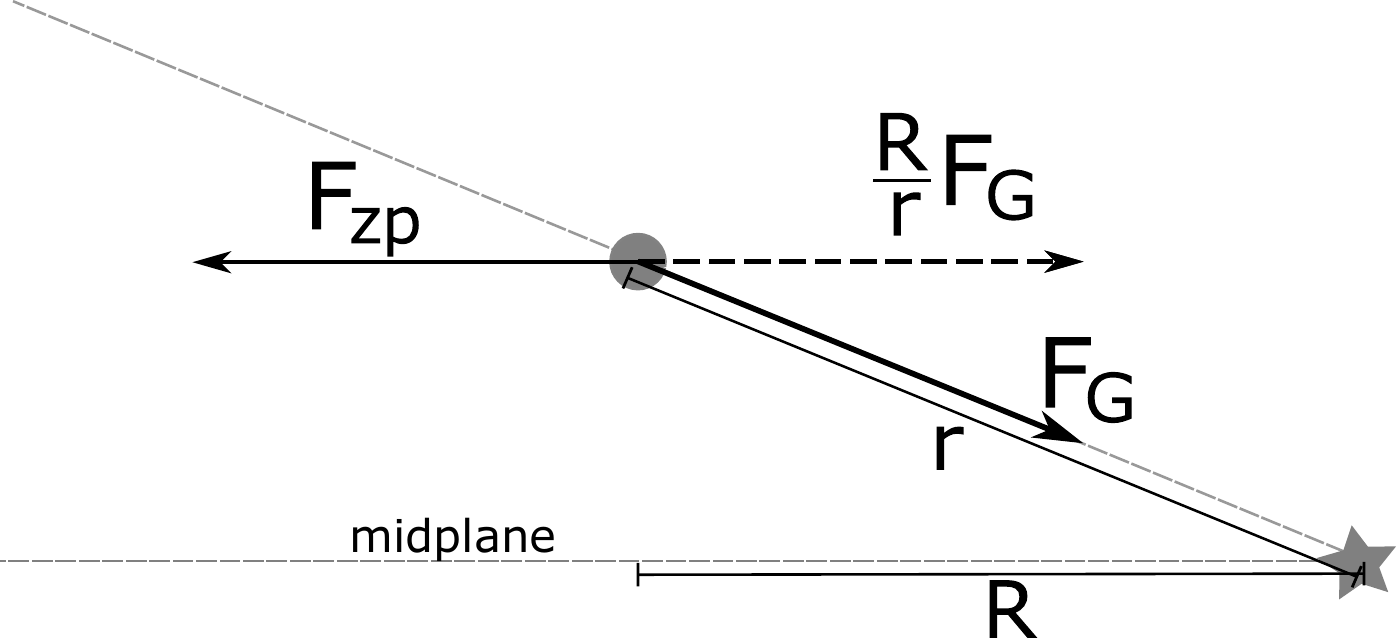}
    \caption{Geometry for calculating the central mass from the Keplerian orbit of gas in a disk around the star. $r$ denotes the radius of the gas, $R$ is the radius projected onto the midplane. $\mathrm{F}_\mathrm{G}$ and $\mathrm{F}_{\mathrm{zp}}$ are the gravitational and centripetal force, respectively.}
    \label{fig:sketch_decomposeFg}
\end{figure}

We note that the different molecules may trace different regions in a disk, because of their different abundance distributions and opacities \citep{pinte2018}. However, assuming the disk as an entity to be Keplerian, these different molecules still trace Keplerian motions in their respective regions.
The mass of a YSO $M_{\star}$ with gas orbiting in Keplerian motion with velocity $v$ and at radius $r$ can be calculated by 
\begin{equation}
\label{equation:kep_motion}
    M_{\star}=\frac{r^2}{G}\frac{r}{R} \cdot \frac{v^2}{R},
\end{equation}
where $G$ is the gravitational constant and $R$ is the radius of the gas projected onto the disk midplane. The term $r/R$ stems from decomposing the gravitational force, to obtain the force component parallel to the midplane. The geometry of the forces acting on the orbiting gas is shown in Figure \ref{fig:sketch_decomposeFg}. The radius of the gas $r$ and its projected radius onto the midplane $R$ are related by the height $h$ of the emitting layer
\begin{equation}
    r^2=h^2+R^2.
\end{equation}
If the height $h$ is small compared to the radius $r$, Equation \ref{equation:kep_motion} can be approximated by 
\begin{equation}
\label{equation:kep_motion_approx}
    M_{\star}=\frac{Rv^2}{G}.
\end{equation}
Therefore, the mass derived with the actual radius of the orbit is different from the mass derived with the above approximation.

Performing the stacking of different molecules while neglecting their different scale heights might introduce additional uncertainties. However, these are typically small.
For example, \citet{pinte2018} found that the $^{13}$CO emission in the disk around HD163296 is originating from a layer at an altitude of \SI{25}{au} when the radius is \SI{200}{au}. This results in an aspect ratio of the scale height of $h/R=0.125$, assuming a linear relation between emitting height and radius. The deviation of the stellar mass calculated with the radius projected onto the disk midplane and calculated with the radius considering the scale height of $h/R=0.125$, is less than 3\% for a face-on disk. 
This effect was further tested on imaging simulations of disks with different inclination angles in Section \ref{sec:robustness and potential bias}.
Due to the smaller optical depth of C$^{18}$O compared to $^{13}$CO, C$^{18}$O is expected to trace regions closer to the midplane than $^{13}$CO \citep{pinte2018}, introducing a smaller error to the stellar mass by neglecting the height of the emitting layer.

Since the uncertainty due to the noise in the data is much larger than the errors due to different scale heights, which are only a few percent, stacking data of different molecular lines is not expected to influence the measurement noticeably. A detailed comparison of the measurements with different molecules is discussed in Section \ref{sec:measurements_indiv_lines}.
The combination of molecular lines resulting in the highest S/N was used to obtain the final measurement of the stellar mass.

\subsection{Measuring Stellar Mass}
\label{sec:analysis_stellar_mass}
In this work, we measured the stellar masses by maximising auto-correction between the image cubes and various Keplerian rotational velocity patterns, as introduced by \citet{Yen2016,Yen2018}. 
The principle of this method is the same as the Keplerian masking technique \citep[e.g.,][]{salinas2017KepMasking,teague2018KepMasking,trapman2020KepMasking}.
It makes use of the Keplerian motion of gas in protoplanetary disks around YSOs and corrects for the Doppler shift of the regions moving towards or away from an observer. 
The spectra at different positions in a disk are shifted by their Keplerian velocity projected onto the line of sight and stacked together. 
Different disk parameters are tested to find the parameters resulting in the stacked spectrum with the highest S/N, which is best corrected for the Doppler shift, and the disk parameters can be measured. Detailed descriptions of this method are given in \citet{Yen2016,Yen2018}. 
Similar methods are also introduced in \citet{teague2018KepMasking}.

In our analysis, we adopted the assumption of a geometrically thin disk structure. 
Thus, the velocity pattern of a disk can be described by the six parameters, inclination angle ($i$), position angle ($PA$), stellar mass ($M_{\star}$), systemic velocity ($v_{sys}$), stellar position, and distance to a star.
The stellar positions of our sample YSOs and $PA$ and $i$ of their disks have been measured with the continuum emission in the same data sets by \citet{Tazzari2017,Ansdell2018} and \citet{long2019taurus}.
Their parallaxes have been measured by the Gaia mission \citep{prusti2016gaia, brown2018gaia} and inverted to obtain their distances, except for Sz123A. 
The distances of our sample stars in the Lupus III cloud ranges from \SIrange{137}{169}{pc} with a median distance of \SI{159}{pc}.
We adopted this median distance to the Lupus III cloud of \SI{159}{pc} for Sz123A in our analysis.
We note that an uncertainty of 10\% in the distance to Sz123A results in an uncertainty of 10\% in its estimated stellar mass, which is not significant compared to the uncertainty due to the noise in the data.

Thus, there are only two unknown parameters, $M_{\star}$ and $v_{sys}$ in our analysis. 
All parameters adopted from the literature for this work are listed in Table \ref{tab:parameter_lupus} and Table \ref{tab:parameter_taurus} for the sources in the Lupus and Taurus star-forming regions, respectively.
We note that actual disks have 3D structures, 
and our assumption of a geometrically thin disk structure could introduce errors in our analysis. 
These possible errors were evaluated by analysing synthetic images in Section \ref{sec:robustness and potential bias}.

For each source, we searched for a combination of $M_{\star}$ and $v_{sys}$ resulting in the velocity aligned stacked spectrum with the highest S/N. 
For each combination of the parameters, 
we generated a velocity aligned stacked spectrum averaged over the area from the center to a radius $R$ in radial direction and from 0 to $2\pi$ in azimuthal direction, 
and measured its S/N. 
To measure the S/N, we first fitted a Gaussian line profile to the velocity aligned stacked spectrum, 
and calculated the S/N of the integrated intensity within a 1$\sigma$ Gaussian line width.
When the integrated intensity was computed, the included velocity channels were weighted with the fitted Gaussian line profile, so that more weight was put on the velocity channels near zero velocity with respect to $v_{sys}$.
This is because when the adopted velocity pattern matches the actual disk rotation, the resultant stacked spectrum is symmetric and centred at zero velocity with respect to $v_{sys}$ \citep{Yen2016}.

This weighting ensures that the narrower, more symmetrical spectrum has a higher S/N than other spectra with the same total flux but with a skewed profile, and is therefore favoured.
The velocity alignment can also cause decorrelation of nearby pixels in an interferometric image, and thereby reduction of noise when large velocity shifts are applied \citep{Yen2016}. Therefore, the noise level of the original, non-aligned spectrum was adopted, when we measured the S/N to find the best-aligned spectrum. 
In this procedure, any velocity aligned stacked spectrum with the S/N at its peak below 4$\sigma$ or with its line width narrower than three times the channel width was excluded from possible solutions to avoid false detections \citep{Yen2018}.

Because the radii of gaseous disks around our sample YSOs are not known, we first assumed $R$ to be 1$^{\prime\prime}$ for all the disks \citep{Ansdell2018} and applied the analysis. 
After finding the parameters, which result in the best aligned spectrum, the radial intensity profile was extracted to measure the radius of the disk. 
We adopted the radius which encloses 95\% of the total flux of the disk as the disk radius.
This process of measuring the mass and determining the radial intensity profile was repeated until the change of the measured radius was less than 10\%. The stellar mass itself is not sensible to the disk radius, but the uncertainty of the measurement of $M_{\star}$ was reduced by adopting this measured disk radius. 
That is because when a radius larger or smaller than the actual disk radius was adopted in the analysis, additional noise or less signal were included. 

We note that there could be cloud absorption in the spectra of our sources, as discussed in \citet{Ansdell2018} and \citet{long2019taurus}. We examined all the spectra analyzed in this work and confirmed that there is no strong absorption in them (Appendix \ref{sec:appendix_lupus_taurus}). In addition, we performed tests on our analysis by arbitrarily masking a range of velocity channels to mimic cloud absorption, and we found that the best-fit results remain unchanged but the uncertainties increase. Thus, we expect that our results are not affected by possible cloud absorption.

\subsection{Error Estimation}
\label{sec:error estimation}
To estimate the uncertainty of our measurements, a series of velocity aligned stacked spectra were generated for each YSO by varying the disk parameters around the measurements. 
To include the uncertainty of the continuum measurements of the inclination and position angles, those parameters were also varied within their uncertainties.
We did not consider the errors in the stellar position and distance, since their contributions are negligible compared to the errors in the inclination and position angles. These test spectra were compared against the best aligned spectrum with the highest S/N. 
We subtracted the test spectra from the best aligned spectrum and computed the $\chi^{2}$-value of the residual. On the assumption that the noise distribution in our spectra follows a Gaussian function, the best aligned and test spectra were considered to be indistinguishable within the uncertainty, when the accumulated probability having a $\chi^{2}$ smaller than the obtained value was 50\%. Then the parameters used to generate these test spectra were also accepted as possible solutions. We defined the error bars as the highest and lowest values of the parameters resulting in a possible solution. This procedure sometimes suggests solutions consisting of very specific combinations of $v_{sys}$ and $M_{\star}$, which are isolated in the parameter space. 
These values were excluded if they had a probability of less than 2.5\% in the number probability distributions of possible solutions.

This error estimation is different from that in \citet{Yen2018}. In \citet{Yen2018}, a Gaussian line profile was fitted to each test spectrum. When the center, peak intensity and width were all consistent with the best aligned spectrum and the difference in the S/Ns of the integrated fluxes was less than $\sqrt{2}$, the test spectrum was claimed to be indistinguishable from the best aligned spectrum. With imaging simulations (Section \ref{sec:robustness and potential bias}), we found that the method in \citet{Yen2018} tends to underestimate the errors, and our error estimate better represents the $1\sigma$ uncertainty of the measurements.

\subsection{Robustness and Potential Bias}
\label{sec:robustness and potential bias}
To test the robustness of our analysis and examine the potential bias due to our assumption of a geometrically thin disk,
we used synthetic images in the $^{13}$CO (3--2) emission of eight disk models with a stellar mass of 1.1~$M_{\odot}$, which were generated by \citet{miotello2016diskmassesDALI} using the physical and chemical code DALI \citep{bruderer2012DALI, bruderer2013DALI}.
The radial density structure was set by the characteristic radius $R_c = 60$~au and the power-law index $\gamma =0.8$. The vertical density structure was varied by using different scale heights ($h_c = 0.1$ and 0.2), and flaring angles ($\psi=$0.1 and 0.2~rad). Several values were adopted for the disk mass (10$^{-1}$, $10^{-3}$, $10^{-5}$ $M_\odot$). More details on the model parameters are given in \citet{miotello2016diskmassesDALI}.
We projected each model at three different inclination angles (20$^\circ$, 45$^\circ$, and 80$^\circ$) and simulated the ALMA observations on the model images to generate synthetic ALMA data cubes.
Then we performed our analysis on these synthetic ALMA data. 
An example of our synthetic data and test is shown in Appendix \ref{sec:appendix_synthetic}.

We found that the stellar mass could be overestimated by 20\% to 30\% with our method, when a disk is highly inclined ($i = 80^{\circ}$).
For face-on disks ($i = 20^{\circ}$), the measured mass could be underestimated by less than 10\%.
For disks with moderate inclination, the measurements obtained from the synthetic data are consistent with the model values. 
This is because the $^{13}$CO (3--2) emission does not originate from the midplane but from an upper layer in a disk. 
This geometrical effect is most significant when a disk is close to face-on or edge-on, 
and our analysis could over- or underestimate the actual distance of the gas in the disk to the center. 
The effect on face-on disks is not notable because the uncertainties of the measurements are larger compared to this effect. 
For the seven YSOs with inclination angles larger than $70^{\circ}$ in our sample, further analysis was done to verify their results (Section \ref{sec:discussion}).

In addition, to test the robustness of our error estimation, we generated 700 synthetic ALMA data cubes from the disk models and included different random noise during the imaging simulations. 
These synthetic data cubes were analyzed with our method, and we obtained the probability distribution of the results. By comparing the distribution of the solutions to the error bars, we found that our error estimate indeed represents the $1\sigma$ uncertainty of our measurements.

\section{Results}
\label{sec:results}

We obtained measurements of the stellar mass for 28 out of 30 sources in the Lupus star forming region and for 17 out of 37 sources in the Taurus region with uncertainties ranging from 0.01$~M_{\odot}$ to 0.33$~M_{\odot}$ ($<$10\% to 30\%), and a mean uncertainty of $0.17~M_{\odot}$ (19\%). 
For the sources for which we couldn't obtain a mass measurement, no alignment of the spectra could be found to meet our requirements (see Section \ref{sec:analysis_stellar_mass}).
The final measurements, using the combination of molecular lines resulting in the highest S/N, are summarised in Tables \ref{tab:final_lupus} and \ref{tab:final_taurus} for the sources in Lupus and Taurus, respectively. The original and best aligned spectra averaged over the disk area from these combined data are shown in Appendix \ref{sec:appendix_lupus_taurus}. The distribution of the measured masses is shown in Appendix \ref{sec:appendix_histograms}.

\begin{deluxetable*}{p{2.8cm}cccc}
\label{tab:final_lupus}
\tablecaption{Final measurements for the sources in the Lupus region}
\tablehead{
\colhead{Name} & \colhead{Molecular Line} & \colhead{$M_{\star}$} & \colhead{$v_{sys}$} & \colhead{Disk Radius} \\
\colhead{} & \colhead{} & \colhead{[$M_{\odot}$]} & \colhead{[\SI{}{\km\per\s}]} & \colhead{[au]}
}
\startdata
EXLup & $^{13}$CO & 0.86$^{+0.36}_{-0.22}$ & 4.15$^{+0.25}_{-0.17}$ & 88 \\ 
J15450887-3417333 & $^{13}$CO+C$^{18}$O & 0.25$\pm$0.12 & 4.84$^{+0.51}_{-0.34}$ & 169\\ 
RXJ1556.1-3655 & $^{13}$CO+C$^{18}$O & 0.75$^{+0.14}_{-0.08}$ & 5.19$^{+0.17}_{-0.59}$ & 106\\ 
J16000236-4222145 & $^{13}$CO+CN+C$^{18}$O & 0.37$\pm$0.04 & 4.09$\pm$0.11 & 244 \\ 
J16083070-3828268 & $^{13}$CO & 1.5$\pm$0.15 & 5.29$^{+0.08}_{-0.25}$ & 284 \\ 
J16085324-3914401 & $^{13}$CO+CN & 0.43$^{+0.11}_{-0.13}$ & 3.72$^{+0.11}_{-0.66}$ &72 \\ 
J16102955-3922144 & $^{13}$CO+C$^{18}$O & 0.19$\pm$0.03 & 3.52$^{+0.09}_{-0.17}$ & 189 \\ 
J16124373-3815031 & $^{13}$CO+CN & 0.77$^{+0.26}_{-0.15}$ & 4.36$^{+0.22}_{-0.44}$ & 90 \\ 
MYLup & $^{13}$CO+CN+C$^{18}$O & 1.49$^{+0.15}_{-0.17}$ & 4.78$^{+0.22}_{-0.44}$ & 187 \\ 
RYLup & $^{13}$CO+C$^{18}$O & 1.5$\pm$0.15 & 3.67$^{+0.68}_{-0.25}$ & 174 \\ 
Sz100 & $^{13}$CO+CN & 0.29$^{+0.04}_{-0.05}$ & 1.76$\pm$0.11 & 92 \\ 
Sz108B & $^{13}$CO+CN+C$^{18}$O & 0.13$^{+0.03}_{-0.05}$ & 2.54$^{+0.11}_{-0.33}$ & 113 \\ 
Sz111 & $^{13}$CO+C$^{18}$O & 0.61$\pm$0.06 & 4.15$^{+0.09}_{-0.17}$ & 203 \\ 
Sz114 & $^{13}$CO+CN+C$^{18}$O & 0.32$^{+0.25}_{-0.14}$ & 4.91$\pm$0.11 & 125 \\ 
Sz123A & $^{13}$CO+C$^{18}$O & 0.46$^{+0.09}_{-0.08}$ & 4.29$^{+0.08}_{-0.34}$ & 122 \\ 
Sz129 & $^{13}$CO+CN+C$^{18}$O & 0.85$^{+0.12}_{-0.14}$ & 4.16$^{+0.11}_{-0.33}$ & 106 \\ 
Sz133 & $^{13}$CO+CN+C$^{18}$O & 1.23$^{+0.15}_{-0.12}$ & 3.92$^{+0.33}_{-0.11}$ & 259 \\ 
Sz65 & $^{13}$CO & 1.14$^{+0.24}_{-0.13}$ & 4.44$^{+0.26}_{-0.34}$ & 161 \\ 
Sz69 & $^{13}$CO & 0.22$\pm$0.12 & 5.5$^{+0.17}_{-0.51}$ & 104 \\ 
Sz71 & $^{13}$CO+CN+C$^{18}$O & 0.46$\pm$0.05 & 3.62$\pm$0.11 & 194 \\ 
Sz73 & $^{13}$CO & 0.86$\pm$0.23 & 4.49$^{+0.08}_{-0.34}$ & 114\\ 
Sz75 & $^{13}$CO & 0.96$\pm$0.1 & 2.76$^{+0.34}_{-0.17}$ & 129 \\ 
Sz76 & $^{13}$CO+C$^{18}$O & 0.14$\pm$0.03 & 3.64$^{+0.09}_{-0.17}$ & 144 \\ 
Sz83 & $^{13}$CO & 1.0$^{+0.63}_{-0.47}$ & 4.68$^{+0.08}_{-0.17}$ & 154 \\ 
Sz84 & $^{13}$CO+C$^{18}$O & 0.96$^{+0.19}_{-0.1}$ & 4.82$^{+0.51}_{-0.26}$ & 158 \\ 
Sz90 & $^{13}$CO+CN & 1.05$^{+0.1}_{-0.45}$ & 5.24$^{+0.44}_{-0.55}$ & 93 \\ 
Sz98 & $^{13}$CO+CN & 1.05$\pm$0.1 & 3.05$^{+0.11}_{-0.22}$ & 206 \\ 
SSTc2dJ160836.2-\newline
    \hspace*{0.3cm}392302 & $^{13}$CO & 1.06$\pm$0.11 & 5.39$\pm$0.08 & 389 \\ 
\enddata
\end{deluxetable*}
\begin{deluxetable}{lcccc}
\label{tab:final_taurus}
\tablecaption{Final measurements for the sources in the Taurus region}
\tablehead{
\colhead{Name} & \colhead{Molecular Line} & \colhead{$M_{\star}$} & \colhead{$v_{sys}$} & \colhead{Disk Radius}\\
\colhead{} & \colhead{} & \colhead{[$M_{\odot}$]} & \colhead{[\SI{}{\km\per\s}]} & \colhead{[au]}
}
\startdata
CIDA9A & $^{13}$CO+C$^{18}$O & 0.78$\pm$0.08 & 6.5$^{+0.08}_{-0.17}$ & 123\\ 
DOTau & $^{13}$CO+C$^{18}$O & 0.54$\pm$0.07 & 6.02$^{+0.08}_{-0.09}$ & 202 \\ 
DQTau & $^{13}$CO+C$^{18}$O & 2.85$^{+0.77}_{-0.72}$ & 9.1$\pm$0.17 & 83 \\ 
DRTau & $^{13}$CO+C$^{18}$O & 1.18$^{+0.59}_{-0.44}$ & 9.9$^{+0.08}_{-0.09}$ & 246 \\ 
DSTau & $^{13}$CO+C$^{18}$O & 1.08$\pm$0.11 & 5.7$\pm$0.17 & 129 \\ 
FTTau & $^{13}$CO+C$^{18}$O & 0.35$^{+0.04}_{-0.06}$ & 7.22$^{+0.09}_{-0.08}$ & 129 \\ 
HKTauA & $^{13}$CO+C$^{18}$O & 0.49$^{+0.06}_{-0.05}$ & 5.98$^{+0.09}_{-0.51}$ & 63 \\ 
HKTauB & $^{13}$CO+C$^{18}$O & 1.23$^{+0.26}_{-0.12}$ & 6.05$^{+0.7}_{-0.37}$ & 69 \\ 
HOTau & $^{13}$CO+C$^{18}$O & 0.44$^{+0.05}_{-0.16}$ & 6.38$^{+0.09}_{-0.51}$ & 102 \\ 
IPTau & $^{13}$CO+C$^{18}$O & 0.8$^{+0.23}_{-0.22}$ & 6.87$^{+0.25}_{-0.6}$ & 49 \\ 
IQTau & $^{13}$CO $J$ = 2--1 & 0.42$^{+0.05}_{-0.15}$ & 5.5$^{+0.34}_{-0.25}$ & 108 \\ 
MWC480 & $^{13}$CO $J$ = 2--1 & 2.16$\pm$0.22 & 5.13$^{+0.09}_{-0.08}$ & 355 \\ 
RWAurA & $^{13}$CO+C$^{18}$O & 1.4$^{+0.28}_{-0.14}$ & 5.84$^{+0.65}_{-0.58}$ & 59 \\ 
TTauN & C$^{18}$O $J$ = 2--1 & 2.06$^{+0.66}_{-0.43}$ & 9.1$^{+0.17}_{-0.42}$ & 152 \\ 
UZTauE & $^{13}$CO $J$ = 2--1 & 1.21$\pm$0.12 & 5.7$^{+0.08}_{-0.17}$ & 140 \\ 
V710Tau & $^{13}$CO+C$^{18}$O & 0.58$^{+0.06}_{-0.07}$ & 6.5$\pm$0.17 & 72 \\ 
V836Tau & $^{13}$CO+C$^{18}$O & 0.92$^{+0.22}_{-0.2}$ & 7.18$^{+0.42}_{-0.43}$ & 71 \\ 
\enddata
\end{deluxetable}

\subsection{Face-on Disks}
\label{sec:faceon_disks}

When a disk is close to face-on, its millimeter continuum intensity distribution could appear nearly circular on the plane of sky, and the difference between its major and minor axes can be hard to detect. Thus, it can be difficult to measure the inclination and position angles of a disk which is nearly face-on with continuum data.
The velocity pattern of a disk carries additional information and provides another way to measure the disk orientation.

In our sample, there are three nearly face-on disks with $i \leq 30^{\circ}$ in the Lupus region, Sz83, Sz113 and Sz114.
In order to exploit the additional information provided by the velocity pattern of the gas, we measured $PA$ and $i$ of these nearly face-on sources using the $^{13}$CO data with our analysis. In addition to the mass and systemic velocity of the YSOs, $PA$ and $i$ were also varied from $0^{\circ}$ to $180^{\circ}$ in steps of $5^{\circ}$ and from $3^{\circ}$ to $30^{\circ}$ in steps of $2^{\circ}$, respectively, to search for the combination to best align their spectra. Then, we followed the same procedure as described above to obtain the measurements, including $i$ and $PA$, and their uncertainties.

Our estimated $i$ and $PA$ for Sz83 deviate from the measurements from the continuum by 13$^{\circ}$ and 44$^{\circ}$, which are 2.2$\sigma$ and 2.7$\sigma$ differences, respectively. The deviations for Sz114 are $<5^{\circ}$ and 11$^{\circ}$ for $i$ and $PA$, respectively, which is less than $1\sigma$ for both quantities. With the disk orientations measured from the molecular line data, we could properly align the spectra, with S/Ns above 17 in Sz83 and Sz114. However, we still could not find any detection in Sz113, even after we considered different disk orientations. 

For the nearly face-on disks in the Taurus region, no significant difference between the orientations measured with continuum or line emission data was found, and the orientations measured with the continuum emission were adopted in our analysis to obtain the measurements of stellar mass and systemic velocity. 
The position and inclination angles measured from the $^{13}$CO data for the nearly face-on sources in the Lupus region are listed in Table \ref{tab:face_on_lupus}.

\begin{deluxetable*}{lccccc}
\label{tab:face_on_lupus}
\tablecaption{Measured parameters for face-on disks in the Lupus region}
\tablehead{
\colhead{Name} & \colhead{RA} & \colhead{Dec} & \colhead{$i$} & \colhead{$PA$} & \colhead{dist} \\
\colhead{} & \colhead{} & \colhead{} & \colhead{[$^{\circ}$]} & \colhead{[$^{\circ}$]} & \colhead{[pc]}
}
\startdata
Sz83 & 15:56:42.29 & -37:49:15.82 & 16.0$^{+6.0}_{-4.0}$ & 120.0$\pm15.0$ & 160 \\
Sz113 & 16:08:57.78 & -39:02:23.21 & $\cdots$ & $\cdots$ & 163 \\
Sz114 & 16:09:01.83 & -39:05:12.79 & 16.0$^{+6.0}_{-4.0}$ & 160.0$^{+25.0}_{-10.0}$ & 162 \\
\enddata
\tablecomments{We measured the inclination and position angles with the molecular-line data for face-on disks whose inclination angles were measured to be $\leq30^{\circ}$ with the continuum data.}
\end{deluxetable*}

\subsection{Dependence on Disk Properties}
As discussed in Section \ref{sec:robustness and potential bias}, when the inclination of a disk is close to face-on or edge-on, our analysis could under- or overestimate the stellar mass. 
Other possible biases could be introduced by larger disk radii, higher disk mass or higher surface density. A flared disk has a larger height at an outer radius, and the opacity of a disk increases with a higher surface density or disk mass. These could cause the emission lines to trace an upper layer of a disk and result in a larger deviation from the geometrically-thin disk approximation. 

\begin{figure*}
    \gridline{\includegraphics[width=0.5\textwidth]{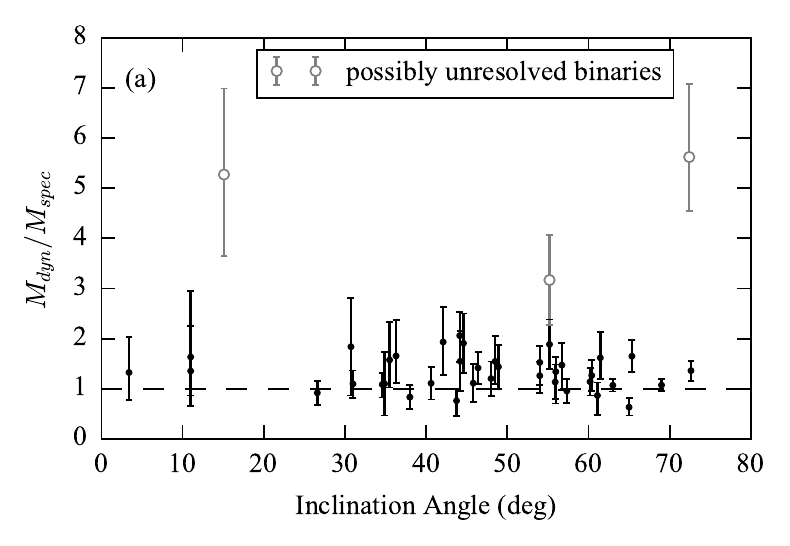}
            \includegraphics[width=0.5\textwidth]{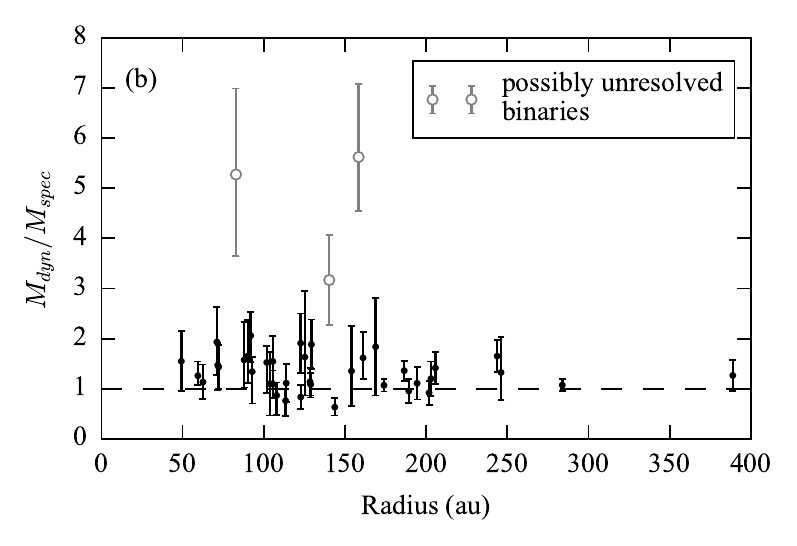}}
    \gridline{\includegraphics[width=0.5\textwidth]{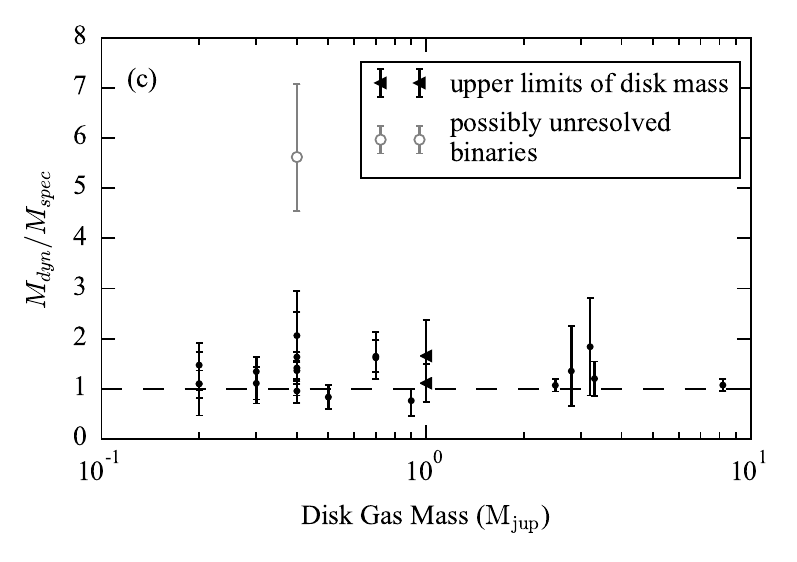}
            \includegraphics[width=0.5\textwidth]{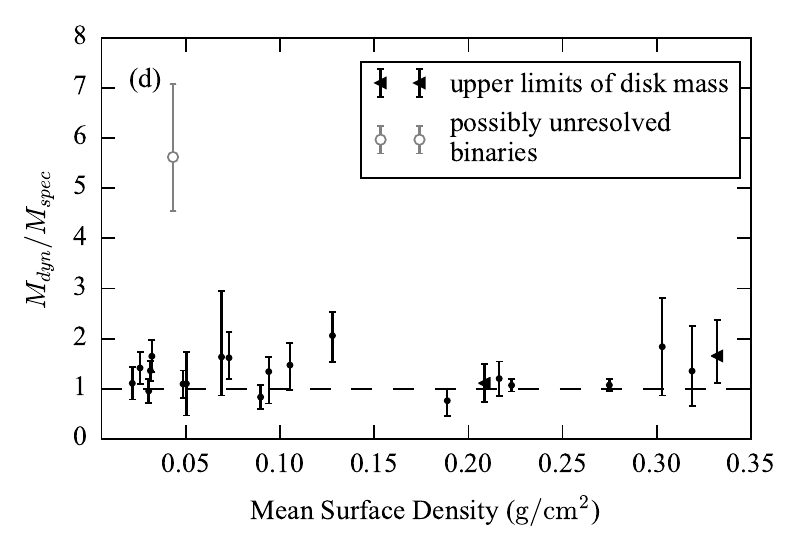}}
    \caption{Ratio between our measurements of dynamical stellar masses $M_{dyn}$ and the spectroscopically determined stellar masses $M_{spec}$ as a function of different disk properties: (a) the inclination angle, (b) the radius, (c) the disk gas mass and (d) the mean surface density. No clear dependence on the disk properties is observable. 
    In Figures (c) and (d) only sources in the Lupus star-forming region are included, for which the disk masses were measured by \citet{Ansdell2016}. 
    The stars, marked with grey circles, are found to be possibly unresolved binaries (see Section \ref{sec:unresolved_binaries}). Triangles indicate upper limits of the disk mass.}
    \label{fig:disk_properties}
\end{figure*}

To examine these possible biases in our measurements, Figure \ref{fig:disk_properties} compares the ratios of our measured dynamical masses to the spectroscopic masses estimated and the stellar evolutionary model by \citet{baraffe2015} with the inclination angles, outer radii, masses, and mean surface densities of the sample disks. 
The outer radius of a disk is defined as the radius enclosing 95\% of the total flux. 
The outer radii of our sample disks were measured in our analysis in Section \ref{sec:analysis_stellar_mass}.
The disk gas masses were measured by \citet{Ansdell2016} using the $^{13}$CO and C$^{18}$O $J$ = 3--2 emission lines.
The mean surface densities were estimated from the disk gas masses and outer radii. 
Because there are no measurements of the disk masses of the Taurus sources available in the literature, we did not include the Taurus sources in the comparison with the disk masses and mean surface densities. 
Details about the stellar evolutionary models are given in Section \ref{sec:discussion}.

There is no clear relation between the ratio of the dynamical and spectroscopic stellar masses and these disk properties in Figure \ref{fig:disk_properties}, suggesting that the effect caused by the geometrically-thin disk approximation is less than the uncertainty of our measurements. 
The comparisons of the disk properties with the ratio of the dynamical and spectroscopic stellar masses were also made for the other stellar evolutionary models mentioned in Section \ref{sec:discussion}. No clear relation was found for any of the models.
Therefore, our comparison and discussion on the dynamical and spectroscopic masses in Section \ref{sec:discussion} are not biased by these effects. 
However, since the last data points at high inclination angles in Figure \ref{fig:disk_properties}a hint at an overestimation of stellar mass, which was also found with the synthetic data, our measurements of the disks with inclination angles larger than 70$^{\circ}$ should be interpreted with caution.

\subsection{Measurements with Individual Lines}
\label{sec:measurements_indiv_lines}
\begin{figure*}
    \centering
    \includegraphics{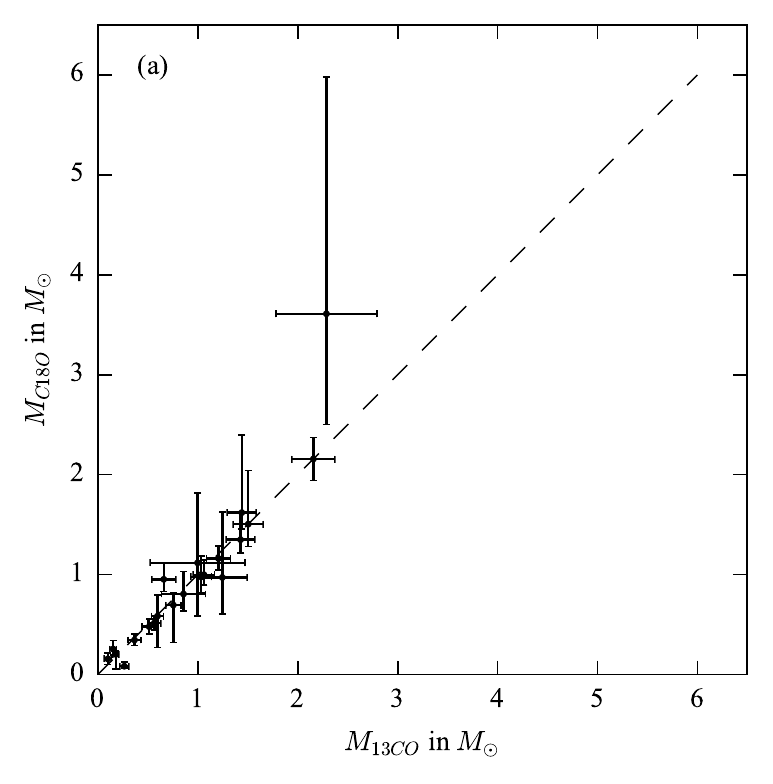}
    \includegraphics{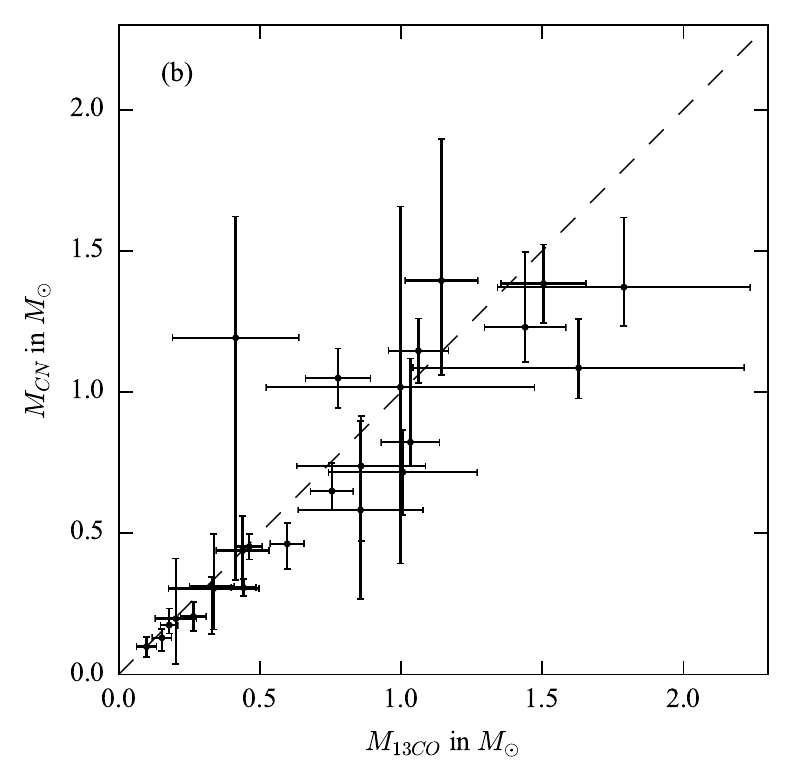}
    \includegraphics{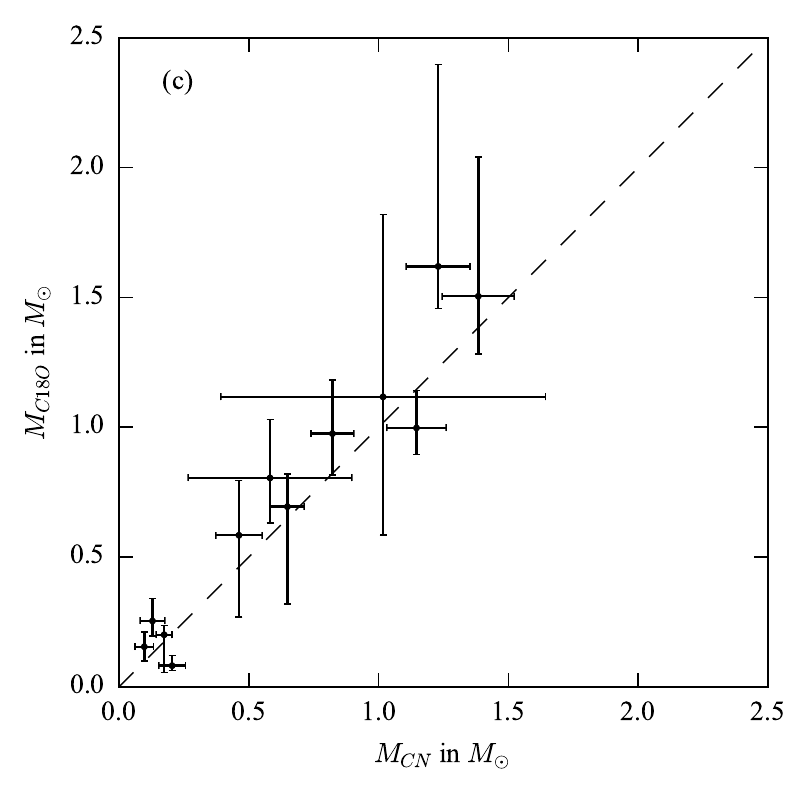}
    \caption{Comparison between the results of the mass measurement using different molecular lines. Comparison of the measurements obtained with (a) the $^{13}$CO and C$^{18}$O data, (b) the $^{13}$CO and CN data and (c) the CN and C$^{18}$O data.}
    \label{fig:comparison_indiv_lines}
\end{figure*}

We obtained 42 and 22 measurements out of the total sample of 67 YSOs using the $^{13}$CO and C$^{18}$O data, respectively. Furthermore, we obtained 26 measurements of the stellar mass with the CN data for the sources in the Lupus region. For HOTau we could only obtain a measurement using the stacked $^{13}$CO+C$^{18}$O data. The results of the measurements using the individual line data for the Lupus region are shown in Table \ref{tab:indv_line_lupus}, and in Table \ref{tab:indv_line_taurus} for the sources in the Taurus region. The results obtained with the different molecular lines are compared in the graphs displayed in Figure \ref{fig:comparison_indiv_lines}. 80\% of the measurements from different line data agree on a level better than 1$\sigma$ and 16\% agree on a level of 1 to 2$\sigma$. Only 4\% of the measurements show deviations on a level between 2$\sigma$ and 3.5$\sigma$. On average, the deviations are within a 0.5$\sigma$ level. Therefore, our measurements obtained with data of different molecular lines do not show significant deviations. This demonstrates that possible differences in emitting regions of these molecular lines in a disk are negligible in our analysis, and stacking different molecular line data does not introduce a bias to our analysis and can increase the S/Ns of our measurements.

\begin{deluxetable*}{l|cc|cc|cc}
\label{tab:indv_line_lupus}
\tablecaption{Measurements of the individual line data for the sources in the Lupus region}
\tablehead{
\colhead{Name} & \multicolumn{2}{|c}{$^{13}$CO} & \multicolumn{2}{|c}{C$^{18}$O} & \multicolumn{2}{|c}{CN} \\ \hline
\colhead{} & \multicolumn{1}{|c}{$M_{\star}$} & \colhead{$v_{sys}$} & \multicolumn{1}{|c}{$M_{\star}$} & \colhead{$v_{sys}$} & \multicolumn{1}{|c}{$M_{\star}$} & \colhead{$v_{sys}$}\\
\colhead{} & \multicolumn{1}{|c}{[$M_{\odot}$]} & \colhead{[\SI{}{\km\per\s}]} & \multicolumn{1}{|c}{[$M_{\odot}$]} & \multicolumn{1}{c}{[\SI{}{\km\per\s}]} & \multicolumn{1}{|c}{[$M_{\odot}$]} & \colhead{[\SI{}{\km\per\s}]}
}
\startdata
EXLup & 0.86$^{+0.36}_{-0.22}$ & 4.15$^{+0.25}_{-0.17}$ & 0.81$^{+0.23}_{-0.17}$ & 3.64$\pm$0.25 & 0.58$\pm$0.31 & 4.59$^{+0.11}_{-0.99}$ \\ 
J15450887-3417333 & 0.2$^{+0.1}_{-0.07}$ & 4.84$^{+0.34}_{-0.17}$ & \nodata & \nodata  & 0.2$^{+0.21}_{-0.16}$ & 4.62$^{+0.22}_{-0.55}$ \\ 
RXJ1556.1-3655 & 0.75$^{+0.14}_{-0.08}$ & 5.19$^{+0.17}_{-0.51}$ & 0.69$^{+0.13}_{-0.38}$ & 5.19$^{+0.08}_{-0.77}$ & 0.65$^{+0.1}_{-0.06}$ & 4.75$^{+0.55}_{-0.22}$ \\ 
J16000236-4222145 & 0.44$\pm$0.04 & 4.01$^{+0.09}_{-0.08}$ & \nodata & \nodata  & 0.31$\pm$0.03 & 4.09$\pm$0.11 \\ 
J16083070-3828268 & 1.5$\pm$0.15 & 5.29$^{+0.08}_{-0.25}$ & 1.5$^{+0.54}_{-0.22}$ & 4.86$^{+0.68}_{-0.34}$ & 1.38$\pm$0.14 & 5.18$^{+0.11}_{-0.33}$ \\ 
J16085324-3914401 & 0.33$^{+0.12}_{-0.08}$ & 3.63$^{+0.58}_{-0.49}$ & \nodata & \nodata  & 0.31$^{+0.03}_{-0.17}$ & 3.38$^{+0.11}_{-0.99}$ \\ 
J16102955-3922144 & 0.18$^{+0.05}_{-0.03}$ & 3.6$^{+0.08}_{-0.17}$ & 0.2$^{+0.04}_{-0.14}$ & 3.86$^{+0.34}_{-0.59}$ & 0.17$^{+0.06}_{-0.03}$ & 3.52$^{+0.22}_{-0.33}$ \\ 
J16124373-3815031 & \nodata & \nodata  & \nodata & \nodata  & 0.66$^{+0.2}_{-0.12}$ & 4.58$^{+0.22}_{-0.77}$ \\ 
MYLup & 1.79$\pm$0.45 & 4.86$^{+0.46}_{-0.57}$ & \nodata & \nodata  & 1.37$^{+0.25}_{-0.14}$ & 4.78$^{+0.22}_{-0.55}$ \\ 
RYLup & 1.44$^{+0.17}_{-0.14}$ & 3.84$^{+0.42}_{-0.34}$ & 1.62$^{+0.78}_{-0.16}$ & 4.18$^{+1.72}_{-1.15}$ & 1.23$^{+0.27}_{-0.12}$ & 3.45$^{+0.44}_{-0.55}$ \\ 
Sz100 & 0.26$^{+0.04}_{-0.05}$ & 1.76$\pm$0.17 & 0.08$^{+0.04}_{-0.02}$ & 1.6$^{+0.17}_{-0.43}$ & 0.21$\pm$0.05 & 1.88$^{+0.11}_{-0.22}$ \\ 
Sz108B & 0.1$\pm$0.04 & 2.46$^{+0.08}_{-0.17}$ & 0.16$\pm$0.06 & 2.46$^{+0.25}_{-0.17}$ & 0.1$\pm$0.04 & 2.54$^{+0.22}_{-0.33}$ \\ 
Sz111 & 0.6$\pm$0.06 & 4.15$^{+0.08}_{-0.17}$ & 0.58$^{+0.21}_{-0.32}$ & 4.32$^{+0.43}_{-0.59}$ & 0.46$^{+0.07}_{-0.09}$ & 3.93$^{+0.11}_{-0.22}$ \\ 
Sz114 & 0.34$^{+0.21}_{-0.16}$ & 4.82$\pm$0.08 & \nodata & \nodata  & 0.3$^{+0.19}_{-0.15}$ & 4.58$\pm$0.11 \\ 
Sz123A & 0.44$^{+0.1}_{-0.09}$ & 4.29$^{+0.08}_{-0.34}$ & \nodata & \nodata  & 0.44$^{+0.12}_{-0.13}$ & 3.96$\pm$0.22 \\ 
Sz129 & 1.01$^{+0.2}_{-0.26}$ & 4.16$^{+0.25}_{-0.42}$ & \nodata & \nodata  & 0.72$\pm$0.15 & 4.16$^{+0.11}_{-0.33}$ \\ 
Sz133 & 1.63$\pm$0.59 & 3.66$^{+0.37}_{-0.75}$ & \nodata & \nodata  & 1.09$^{+0.17}_{-0.11}$ & 3.92$^{+0.44}_{-0.11}$ \\ 
Sz65 & 1.14$^{+0.24}_{-0.13}$ & 4.44$^{+0.26}_{-0.34}$ & \nodata & \nodata  & 1.39$^{+0.5}_{-0.33}$ & 4.56$^{+0.47}_{-0.62}$ \\ 
Sz69 & 0.22$\pm$0.12 & 5.5$^{+0.17}_{-0.51}$ & \nodata & \nodata  & \nodata & \nodata  \\ 
Sz71 & 0.46$\pm$0.05 & 3.62$\pm$0.08 & \nodata & \nodata  & 0.45$\pm$0.05 & 3.52$\pm$0.11 \\ 
Sz73 & 0.86$\pm$0.23 & 4.49$^{+0.08}_{-0.34}$ & \nodata & \nodata  & 0.74$^{+0.18}_{-0.27}$ & 3.72$\pm$0.33 \\ 
Sz75 & 0.96$\pm$0.1 & 2.76$^{+0.34}_{-0.17}$ & \nodata & \nodata  & \nodata & \nodata  \\ 
Sz76 & 0.15$\pm$0.03 & 3.64$^{+0.08}_{-0.17}$ & 0.25$^{+0.09}_{-0.06}$ & 3.81$^{+0.34}_{-0.43}$ & 0.13$^{+0.03}_{-0.05}$ & 3.53$^{+0.11}_{-0.22}$ \\ 
Sz83 & 1.0$^{+0.63}_{-0.47}$ & 4.68$^{+0.08}_{-0.17}$ & 1.12$^{+0.7}_{-0.53}$ & 4.6$^{+0.08}_{-0.42}$ & 1.02$^{+0.64}_{-0.62}$ & 4.35$\pm$0.22 \\ 
Sz84 & 1.03$^{+0.18}_{-0.1}$ & 5.16$^{+0.25}_{-0.51}$ & 0.98$^{+0.21}_{-0.16}$ & 4.82$^{+0.34}_{-0.42}$ & 0.82$^{+0.3}_{-0.08}$ & 5.04$^{+0.33}_{-0.55}$ \\ 
Sz90 & 0.41$^{+0.04}_{-0.22}$ & 5.5$^{+0.6}_{-0.25}$ & \nodata & \nodata  & 1.19$^{+0.43}_{-0.86}$ & 5.24$^{+0.77}_{-0.44}$ \\ 
Sz98 & 0.78$^{+0.14}_{-0.11}$ & 2.96$\pm$0.08 & \nodata & \nodata  & 1.05$\pm$0.1 & 3.05$^{+0.11}_{-0.33}$ \\ 
SSTc2dJ160836.2-392302 & 1.06$\pm$0.11 & 5.39$\pm$0.08 & 1.0$^{+0.14}_{-0.1}$ & 5.3$^{+0.17}_{-0.08}$ & 1.15$\pm$0.11 & 5.5$\pm$0.11 \\ 
\enddata
\end{deluxetable*}
\begin{deluxetable}{l|cc|cc}
\label{tab:indv_line_taurus}
\tablecaption{Measurements of the individual line data for the sources in the Taurus region}
\tablehead{
\colhead{Name} & \multicolumn{2}{|c}{$^{13}$CO $J$ = 2--1} & \multicolumn{2}{|c}{C$^{18}$O $J$ = 2--1} \\ \hline
\colhead{} & \multicolumn{1}{|c}{$M_{\star}$} & \colhead{$v_{sys}$} & \multicolumn{1}{|c}{$M_{\star}$} & \colhead{$v_{sys}$} \\
\colhead{} & \multicolumn{1}{|c}{[$M_{\odot}$]} & \colhead{[\SI{}{\km\per\s}]} & \multicolumn{1}{|c}{[$M_{\odot}$]} & \colhead{[\SI{}{\km\per\s}]}
}
\startdata
CIDA9A & 0.78$\pm$0.08 & 6.50$^{+0.09}_{-0.17}$  & 0.95$^{+0.17}_{-0.12}$ & 6.33$\pm{0.17}$  \\ 
DOTau & 0.51$\pm$0.07 & 6.02$^{+0.08}_{-0.09}$ & 0.48$^{+0.08}_{-0.07}$ & 6.02$^{+0.08}_{-0.09}$ \\ 
DQTau & 2.29$^{+0.72}_{-0.51}$ & 9.1$\pm$0.34 & 3.61$^{+2.37}_{-1.11}$ & 9.27$^{+0.6}_{-0.17}$ \\ 
DRTau & 1.24$\pm$0.25 & 9.9$^{+0.08}_{-0.09}$ & 0.97$^{+0.65}_{-0.36}$ & 9.9$^{+0.08}_{-0.09}$ \\ 
DSTau & 0.99$\pm$0.1 & 5.7$\pm$0.17 & \nodata & \nodata  \\
FTTau & 0.37$^{+0.04}_{-0.06}$ & 7.22$^{+0.09}_{-0.08}$ & 0.34$\pm$0.06 & 7.22$^{+0.09}_{-0.17}$ \\ 
HKTauA & 0.52$^{+0.05}_{-0.06}$ & 5.98$^{+0.09}_{-0.34}$ & \nodata & \nodata  \\
HKTauB & 1.18$^{+0.29}_{-0.12}$ & 6.9$^{+0.26}_{-0.59}$ & \nodata & \nodata\\
HOTau & \nodata & \nodata  & \nodata & \nodata  \\ 
IPTau & 0.4$^{+0.13}_{-0.05}$ & 7.04$^{+0.08}_{-0.51}$ & \nodata & \nodata  \\ 
IQTau & 0.42$^{+0.05}_{-0.15}$ & 5.5$^{+0.34}_{-0.25}$ & \nodata & \nodata  \\ 
MWC480 & 2.16$\pm$0.22 & 5.13$^{+0.09}_{-0.08}$ & 2.16$\pm$0.22 & 5.04$^{+0.09}_{-0.08}$ \\ 
RWAurA & 1.43$^{+0.2}_{-0.14}$ & 5.5$^{+0.46}_{-0.38}$ & 1.35$^{+0.28}_{-0.13}$ & 6.86$^{+0.38}_{-0.76}$ \\ 
TTauN & \nodata & \nodata  & 2.06$^{+0.66}_{-0.43}$ & 9.1$^{+0.17}_{-0.42}$ \\ 
UZTauE & 1.21$\pm$0.12 & 5.7$^{+0.08}_{-0.17}$ & 1.16$\pm$0.12 & 5.7$^{+0.25}_{-0.17}$ \\ 
V710Tau & 0.56$\pm$0.06 & 6.5$^{+0.17}_{-0.25}$ & 0.51$\pm$0.07 & 6.5$^{+0.08}_{-0.17}$ \\ 
V836Tau & 0.66$^{+0.2}_{-0.12}$ & 7.18$^{+0.42}_{-0.43}$ & \nodata & \nodata\\
\enddata
\end{deluxetable}

\subsection{Comparison with Literature}
\citet{Yen2018} applied the same analysis on the $^{13}$CO and C$^{18}$O $J$ = 3--2 data obtained with the same ALMA observations. The Band 6 data were not available yet. 
By stacking the $^{13}$CO and C$^{18}$O (3--2 \& 2--1) and CN data, we obtained six more measurements, compared to \citet{Yen2018}. Although the new error estimation is less restrictive than the one used in \citet{Yen2018}, the uncertainty decreased on average by 18\%. Sz68, which was detected in \citet{Yen2018}, could not be detected in this work with higher S/N data, and thus it could be a false detection in \citet{Yen2018}. For the sources detected in both works, all measurements are consistent within the 2$\sigma$ uncertainty, except Sz84, for which the deviation is 3.4$\sigma$. 

\begin{figure}
    \centering
    \includegraphics{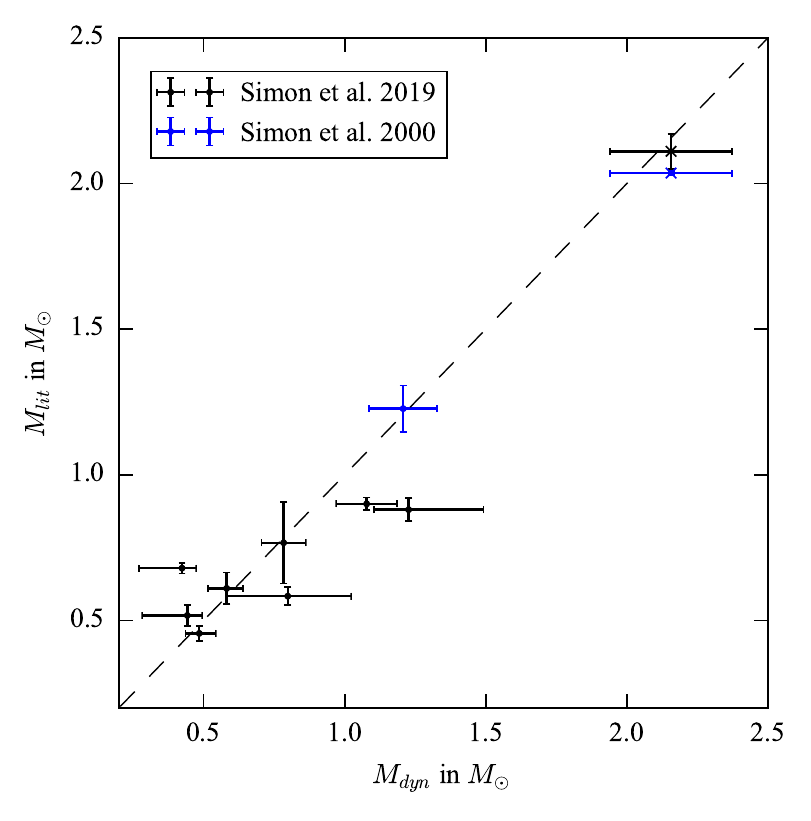}
    \caption{Comparison of our results $M_{dyn}$ and the dynamical mass measurements in the literature $M_{lit}$ \citep{simon2000dynamical, simon2019masses}. The data points marked with crosses denote MW480, which was measured in both papers.}
    \label{fig:Comparison_Simon2019_VAST}
\end{figure}

Eleven of our detected sources were also studied by \citet{simon2000dynamical,simon2019masses}. 
In those studies, disk models were fitted to the velocity channel maps in the visibility domain to measure the dynamical stellar mass. The results depend on the adopted inclination angle and distance. To compare our results with the literature, the measured stellar masses in the literature were scaled with the distances and inclination angles adopted in our analysis. The relation of stellar mass $M_{\star}$, inclination angle $i$ and distance $D$ is as follows \citep{simon2000dynamical}:
\begin{equation}
    M_{\star} \propto \frac{D}{\sin^2i}
\end{equation}
The stellar masses obtained from the literature after scaling are listed in Table \ref{tab:Simon2019+2000}.  
For seven out of the eleven (64\%) sources, our results agree with the literature on a level better than $1\sigma$. There are two and one YSOs with their results consistent with the literature within 2$\sigma$ and 3$\sigma$, respectively. The only exception is IQTau, where our estimated stellar mass is 50\% different from the literature, which is 4.7$\sigma$ difference. The comparisons of our results with the literature are displayed in Figure \ref{fig:Comparison_Simon2019_VAST}. By calculating the percental deviations of our results and the results of \citet{simon2000dynamical,simon2019masses}, we found an average deviation of $0.33_{-7.5}^{+4.6}$\%. Thus, our analysis on the data with much lower S/Ns to measure the stellar mass dynamically yields consistent results.

\begin{deluxetable}{lccccc}
\label{tab:Simon2019+2000}
\tablecaption{Dynamically determined stellar masses obtained from the literature and re-scaled values}
\tablehead{
\colhead{Name} & \colhead{$M_{pub}$} & \colhead{$i_{pub}$} &\colhead{dist$_{pub}$} & \colhead{$M_{\star}$} & \colhead{Ref.} \\
\colhead{} & \colhead{[$M_{\odot}$]} & \colhead{[$^{\circ}$]} & \colhead{[pc]} & \colhead{[$M_{\odot}$]} &\colhead{}
}
\startdata
\multicolumn{6}{l}{\citet{simon2000dynamical}} \\
MWC480 & 1.65$\pm$0.07 & 38 & 140 & 2.035$\pm$0.007 & \\
UZTauE & 1.31$\pm$0.08 & 56 & 140 & 1.23$\pm$0.08 & \\ \hline
\multicolumn{6}{l}{\citet{simon2019masses}} \\
IPTau   & 0.94$\pm$0.05 & 34 & 130 & 0.58$\pm$0.03 & 1 \\
IQTau   & 0.74$\pm$0.02 & 57.9 & 131 & 0.68$\pm$0.02 & 2 \\
HKTauA  & 0.53$\pm$0.03 & 51 & 133 & 0.46$\pm$0.03 & 1 \\
HKTauB  & 0.89$\pm$0.04 & 81 & 133 & 0.88$\pm$0.04 & 1 \\
V710Tau & 0.67$\pm$0.06 & 46 & 142 & 0.61$\pm$0.05 & 1 \\
HOTau   & 0.43$\pm$0.03 & 64 & 161 & 0.52$\pm$0.04 & 1 \\
DSTau   & 0.83$\pm$0.02 & 71 & 158 & 0.9$\pm$0.02 & 1 \\
MWC480  & 2.11$\pm$0.06 & \nodata & 161& 2.11$\pm$0.06 &  3 \\
CIDA9A  & 1.32$\pm$0.24 & 33 & 171 & 0.77$\pm$0.14 & 1 \\
\enddata
\tablecomments{$M_{pub}$, $i_{pub}$, dist$_{pub}$ are dynamical stellar mass, inclination angle, and distance in the literature, respectively. $M_\star$ is the stellar mass, scaled to have the same inclination angle and distance as those in our study. References for the inclination angles adopted in \citet{simon2019masses}: 1) \citet{simon2017dynamical}, 2) \citet{guilloteau2014}, 3) the originally adopted inclination angle could not be found in the literature.}
\end{deluxetable}

\section{Discussion}
\label{sec:discussion}

\subsection{Unresolved Binaries}
\label{sec:unresolved_binaries}
In order to examine the stellar evolutionary models, we obtained the dynamical stellar masses of 45 YSOs in the Lupus and Taurus star-forming regions. By comparing their dynamically and spectroscopically determined stellar masses, we can assess the accuracy of the stellar evolutionary models.
We note that such a comparison can be biased if unresolved binaries are included in our sample.
In the case of an unresolved binary system, the dynamical mass measurement includes the total mass of both stars.
The estimated spectroscopic mass depends on how the temperature and luminosity of the stars add up. The measured luminosity of an unresolved binary system is the combined luminosity of both stars, while the spectral types, and therefore the effective temperatures of the stars, do not add up directly. When fitting the binary system's spectrum with a template spectrum of a single star, the typical error in the estimated effective temperature of the primary star is around \SI{300}{K}, and the error depends on the mass ratio of primary and secondary stars \citep{el2018unresolved_binaries}. Since the evolutionary tracks for low mass stars are close to vertical in the HR diagram, the errors in the effective temperature directly transfer to be the errors in the derived stellar masses.
Thus, the dynamical mass can be significantly higher than the spectroscopic mass for unresolved binaries, in addition to the inaccuracies in the stellar evolutionary model.
In order to identify possible unresolved binaries in our sample, we first divided the effective temperatures in our sample into five ranges and calculated the median dynamical stellar mass per section. If the dynamical mass of a star in our sample deviates from the median mass by more than three times the median absolute deviation (mad), we consider this source a possible unresolved binary system. 

\begin{figure}[htb!]
    \centering
    \includegraphics[width=1.1\linewidth]{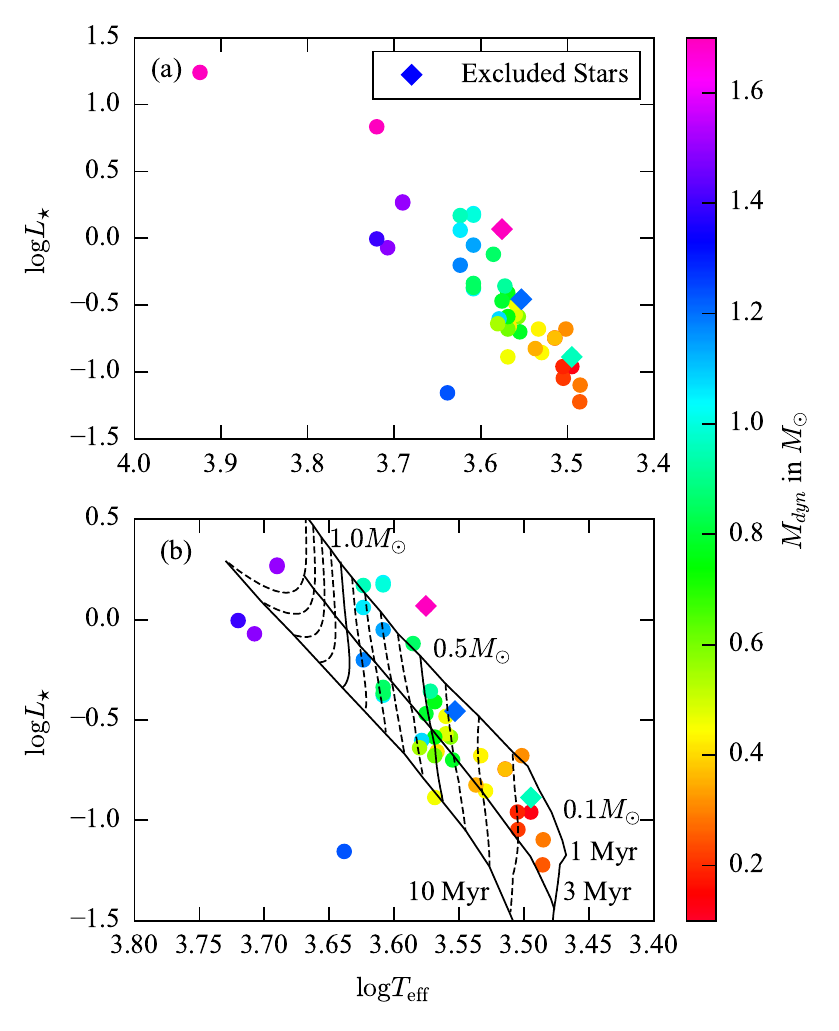}
    \caption{Locations of our sample sources in the HR diagram. Colours of the symbols present their dynamical stellar masses. Diamonds denote the candidates of unresolved binaries. In (a), the complete sample is shown, and (b) enlarges the region of $\sim$\SIrange{2500}{6300}{K} and shows exemplarily the 1, 3 and 10~Myr isochrones of the evolutionary model of \citet{baraffe2015}.}
    \label{fig:HR_diagram}
\end{figure}

In Figure \ref{fig:HR_diagram}, the locations of our sample sources in the HR diagram are shown. 
Three sources in our sample were found to have a large deviation from the median stellar mass. Sz84 deviates by \SI{11.0}{mad} from the median stellar mass, which corresponds to a deviation of 278\%. This object is highly inclined ($i = 73.99 \pm 1.56$), which might cause an additional overestimation of the stellar mass (see Section \ref{sec:robustness and potential bias}). UZTauE and DQTau deviate by \SI{3.7}{mad} (103\%) and \SI{13.6}{mad} (379\%), respectively. These candidates of unresolved binaries are marked with diamond shaped markers in Figure \ref{fig:HR_diagram}. Indeed, we note that other observations in the literature have confirmed that UZTauE and DQTau are binary systems \citep{prato2002UZTauEbinary,mathieu1997DQTaubinary}. Thus, these three sources are excluded in the following comparison of the spectroscopic and dynamical mass measurements.

\subsection{Comparison between Dynamical and Spectroscopic Mass Measurements}
\label{sec:comparison_Mspec}

We compare our measured dynamical stellar masses with spectroscopic stellar masses estimated with seven different stellar evolutionary models, developed by \citet{palla1999star,palla1993pre}, \citet{siess2000}, \citet{bressan2012parsec}, \citet{chen2014improvingPARSEC} and \citet{baraffe2015}, hereafter \citetalias{palla1999star}, \citetalias{siess2000}, \citetalias{bressan2012parsec}, \citetalias{chen2014improvingPARSEC} and \citetalias{baraffe2015}, respectively, as well as by \citet{feiden2016magnetic}. \citet{feiden2016magnetic} includes two evolutionary models, one without and one with the influence of magnetic fields, hereafter \citetalias{feiden2016magnetic} and the magnetic \citetalias{feiden2016magnetic}, respectively.
The spectroscopic masses were calculated using the values given in Table \ref{tab:parameter_lupus} and \ref{tab:parameter_taurus}. For \citetalias{palla1999star} and \citetalias{siess2000} an uncertainty of 15\% was adopted. 
For the other evolutionary models the uncertainty was estimated by calculating the probability distribution of the derived stellar mass, considering the errors of luminosity and effective temperature. The mass range around the median containing 68\% of the distribution was adopted as the uncertainty.
The typical uncertainty of 20\% in luminosity and 2.5\% for an effective temperature $T_{\mathrm{eff}}<3500$~K and 4.5\% for $T_{\mathrm{eff}}>3500$~K was adopted in the calculations. 
The mean uncertainty of the spectroscopic mass in our sample is 0.1~$M_{\odot}$. All calculated values for the spectroscopic models are given in Table \ref{tab:Mspec_lupus} and Table \ref{tab:Mspec_taurus} in Appendix \ref{sec:appendix_Mspec}.
As already noted in several works \citep[e.g.,][]{alcala2014xshootersurvey, alcala2017xshootersurveyMspec}, a small number of stars are found on the HR diagram to be well below the \SI{10}{Myr} isochrone. Several reasons can lead to these effects, but most probably this is due to dust obscuration by an edge-on disk \citep{alcala2014xshootersurvey, alcala2017xshootersurveyMspec}

The adopted stellar evolutionary models are often used in the literature (e.g. \citet{hillenbrand2004, stassun2014empirical,herczeg2015empiricalHR,manara2017XshootersurveyMspec}). These models differ in the adopted surface boundary conditions and equation of state. \citetalias{chen2014improvingPARSEC}, \citetalias{feiden2016magnetic} and \citetalias{baraffe2015} adopted non-gray boundary conditions, while \citetalias{palla1999star} and \citetalias{bressan2012parsec} adopted gray boundary conditions. Gray boundary conditions assume constant opacity and neglect the frequency dependence. This approximation can be used for higher mass stars, but causes inaccuracies for lower mass stars. \citetalias{siess2000} adopted semi-gray boundary conditions, which is an analytical fit to the thermal structure of a non-grey atmosphere. All models use the mixing length theory to describe convection, but adopt different mixing length parameters.

\begin{figure*}[htb!]
    \centering
    \includegraphics[width=\linewidth]{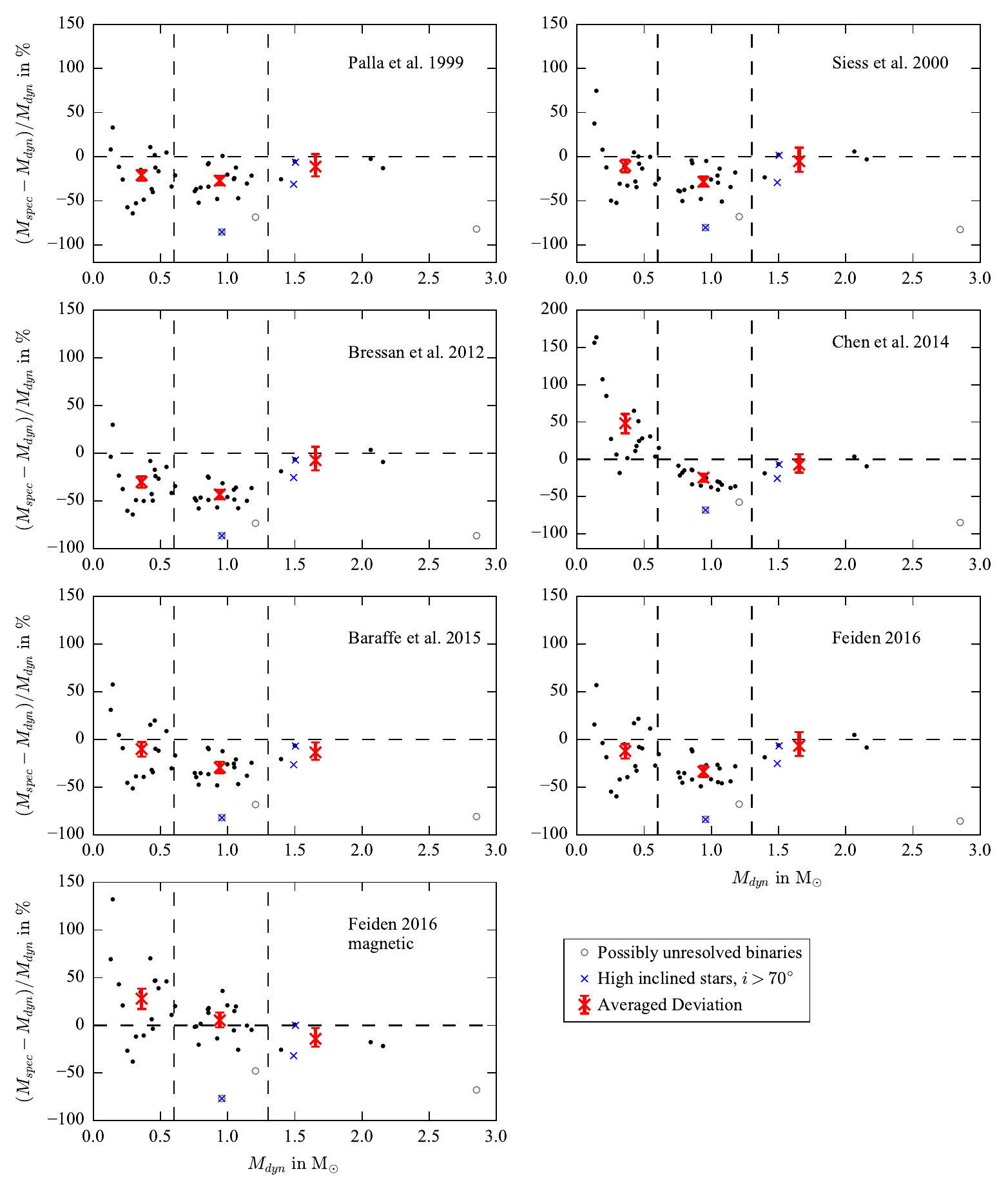}
    \caption{Deviation between the spectroscopic and dynamical stellar masses. The name of the model is labelled in the upper right corner of each diagram. Red crosses mark the mean deviations of the measurements in the respective mass ranges, and dashed vertical lines denote the three mass ranges. The calculated mean deviations are shown in Table \ref{tab:comparison_Mvast_Mspec_highinclined-excluded}. The stars, marked with grey circles, are candidates of unresolved binaries and are excluded from the comparisons. Stars marked with blue crosses are highly inclined sources ($i>70^{\circ}$). The error bars for individual sources are not plotted here for clarity. The error bars range from 3\% to 53\% for all models except for the magnetic \citetalias{feiden2016magnetic} and \citetalias{chen2014improvingPARSEC}, for which the uncertainties go up to 80\% and 108\%, respectively. The shown uncertainties in the averaged deviations were calculated by propagating the uncertainties of the individual sources.}
    \label{fig:comparison_Mvast_Mspec}
\end{figure*}

Figure \ref{fig:comparison_Mvast_Mspec} presents the difference between the spectroscopic and dynamical stellar masses in ratio for different stellar evolutionary models. For individual stars, we found that the spectroscopic measurements are consistent with the dynamical masses within $3\sigma$ for most stars. No star shows a difference between spectroscopic and dynamical masses more than $5\sigma$, except for the stars which are possibly unresolved binaries (see Section \ref{sec:unresolved_binaries}). Nevertheless, possible systematical trends are seen in Figure \ref{fig:comparison_Mvast_Mspec}. To examine if there is any significant trend between the spectroscopic and dynamical masses in our sample, we separated our sample sources into three different mass ranges, low ($M_{\star}\leq 0.6~M_{\odot}$), intermediate ($0.6~M_{\odot}\leq M_{\star} \leq 1.3~M_{\odot}$) and high ($M_{\star}\geq 1.3~M_{\odot}$) mass stars. 
These three mass ranges contain 17, 19 and 6 stars, respectively. The mass ranges are chosen based on Figure \ref{fig:comparison_Mvast_Mspec} and possibly exhibit the largest deviations between spectroscopic and dynamical masses for most of the stellar evolutionary models. Then we computed the mean differences between the spectroscopic and dynamical masses in these mass ranges as well as the uncertainties of the mean differences. The uncertainties of the mean differences were calculated by propagating the errors of the individual measurements of dynamical and spectroscopic masses. The results are listed in Table \ref{tab:comparison_Mvast_Mspec}.

\begin{deluxetable}{lccc}
\label{tab:comparison_Mvast_Mspec}
\tablecaption{Mean deviations between the spectroscopic and dynamical measurements in the respective mass ranges, including stars with $i>70^{\circ}$}
\tablehead{
\colhead{} & \colhead{0--0.6 $M_{\odot}$} & \colhead{0.6--1.3 $M_{\odot}$} & \colhead{1.3--3.0 $M_{\odot}$} \\
\colhead{Model} & \colhead{Deviation [\%]} & \colhead{Deviation [\%]} & \colhead{Deviation [\%]}
}
\startdata
\citetalias{palla1999star} & -21$\pm$6 & -27$^{+6}_{-5}$ & -14$^{+10}_{-8}$ \\
\citetalias{siess2000} & -11$\pm7$ & -28$^{+6}_{-5}$ & -9$^{+10}_{-8}$\\
\citetalias{bressan2012parsec} & -30$\pm6$ & -43$\pm5$ & -11$^{+9}_{-7}$ \\
\citetalias{chen2014improvingPARSEC} & 48$^{+12}_{-13}$ & -25$^{+6}_{-5}$ & -11$^{+9}_{-7}$ \\
\citetalias{baraffe2015} & -10$\pm8$ & -30$\pm6$ & -15$^{+6}_{-5}$ \\
\citetalias{feiden2016magnetic} & -12$^{+7}_{-8}$ & -34$\pm6$ & -11$^{+9}_{-7}$ \\
magnetic \citetalias{feiden2016magnetic} & 28$^{+10}_{-11}$ & 5$^{+8}_{-7}$ & -15$^{+8}_{-6}$ \\
\enddata
\tablecomments{Negative values for the mean deviations where $M_{spec}<M_{dyn}$}
\end{deluxetable}
\begin{deluxetable}{lccc}
\label{tab:comparison_Mvast_Mspec_highinclined-excluded}
\tablecaption{Mean deviations between the spectroscopic and dynamical measurements in the respective mass ranges, excluding stars with $i>70^{\circ}$}
\tablehead{
\colhead{} & \colhead{0--0.6 $M_{\odot}$} & \colhead{0.6--1.3 $M_{\odot}$} & \colhead{1.3--3.0 $M_{\odot}$} \\
\colhead{Model} & \colhead{Deviation [\%]} & \colhead{Deviation [\%]} & \colhead{Deviation [\%]}
}
\startdata
\citetalias{palla1999star} & -21$\pm6$ & -27$^{+6}_{-5}$ & -11$^{+14}_{-11}$ \\
\citetalias{siess2000} & -11$\pm7$ & -28$^{+6}_{-5}$ & -5$^{+15}_{-12}$\\
\citetalias{bressan2012parsec} & -30$\pm6$ & -43$\pm5$ & -7$^{+14}_{-10}$ \\
\citetalias{chen2014improvingPARSEC} & 48$^{+12}_{-13}$ & -25$^{+6}_{-5}$ & -7$^{+14}_{-10}$ \\
\citetalias{baraffe2015} & -10$\pm8$ & -30$\pm6$ & -14$^{+10}_{-8}$ \\
\citetalias{feiden2016magnetic} & -12$^{+7}_{-8}$ & -34$\pm6$ & -7$^{+14}_{-11}$ \\
magnetic \citetalias{feiden2016magnetic} & 28$^{+10}_{-11}$ & 5$^{+8}_{-7}$ & -14$^{+11}_{-8}$ \\
\enddata
\tablecomments{Negative values for the mean deviations where $M_{spec}<M_{dyn}$}
\end{deluxetable}

In the low mass range, 
the spectroscopic masses from the stellar evolutionary models of \citetalias{siess2000}, \citetalias{baraffe2015} and \citetalias{feiden2016magnetic} are consistent with the dynamical masses within the 1$\sigma$ to 2$\sigma$ uncertainties. 
On the contrary, magnetic \citetalias{feiden2016magnetic} results in a deviation of 2.6$\sigma$. Compared to the dynamical measurement, magnetic \citetalias{feiden2016magnetic} overestimates the stellar mass by 28$^{+10}_{-11}$\%.
\citetalias{chen2014improvingPARSEC} and \citetalias{palla1999star} result in deviations between 3$\sigma$ to 4$\sigma$. \citet{palla1999star} underestimates the stellar mass by 21$\pm6$\%, while \citet{chen2014improvingPARSEC} overestimates the stellar mass by 48$^{+12}_{-13}$\%.
\citetalias{bressan2012parsec} shows the largest deviation of 5$\sigma$ in this mass range, which underestimates the stellar masses by 30$\pm6$\%.

In the intermediate mass range, the magnetic \citetalias{feiden2016magnetic} is the only model providing the stellar mass estimate consistent with the dynamical measurements within the 1$\sigma$ uncertainty. 
All the other models included in the present paper result in spectroscopic masses deviating from the dynamical masses with mean deviations of more than 4$\sigma$. 
Compared to the dynamical measurement, 
these stellar evolutionary models underestimate the stellar mass by $\sim$20\% to $\sim$40\%.
In the high mass range, the spectroscopic masses estimated with the model of \citetalias{siess2000} best match the dynamical masses, with deviations of $\sim$1$\sigma$ significance. 
Magnetic \citetalias{feiden2016magnetic} and \citetalias{baraffe2015} show deviations of 2$\sigma$ to 3$\sigma$ significance, while all other models show deviations between 1$\sigma$ and 2$\sigma$.
However, only six stars in our sample are in this mass range, and thus, the result of this mass range may not be fully reliable due to the small sample size.

We note that the mass of a star with a highly inclined disk can be overestimated with our analysis. The sources with an inclination angle $i>70^{\circ}$ are marked by blue crosses in Figure \ref{fig:comparison_Mvast_Mspec}. We also compared our dynamical measurement with the mass estimate from the stellar evolutionary models by excluding those sources. The results are shown in Table \ref{tab:comparison_Mvast_Mspec_highinclined-excluded}. The results in the low mass range do not change because there are no highly inclined sources in this mass range.  The only high inclined star in the intermediate mass range, Sz84, is also a candidate for a possibly unresolved binary system, and is already excluded from our comparison.
In the high mass range, the deviations between the spectroscopic and dynamical measurements decrease by about 20\% to 40\% and become smaller than $1\sigma$ for all models after the highly inclined sources are excluded, expect for \citetalias{baraffe2015} and the magnetic \citetalias{feiden2016magnetic} model. However, our results of the high mass range are uncertain due to the small sample in this mass range. 

In summary, our results show that the stellar evolutionary models of \citetalias{baraffe2015}, \citetalias{feiden2016magnetic} and \citetalias{siess2000} are in good agreement (1.4$\sim$1.6$\sigma$) with the dynamical measurements in the low mass range of $M_{\star} \leq 0.6~M_{\odot}$, while the other four models tend to under- or overestimate the stellar mass by 20\% to 30\%.
In the intermediate mass range of $0.6~M_{\odot}\leq M_{\star} \leq 1.3~M_{\odot}$, our results suggest that all spectroscopic models systematically underestimate the stellar mass by 20\% to 40\%, except for the magnetic \citetalias{feiden2016magnetic}, which is in good agreement with the dynamical mass measurement with a deviation smaller than 0.7$\sigma$. In general, the agreement of the spectroscopic and dynamical measurement becomes better in the mass range of $M_{\star}\geq1.3~M_{\odot}$, except for the magnetic \citetalias{feiden2016magnetic}. However, there are only a few sources in this mass range, and therefore we cannot make a reliable comparison.

\subsection{Comparison with Literature}

\citet{hillenbrand2004} assembled a data set consisting of 27 PMS stars and 88 MS stars, where the stellar masses were both spectroscopically and dynamically determined. The methods used to measure the dynamical stellar masses included eclipsing binary systems, astrometric and radial velocity orbits, disk kinematics and double-lined spectroscopic binaries. In their work, the spectroscopic masses were obtained using the evolutionary models of \citetalias{siess2000} and \citetalias{palla1999star}, among others. The masses of the PMS stars included in their sample range from 0.3 to 1.9~$M_{\odot}$, the masses of the MS stars range from 0.1 to 2 $M_{\odot}$. 
Four, ten, and eight PMS stars were included with masses below  0.5~$M_{\odot}$, between 0.5~$M_{\odot}$ and 1.2~$M_{\odot}$, and above 1.2~$M_{\odot}$, respectively.
For PMS stars with masses above $1.2~M_{\odot}$ and in the range of 0.3 to 0.5~$M_{\odot}$, all models were found to be in good agreement with the spectroscopic measurement with a mean deviation $<1\sigma$. On the contrary, it was found that most evolutionary models underestimate stellar masses by $\sim25\%$ in the range of $0.5~M_{\odot}$ to $1.2~M_{\odot}$ at 1-2$\sigma$ significance.
Thus, our results are consistent with the findings by \citet{hillenbrand2004} for the stellar evolutionary model of \citetalias{siess2000}. In our results, a larger deviation was found for the evolutionary model of \citetalias{palla1999star} in the lowest mass range. However, this might be explained by the larger number of stars with masses smaller $0.5~M_{\odot}$, included in the present paper.

\citet{simon2019masses} present the dynamical measurements of 29 stars from the Taurus and 3 in the Ophiuchus star-forming regions, using the rotation in the disks around the stars. The dynamical masses were compared to the spectroscopic masses from the non-magnetic evolutionary models by \citetalias{baraffe2015} and \citetalias{feiden2016magnetic}, as well as the magnetic evolutionary model by \citet{feiden2016magnetic}. 
The mass range of the sample in \citet{simon2019masses} is from 0.4 to 1.4~$M_{\odot}$. 
\citet{simon2019masses} found that the non-magnetic evolutionary models underestimate the stellar mass by $\sim$30\% in this mass range.
The magnetic \citetalias{feiden2016magnetic} yields spectroscopic masses in good agreement with the dynamical measurements with deviations of only 0.01$\pm$0.02~$M_{\odot}$. 
Our results of the intermediate mass range also show the same tendency. 
To make a more detailed comparison, we separated the results in \citet{simon2019masses} into the same mass ranges as those in Section \ref{sec:comparison_Mspec} and calculated the mean deviations in the respective mass ranges. Three and seven stars were found in the low and intermediate mass range, respectively. For stars with masses $M_{\star}<0.6M_{\odot}$ deviations of -14$\pm$6\% for the non-magnetic and 21$\pm$6\% for the magnetic model were found. For the intermediate mass stars ($0.6~M_{\odot}\leq M_{\star} \leq 1.3~M_{\odot}$) -24$\pm$3\% and -4$\pm$3\% were found, for the non-magnetic and the magnetic model, respectively. 
Thus, the trends in our results are consistent with those in \citet{simon2019masses} for the same stellar evolutionary models. 

Our results and those in the literature show the same tendency that most of the studied spectroscopic models tend to underestimate the masses of stars in the mass range of $0.4\mbox{--}0.6~M_{\odot}\lesssim M_{\star} \lesssim 1.2\mbox{--}1.4~M_{\odot}$ by 20\%--40\%, and only the magnetic \citetalias{feiden2016magnetic} model provides an accurate mass estimate.
Compared to the non-magnetic evolutionary models studied in the present paper, the magnetic \citetalias{feiden2016magnetic} model suggests a factor of 2 older ages of a star because of the slower contraction due to additional magnetic pressure. 
This could imply a longer T Tauri phase and thus longer time scales of star and planet formation \citep{simon2019masses,feiden2016magnetic}. 
On the other hand, our results with a larger sample of low-mass stars than those in the literature further show that the stellar masses estimated with the magnetic \citetalias{feiden2016magnetic} model are inconsistent with the dynamical masses for the low-mass stars with masses $\leq 0.6~M_\odot$ at a level of more than 2$\sigma$, while the stellar masses estimated with the other non-magnetic models by \citetalias{baraffe2015}, \citetalias{feiden2016magnetic} and \citetalias{siess2000} are in a better agreement with the dynamical masses.
Therefore, our study suggests that none of the stellar evolutionary models studied in the present paper can provide an accurate estimate of stellar mass over a wide mass range from 0.1~$M_\odot$ to 1.4~$M_\odot$.
This result could hint at the mass dependence of physical processes which is possibly not accurately described in the stellar evolutionary models.
Nevertheless, the accuracy of our dynamical mass measurements is limited by the S/N of the data. 
Higher S/N data and a larger sample covering a wide mass range from 0.1 to 1~$M_{\odot}$ are needed to further reveal the mass dependence of the accuracy of the stellar evolutionary models.
\section{Summary}
\label{sec:summary}

We analysed the ALMA archival data of $^{13}$CO ($J$ = 2--1 \& 3--2), C$^{18}$O ($J$ = 2--1 \& 3--2), and CN ($N$ = 3--2, $J$ = 7/2--5/2) of 30 PMS stars in the Lupus star-forming region and of $^{13}$CO ($J$ = 2--1) and C$^{18}$O ($J$ = 2--1) of 37 PMS stars in the Taurus star-forming region. For individual stars, we stacked the different molecular line data to enhance the S/Ns and measured the dynamical stellar masses from Keplerian rotation of their surrounding disks. 
The stellar masses were measured by maximizing the S/Ns by multiplying image cubes by various Keplerian rotational velocity patterns. After stacking different molecular lines, we obtained measurements of stellar masses for 45 out of 67 PMS stars with our method. We also obtained measurements when only using the data of individual lines, and the measured stellar masses from individual lines are consistent with each other within the uncertainties. 
In addition, we tested our method with synthetic ALMA images of chemical and physical disk models generated with the DALI code. Our method indeed can provide robust estimates of stellar mass.
We compared the measured dynamical masses with the spectroscopically determined stellar masses with different stellar evolutionary models to examine the accuracy of these models.

For individual stars, most of our measured dynamical masses are consistent with the spectroscopic masses within a 3$\sigma$ level for all the stellar evolutionary models due to the limited S/Ns of our data. 
Given our sufficiently large sample, we separated our sample into three mass ranges and computed the mean differences between the dynamical and spectroscopic masses and the uncertainties of the mean differences in these mass ranges.
Our results show that in the low mass range of $<$0.6~$M_{\odot}$, the spectroscopic masses estimated with the stellar evolutionary models by \citet{baraffe2015}, \citet{feiden2016magnetic} (without magnetic field) and \citet{siess2000} are in good agreement (1.4$\sim$1.6$\sigma$) with the dynamical masses, while the other four models, \citet{palla1999star}, \citet{chen2014improvingPARSEC}, \citet{bressan2012parsec}, and \citet{feiden2016magnetic} (with magnetic field) tend to under- or overestimate the stellar mass by 20\% to 30\%.
In the intermediate mass range of $0.6~M_{\odot}\leq M_{\star} \leq 1.3~M_{\odot}$, only the stellar evolutionary model with magnetic field by \citet{feiden2016magnetic} provides a stellar mass estimate in good agreement with the dynamical mass measurement ($<0.7\sigma$). All the other models underestimate the stellar mass by 20\% to 40\% compared to the dynamical masses. 
After excluding the high-inclined sources, all models show deviations $<$1$\sigma$ in the mass range of $\geq1.3~M_{\odot}$, except for the model of \citet{baraffe2015} and the magnetic \citet{feiden2016magnetic} including magnetic fields, which show deviations of 2$\sigma$.
However, our sample only contains a few sources in this mass range, so we cannot make a fully reliable comparison for this mass range. 

Therefore, our study suggests that none of the stellar evolutionary models studied in the present paper can provide an accurate estimate of stellar mass over a wide mass range from 0.1~$M_\odot$ to 1.4~$M_\odot$.
This result could hint at the mass dependence of physical processes which is possibly not accurately described in the stellar evolutionary models.

\acknowledgments
{We thank Patrick D. Sheehan for his help and the discussions on the stellar mass estimate with the spectroscopic measurements using the stellar evolutionary models.
This paper makes use of the following ALMA data: ADS/JAO.ALMA\#2013.1.00220.S, ADS/JAO.ALMA\#2015.1.00222.S, ADS/JAO.ALMA \#2016.1.01239.S, ADS/JAO.ALMA\#2016.1.01164.S. ALMA is a partnership of ESO (representing its member states), NSF (USA) and NINS (Japan), together with NRC (Canada), MOST and ASIAA (Taiwan), and KASI (Republic of Korea), in cooperation with the Republic of Chile. The Joint ALMA Observatory is operated by ESO, AUI/NRAO and NAOJ.
This work has made use of data from the European Space Agency (ESA) mission {\it Gaia} (\url{https://www.cosmos.esa.int/gaia}), processed by the {\it Gaia} Data Processing and Analysis Consortium (DPAC, \url{https://www.cosmos.esa.int/web/gaia/dpac/consortium}). Funding for the DPAC has been provided by national institutions, in particular the institutions participating in the {\it Gaia} Multilateral Agreement.
This project has received funding from the European Union's Horizon 2020 research and innovation programme under the Marie Sklodowska-Curie grant agreement No 823823 (DUSTBUSTERS).
This work was partly supported by the Deutsche Forschungsgemeinschaft (DFG, German Research Foundation) - Ref no. FOR 2634/1 TE 1024/1-1, by the Italian Ministero dell’Istruzione,Università e Ricerca (MIUR) through the grantProgettiPremiali 2012 iALMA (CUP C52I13000140001), 
and by the DFG cluster of excellence Origin and Structure of the Universe (www.universe-cluster.de).
H.-W.Y. acknowledges support from MOST 108-2112-M-001-003-MY2. 
P.M.K. acknowledges support from MOST 108-2112-M-001-012 and MOST 109-2112-M-001-022, and from an Academia Sinica Career Development Award.
}

\bibliography{references}

\begin{thebibliography}{}
\expandafter\ifx\csname natexlab\endcsname\relax\def\natexlab#1{#1}\fi
\providecommand{\url}[1]{\href{#1}{#1}}
\providecommand{\dodoi}[1]{doi:~\href{http://doi.org/#1}{\nolinkurl{#1}}}
\providecommand{\doeprint}[1]{\href{http://ascl.net/#1}{\nolinkurl{http://ascl.net/#1}}}
\providecommand{\doarXiv}[1]{\href{https://arxiv.org/abs/#1}{\nolinkurl{https://arxiv.org/abs/#1}}}

\bibitem[{Alcal{\'a} {et~al.}(2019)Alcal{\'a}, Manara, France, Schneider,
  Arulanantham, Miotello, G{\"u}nther, \& Brown}]{alcala2019updatedTeffL}
Alcal{\'a}, J., Manara, C., France, K., {et~al.} 2019, Astronomy \&
  Astrophysics, 629, A108

\bibitem[{Alcal{\'a} {et~al.}(2014)Alcal{\'a}, Natta, Manara, Spezzi, Stelzer,
  Frasca, Biazzo, Covino, Randich, Rigliaco,
  {et~al.}}]{alcala2014xshootersurvey}
Alcal{\'a}, J., Natta, A., Manara, C., {et~al.} 2014, Astronomy \&
  Astrophysics, 561, A2

\bibitem[{Alcal{\'a} {et~al.}(2017)Alcal{\'a}, Manara, Natta, Frasca, Testi,
  Nisini, Stelzer, Williams, Antoniucci, Biazzo,
  {et~al.}}]{alcala2017xshootersurveyMspec}
Alcal{\'a}, J., Manara, C., Natta, A., {et~al.} 2017, Astronomy \&
  Astrophysics, 600, A20

\bibitem[{Ansdell {et~al.}(2017)Ansdell, Williams, Manara, Miotello, Facchini,
  van~der Marel, Testi, \& van Dishoeck}]{ansdell2017almasurvey}
Ansdell, M., Williams, J.~P., Manara, C.~F., {et~al.} 2017, The Astronomical
  Journal, 153, 240

\bibitem[{Ansdell {et~al.}(2016)Ansdell, Williams, van~der Marel, Carpenter,
  Guidi, Hogerheijde, Mathews, Manara, Miotello, Natta, Oliveira, Tazzari,
  Testi, van Dishoeck, \& van Terwisga}]{Ansdell2016}
Ansdell, M., Williams, J.~P., van~der Marel, N., {et~al.} 2016, The
  Astrophysical Journal, 828, 46, \dodoi{10.3847/0004-637x/828/1/46}

\bibitem[{{Ansdell} {et~al.}(2018){Ansdell}, {Williams}, {Trapman}, {van
  Terwisga}, {Facchini}, {Manara}, {van der Marel}, {Miotello}, {Tazzari},
  {Hogerheijde}, {Guidi}, {Testi}, \& {van Dishoeck}}]{Ansdell2018}
{Ansdell}, M., {Williams}, J.~P., {Trapman}, L., {et~al.} 2018, The
  Astrophysical Journal, 859, 21, \dodoi{10.3847/1538-4357/aab890}

\bibitem[{Baraffe {et~al.}(1998)Baraffe, Chabrier, Allard, \&
  Hauschildt}]{baraffe1998}
Baraffe, I., Chabrier, G., Allard, F., \& Hauschildt, P. 1998, arXiv preprint
  astro-ph/9805009

\bibitem[{Baraffe {et~al.}(2009)Baraffe, Chabrier, \&
  Gallardo}]{baraffe2009accretion}
Baraffe, I., Chabrier, G., \& Gallardo, J. 2009, The Astrophysical Journal
  Letters, 702, L27

\bibitem[{Baraffe {et~al.}(2015)Baraffe, Homeier, Allard, \&
  Chabrier}]{baraffe2015}
Baraffe, I., Homeier, D., Allard, F., \& Chabrier, G. 2015, Astronomy \&
  Astrophysics, 577, A42

\bibitem[{Barenfeld {et~al.}(2016)Barenfeld, Carpenter, Ricci, \&
  Isella}]{barenfeld2016almaSurvey_disks}
Barenfeld, S.~A., Carpenter, J.~M., Ricci, L., \& Isella, A. 2016, The
  Astrophysical Journal, 827, 142

\bibitem[{Bressan {et~al.}(2012)Bressan, Marigo, Girardi, Salasnich, Dal~Cero,
  Rubele, \& Nanni}]{bressan2012parsec}
Bressan, A., Marigo, P., Girardi, L., {et~al.} 2012, Monthly Notices of the
  Royal Astronomical Society, 427, 127

\bibitem[{Bruderer(2013)}]{bruderer2013DALI}
Bruderer, S. 2013, Astronomy \& Astrophysics, 559, A46

\bibitem[{Bruderer {et~al.}(2012)Bruderer, van Dishoeck, Doty, \&
  Herczeg}]{bruderer2012DALI}
Bruderer, S., van Dishoeck, E.~F., Doty, S.~D., \& Herczeg, G.~J. 2012,
  Astronomy \& Astrophysics, 541, A91

\bibitem[{Burkert(2004)}]{burkert2004stellar_feedback}
Burkert, A. 2004, arXiv preprint astro-ph/0404015

\bibitem[{Cassisi(2012)}]{cassisi2012uncertainties_EvModels}
Cassisi, S. 2012, in Red Giants as Probes of the Structure and Evolution of the
  Milky Way (Springer), 57--68

\bibitem[{Cazzoletti {et~al.}(2019)Cazzoletti, Manara, Liu, Van~Dishoeck,
  Facchini, Alcal{\`a}, Ansdell, Testi, Williams, Carrasco-Gonz{\'a}lez,
  {et~al.}}]{cazzoletti2019almasurvey}
Cazzoletti, P., Manara, C., Liu, H.~B., {et~al.} 2019, Astronomy \&
  Astrophysics, 626, A11

\bibitem[{Ceverino \& Klypin(2009)}]{ceverino2009stellar_feedback}
Ceverino, D., \& Klypin, A. 2009, The Astrophysical Journal, 695, 292

\bibitem[{Chen {et~al.}(2014)Chen, Girardi, Bressan, Marigo, Barbieri, \&
  Kong}]{chen2014improvingPARSEC}
Chen, Y., Girardi, L., Bressan, A., {et~al.} 2014, Monthly Notices of the Royal
  Astronomical Society, 444, 2525

\bibitem[{D'Antona \& Mazzitelli(1997)}]{dantona1997}
D'Antona, F., \& Mazzitelli, I. 1997, Memorie della Societa Astronomica
  Italiana, 68, 807

\bibitem[{Eisner {et~al.}(2018)Eisner, Arce, Ballering, Bally, Andrews, Boyden,
  Di~Francesco, Fang, Johnstone, Kim, {et~al.}}]{eisner2018alma_survey}
Eisner, J.~A., Arce, H., Ballering, N., {et~al.} 2018, The Astrophysical
  Journal, 860, 77

\bibitem[{El-Badry {et~al.}(2018)El-Badry, Rix, Ting, Weisz, Bergemann,
  Cargile, Conroy, \& Eilers}]{el2018unresolved_binaries}
El-Badry, K., Rix, H.-W., Ting, Y.-S., {et~al.} 2018, Monthly Notices of the
  Royal Astronomical Society, 473, 5043

\bibitem[{Feiden(2016)}]{feiden2016magnetic}
Feiden, G.~A. 2016, Astronomy \& Astrophysics, 593, A99

\bibitem[{Gaia~Collaboration {et~al.}(2018)Gaia~Collaboration, Vallenari,
  Prusti, De~Bruijne, Babusiaux, Bailer-Jones, Biermann, Evans, Eyer, Jansen,
  {et~al.}}]{brown2018gaia}
Gaia~Collaboration, Brown, A., Vallenari, A., Prusti, T., {et~al.} 2018,
  Astronomy \& Astrophysics, 616, A1

\bibitem[{Gaia~Collaboration {et~al.}(2016)Gaia~Collaboration, De~Bruijne,
  Brown, Vallenari, Babusiaux, Bailer-Jones, Bastian, Biermann, Evans, Eyer,
  {et~al.}}]{prusti2016gaia}
Gaia~Collaboration, Prusti, T., De~Bruijne, J., Brown, A.~G., {et~al.} 2016,
  Astronomy \& astrophysics, 595, A1

\bibitem[{Guilloteau {et~al.}(2014)Guilloteau, Simon, Pi{\'e}tu, Di~Folco,
  Dutrey, Prato, \& Chapillon}]{guilloteau2014}
Guilloteau, S., Simon, M., Pi{\'e}tu, V., {et~al.} 2014, Astronomy \&
  Astrophysics, 567, A117

\bibitem[{Herczeg \& Hillenbrand(2014)}]{herczeg2014specSurvey}
Herczeg, G.~J., \& Hillenbrand, L.~A. 2014, The Astrophysical Journal, 786, 97

\bibitem[{Herczeg \& Hillenbrand(2015)}]{herczeg2015empiricalHR}
---. 2015, The Astrophysical Journal, 808, 23

\bibitem[{Hillenbrand \& White(2004)}]{hillenbrand2004}
Hillenbrand, L.~A., \& White, R.~J. 2004, The Astrophysical Journal, 604, 741

\bibitem[{Hopkins(2018)}]{hopkins2018_IMF}
Hopkins, A. 2018, Publications of the Astronomical Society of Australia, 35

\bibitem[{Jeffries(2012)}]{jeffries2012measuring_IMF}
Jeffries, R. 2012, European Astronomical Society Publications Series, 57, 45

\bibitem[{Long {et~al.}(2018)Long, Pinilla, Herczeg, Harsono, Dipierro,
  Pascucci, Hendler, Tazzari, Ragusa, Salyk, {et~al.}}]{long2018taurus}
Long, F., Pinilla, P., Herczeg, G.~J., {et~al.} 2018, The Astrophysical
  Journal, 869, 17

\bibitem[{Long {et~al.}(2019)Long, Herczeg, Harsono, Pinilla, Tazzari, Manara,
  Pascucci, Cabrit, Nisini, Johnstone, {et~al.}}]{long2019taurus}
Long, F., Herczeg, G.~J., Harsono, D., {et~al.} 2019, The Astrophysical
  Journal, 882, 49

\bibitem[{Manara {et~al.}(2017)Manara, Testi, Herczeg, Pascucci, Alcal{\'a},
  Natta, Antoniucci, Fedele, Mulders, Henning,
  {et~al.}}]{manara2017XshootersurveyMspec}
Manara, C., Testi, L., Herczeg, G., {et~al.} 2017, Astronomy \& Astrophysics,
  604, A127

\bibitem[{Manara {et~al.}(2019)Manara, Tazzari, Long, Herczeg, Lodato, Rota,
  Cazzoletti, van~der Plas, Pinilla, Dipierro, {et~al.}}]{manara2019taurus}
Manara, C., Tazzari, M., Long, F., {et~al.} 2019, arXiv preprint
  arXiv:1907.03846

\bibitem[{Mathieu {et~al.}(1997)Mathieu, Stassun, Basri, Jensen, Johns-Krull,
  Valenti, \& Hartmann}]{mathieu1997DQTaubinary}
Mathieu, R.~D., Stassun, K., Basri, G., {et~al.} 1997, The Astronomical
  Journal, 113, 1841

\bibitem[{Matr{\`a} {et~al.}(2017)Matr{\`a}, MacGregor, Kalas, Wyatt, Kennedy,
  Wilner, Duchene, Hughes, Pan, Shannon, {et~al.}}]{matra2017aligningSpectra}
Matr{\`a}, L., MacGregor, M.~A., Kalas, P., {et~al.} 2017, The Astrophysical
  Journal, 842, 9

\bibitem[{Miotello {et~al.}(2016)Miotello, van Dishoeck, Kama, \&
  Bruderer}]{miotello2016diskmassesDALI}
Miotello, A., van Dishoeck, E.~F., Kama, M., \& Bruderer, S. 2016, Astronomy \&
  Astrophysics, 594, A85

\bibitem[{Palla \& Stahler(1993)}]{palla1993pre}
Palla, F., \& Stahler, S.~W. 1993, The Astrophysical Journal, 418, 414

\bibitem[{Palla \& Stahler(1999)}]{palla1999star}
---. 1999, The Astrophysical Journal, 525, 772

\bibitem[{Pinte {et~al.}(2018)Pinte, M{\'e}nard, Duch{\^e}ne, Hill, Dent,
  Woitke, Maret, van~der Plas, Hales, Kamp, {et~al.}}]{pinte2018}
Pinte, C., M{\'e}nard, F., Duch{\^e}ne, G., {et~al.} 2018, Astronomy \&
  Astrophysics, 609, A47

\bibitem[{Prato {et~al.}(2002)Prato, Simon, Mazeh, Zucker, \&
  McLean}]{prato2002UZTauEbinary}
Prato, L., Simon, M., Mazeh, T., Zucker, S., \& McLean, I. 2002, The
  Astrophysical Journal Letters, 579, L99

\bibitem[{Rigliaco {et~al.}(2012)Rigliaco, Natta, Testi, Randich, Alcala,
  Covino, \& Stelzer}]{rigliaco2012xShooter}
Rigliaco, E., Natta, A., Testi, L., {et~al.} 2012, Astronomy \& Astrophysics,
  548, A56

\bibitem[{Rizzuto {et~al.}(2019)Rizzuto, Dupuy, Ireland, \&
  Kraus}]{rizzuto2019dynamical}
Rizzuto, A.~C., Dupuy, T.~J., Ireland, M.~J., \& Kraus, A.~L. 2019, arXiv
  preprint arXiv:1911.12378

\bibitem[{Rizzuto {et~al.}(2016)Rizzuto, Ireland, Dupuy, \&
  Kraus}]{rizzuto2016dynamical}
Rizzuto, A.~C., Ireland, M.~J., Dupuy, T.~J., \& Kraus, A.~L. 2016, The
  Astrophysical Journal, 817, 164

\bibitem[{Salinas {et~al.}(2017)Salinas, Hogerheijde, Mathews, {\"O}berg, Qi,
  Williams, \& Wilner}]{salinas2017KepMasking}
Salinas, V., Hogerheijde, M., Mathews, G., {et~al.} 2017, Astronomy \&
  Astrophysics, 606, A125

\bibitem[{Sheehan {et~al.}(2019)Sheehan, Wu, Eisner, \&
  Tobin}]{sheehan2019high}
Sheehan, P.~D., Wu, Y.-L., Eisner, J.~A., \& Tobin, J.~J. 2019, The
  Astrophysical Journal, 874, 136

\bibitem[{Siess {et~al.}(2000)Siess, Dufour, \& Forestini}]{siess2000}
Siess, L., Dufour, E., \& Forestini, M. 2000, arXiv preprint astro-ph/0003477

\bibitem[{Simon {et~al.}(2000)Simon, Dutrey, \&
  Guilloteau}]{simon2000dynamical}
Simon, M., Dutrey, A., \& Guilloteau, S. 2000, The Astrophysical Journal, 545,
  1034

\bibitem[{Simon {et~al.}(2017)Simon, Guilloteau, Di~Folco, Dutrey, Grosso,
  Pi{\'e}tu, Chapillon, Prato, Schaefer, Rice, {et~al.}}]{simon2017dynamical}
Simon, M., Guilloteau, S., Di~Folco, E., {et~al.} 2017, The Astrophysical
  Journal, 844, 158

\bibitem[{Simon {et~al.}(2019)Simon, Guilloteau, Beck, Chapillon, Di~Folco,
  Dutrey, Feiden, Grosso, Pi{\'e}tu, Prato, {et~al.}}]{simon2019masses}
Simon, M., Guilloteau, S., Beck, T.~L., {et~al.} 2019, The Astrophysical
  Journal, 884, 42

\bibitem[{Stassun {et~al.}(2014)Stassun, Feiden, \&
  Torres}]{stassun2014empirical}
Stassun, K.~G., Feiden, G.~A., \& Torres, G. 2014, New Astronomy Reviews, 60, 1

\bibitem[{{Tazzari} {et~al.}(2017){Tazzari}, {Testi}, {Natta}, {Ansdell},
  {Carpenter}, {Guidi}, {Hogerheijde}, {Manara}, {Miotello}, {van der Marel},
  {van Dishoeck}, \& {Williams}}]{Tazzari2017}
{Tazzari}, M., {Testi}, L., {Natta}, A., {et~al.} 2017, Astronomy and
  Astrophysics, 606, A88, \dodoi{10.1051/0004-6361/201730890}

\bibitem[{Teague {et~al.}(2018{\natexlab{a}})Teague, Bae, Bergin, Birnstiel, \&
  Foreman-Mackey}]{teague2018stacking}
Teague, R., Bae, J., Bergin, E.~A., Birnstiel, T., \& Foreman-Mackey, D.
  2018{\natexlab{a}}, The Astrophysical Journal Letters, 860, L12

\bibitem[{Teague {et~al.}(2018{\natexlab{b}})Teague, Henning, Guilloteau,
  Bergin, Semenov, Dutrey, Flock, Gorti, \& Birnstiel}]{teague2018KepMasking}
Teague, R., Henning, T., Guilloteau, S., {et~al.} 2018{\natexlab{b}}, The
  Astrophysical Journal, 864, 133

\bibitem[{Trapman {et~al.}(2020)Trapman, Ansdell, Hogerheijde, Facchini,
  Manara, Miotello, Williams, \& Bruderer}]{trapman2020KepMasking}
Trapman, L., Ansdell, M., Hogerheijde, M., {et~al.} 2020, arXiv preprint
  arXiv:2004.07257

\bibitem[{van Terwisga {et~al.}(2018)van Terwisga, van Dishoeck, Ansdell,
  van~der Marel, Testi, Williams, Facchini, Tazzari, Hogerheijde, Trapman,
  {et~al.}}]{vanterwisga}
van Terwisga, S., van Dishoeck, E., Ansdell, M., {et~al.} 2018, arXiv preprint
  arXiv:1805.03221

\bibitem[{van Terwisga {et~al.}(2020)van Terwisga, van Dishoeck, Mann,
  Di~Francesco, van~der Marel, Meyer, Andrews, Carpenter, Eisner, Manara,
  {et~al.}}]{vanterwisga2020almasurvey_disks}
van Terwisga, S.~E., van Dishoeck, E.~F., Mann, R.~K., {et~al.} 2020, arXiv
  preprint arXiv:2004.13551

\bibitem[{Williams {et~al.}(2019)Williams, Cieza, Hales, Ansdell,
  Ruiz-Rodriguez, Casassus, Perez, \& Zurlo}]{williams2019ophiuchus_almasurvey}
Williams, J.~P., Cieza, L., Hales, A., {et~al.} 2019, The Astrophysical Journal
  Letters, 875, L9

\bibitem[{{Yen} {et~al.}(2016){Yen}, {Koch}, {Liu}, {Puspitaningrum}, {Hirano},
  {Lee}, \& {Takakuwa}}]{Yen2016}
{Yen}, H.-W., {Koch}, P.~M., {Liu}, H.~B., {et~al.} 2016, The Astrophysical
  Journal, 832, 204, \dodoi{10.3847/0004-637X/832/2/204}

\bibitem[{{Yen} {et~al.}(2018){Yen}, {Koch}, {Manara}, {Miotello}, \&
  {Testi}}]{Yen2018}
{Yen}, H.-W., {Koch}, P.~M., {Manara}, C.~F., {Miotello}, A., \& {Testi}, L.
  2018, Astronomy and Astrophysics, 616, A100,
  \dodoi{10.1051/0004-6361/201732196}

\end{thebibliography}

\appendix

\section{Distribution of the Stellar Parameters}
\label{sec:appendix_histograms}

\begin{figure}[hbt!]
\centering
    \includegraphics[width=0.45\linewidth]{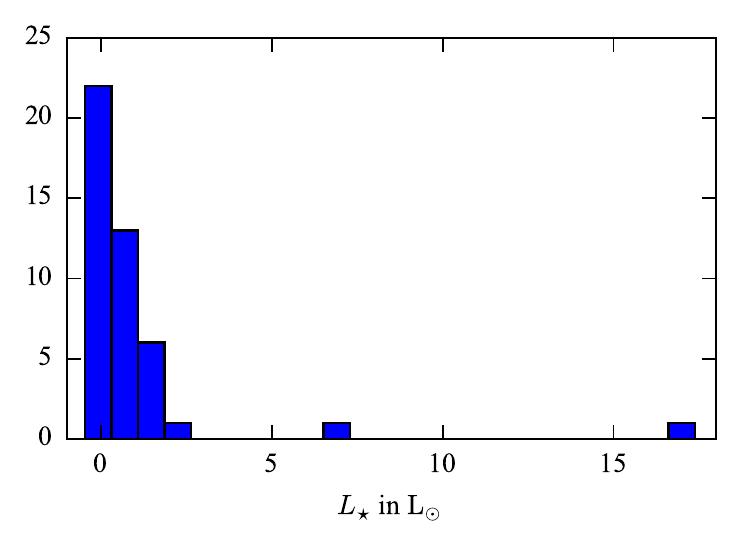}
    \includegraphics[width=0.45\linewidth]{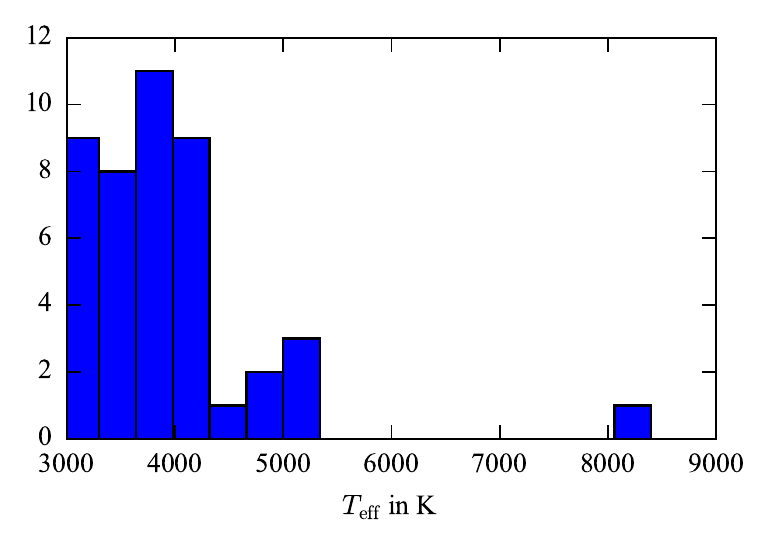}
    \includegraphics[width=0.45\linewidth]{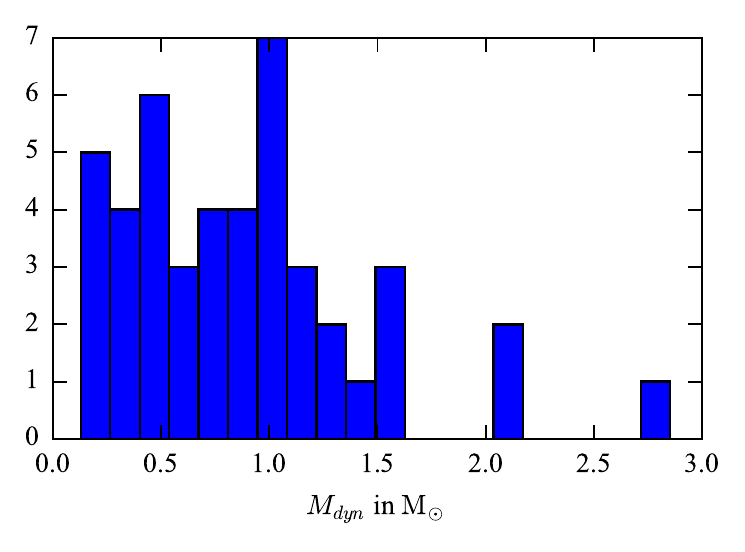}
\caption{Distribution of the luminosity ($L_{\star}$), effective temperature ($T_{\mathrm{eff}}$), and dynamical stellar mass ($M_{dyn}$) of the stars included in this paper}
\label{fig:histograms}
\end{figure}

Histograms of the luminosity, effective temperature and stellar mass are shown in Figure \ref{fig:histograms}, to give an overview over the distribution of the parameters of the stars included in this paper.

\FloatBarrier

\section{Aligned Spectra for the Sources in the Lupus and Taurus Star-forming Regions}
\label{sec:appendix_lupus_taurus}

\begin{figure}[htb!]
\centering
\gridline{ \rotatefig{180}{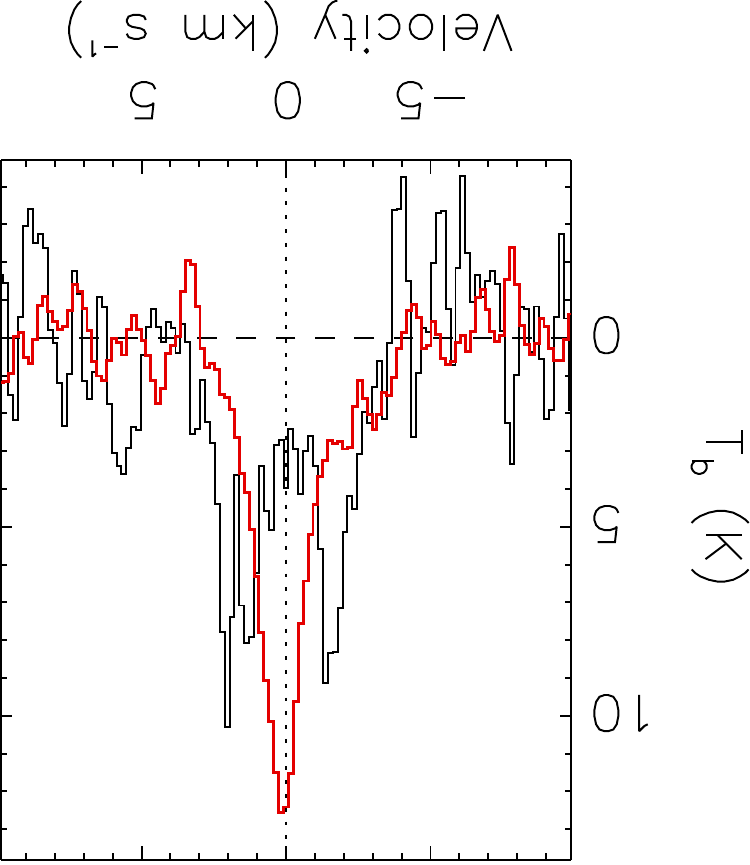}{0.3\textwidth}{EXLup, $^{13}$CO}
           \rotatefig{180}{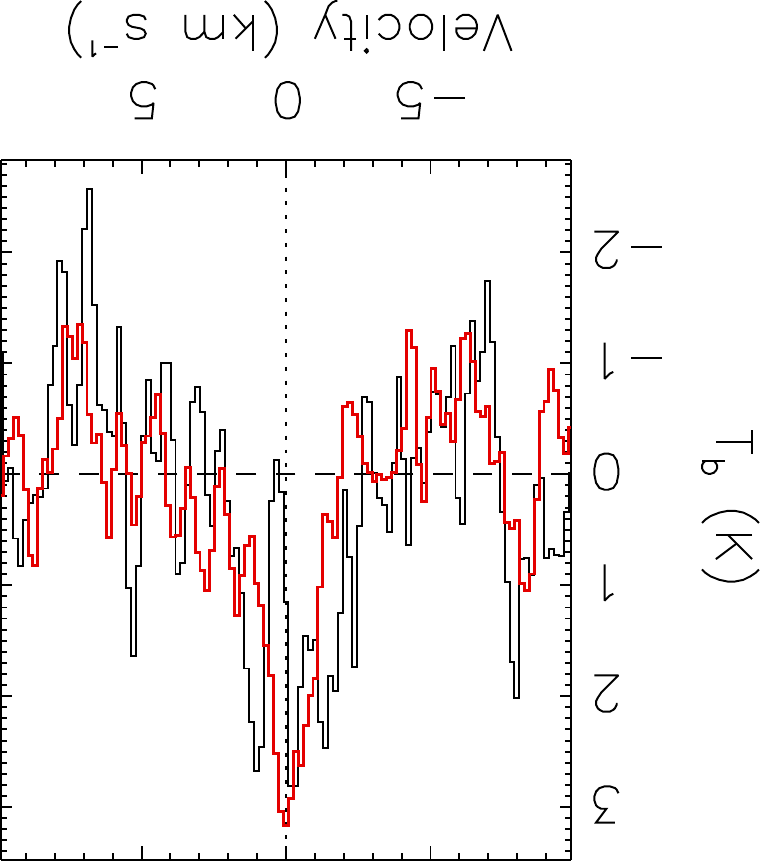}{0.3\textwidth}{J15450887-3417333, $^{13}$CO+C$^{18}$O}
           \rotatefig{180}{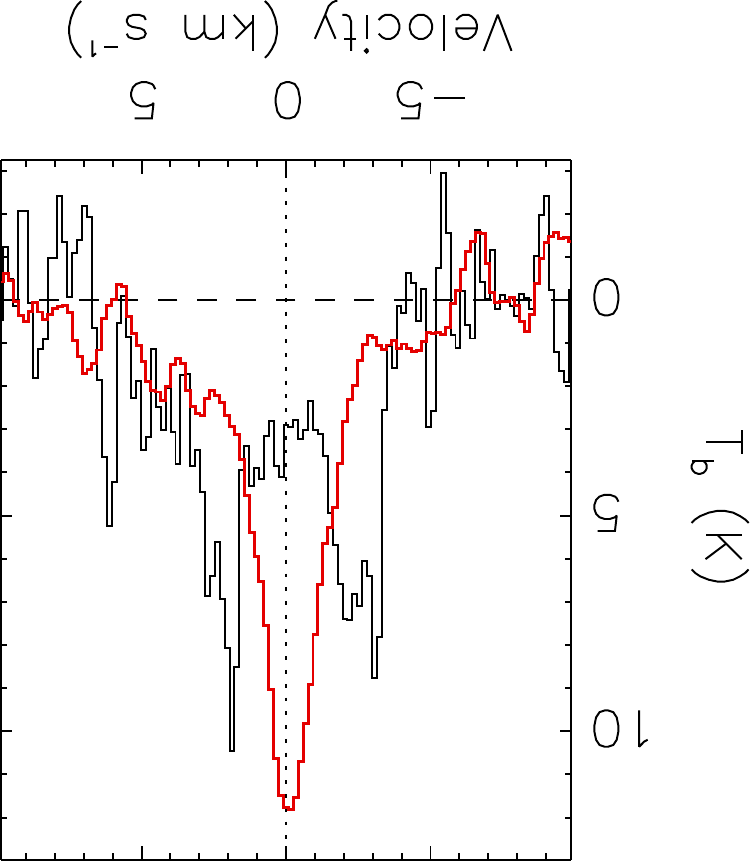}{0.3\textwidth}{RXJ1556.1-3655, $^{13}$CO+C$^{18}$O}}
\gridline{ \rotatefig{180}{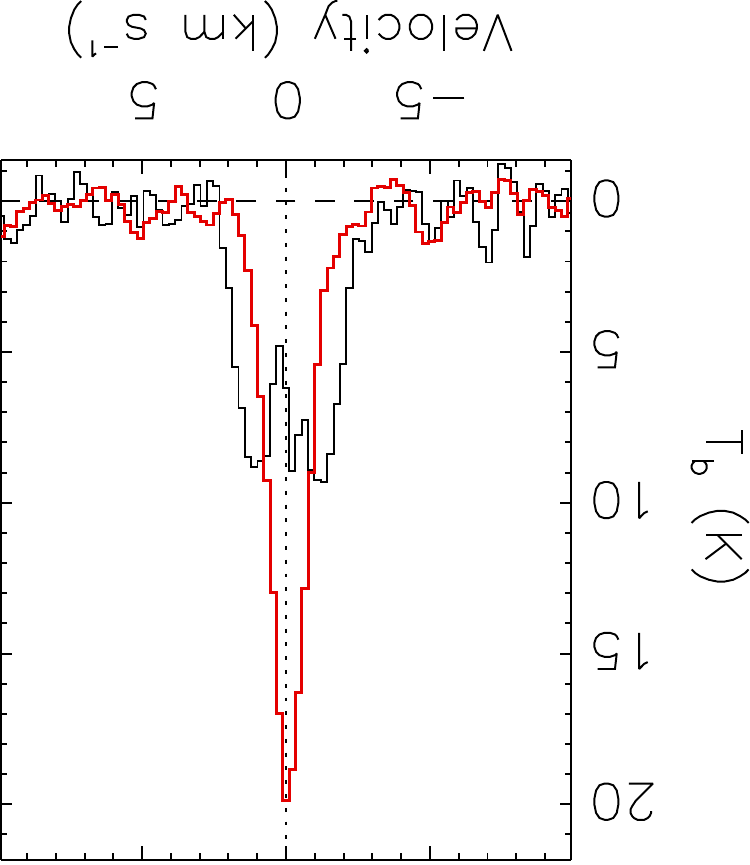}{0.3\textwidth}{J16000236-4222145, $^{13}$CO+CN+C$^{18}$O}
           \rotatefig{180}{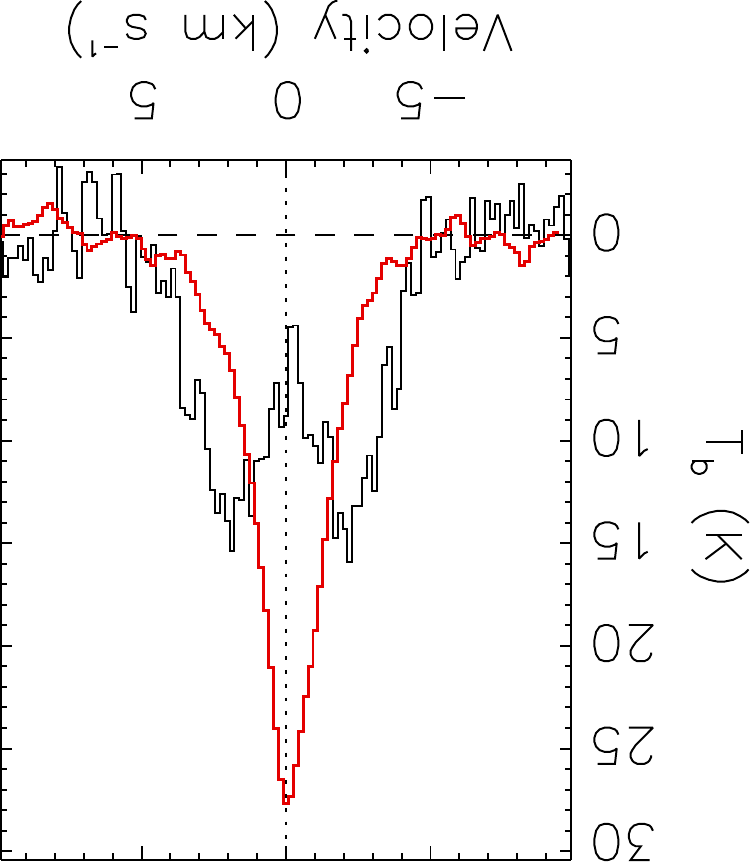}{0.3\textwidth}{J16083070-3828268, $^{13}$CO}
           \rotatefig{180}{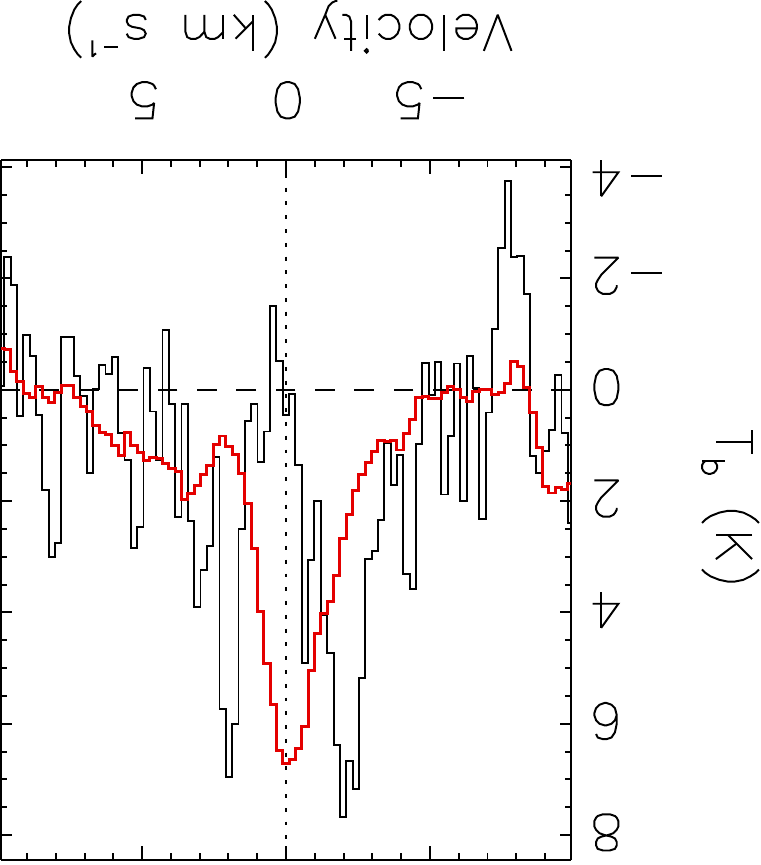}{0.3\textwidth}{J16085324-3914401, $^{13}$CO+CN}}
\gridline{ \rotatefig{180}{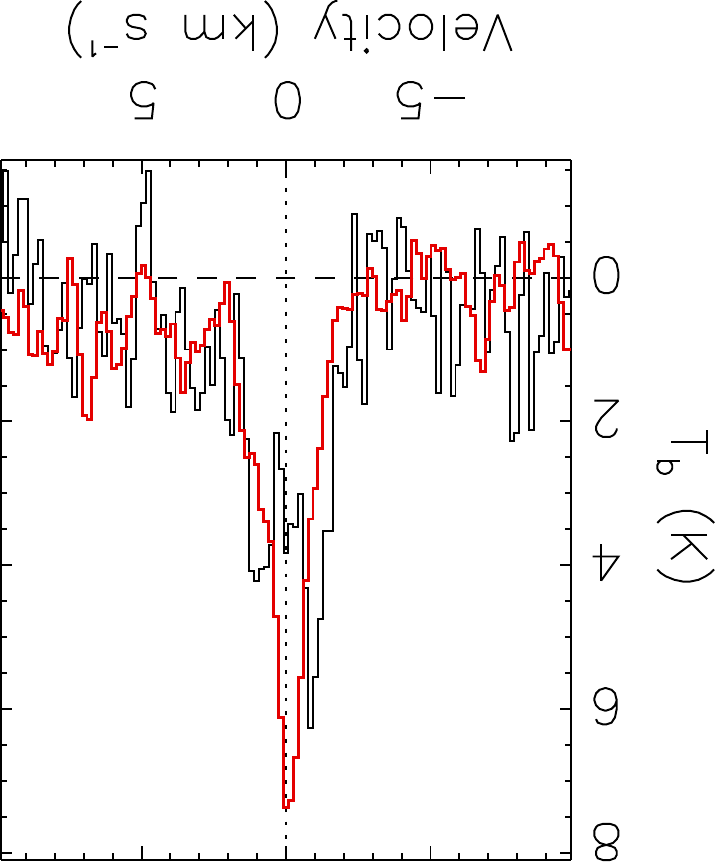}{0.3\textwidth}{J16102955-3922144, $^{13}$CO+C$^{18}$O}
           \rotatefig{180}{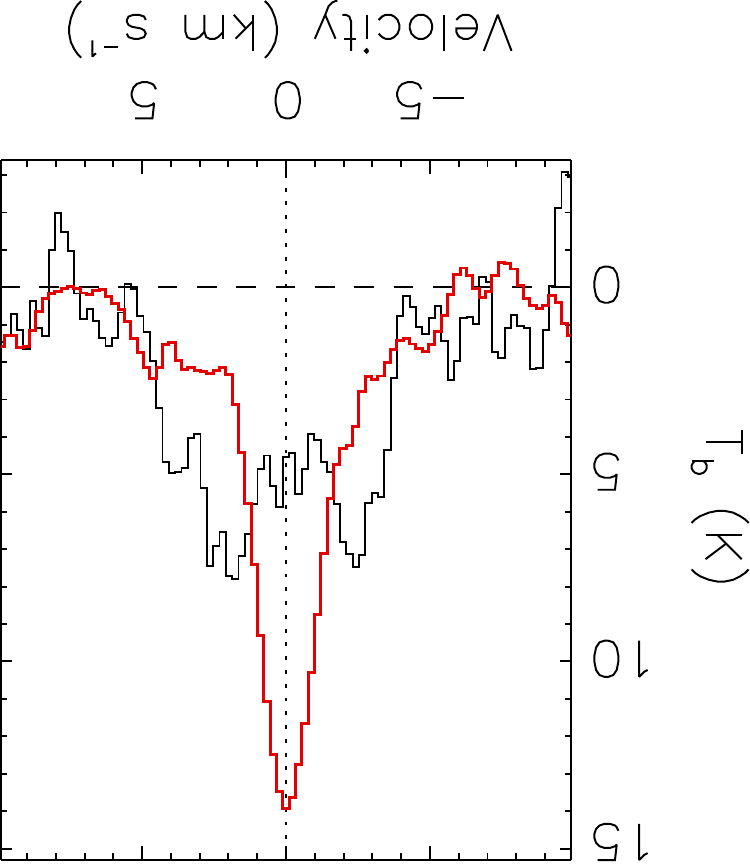}{0.3\textwidth}{J16124373-3815031, $^{13}$CO+CN}
           \rotatefig{180}{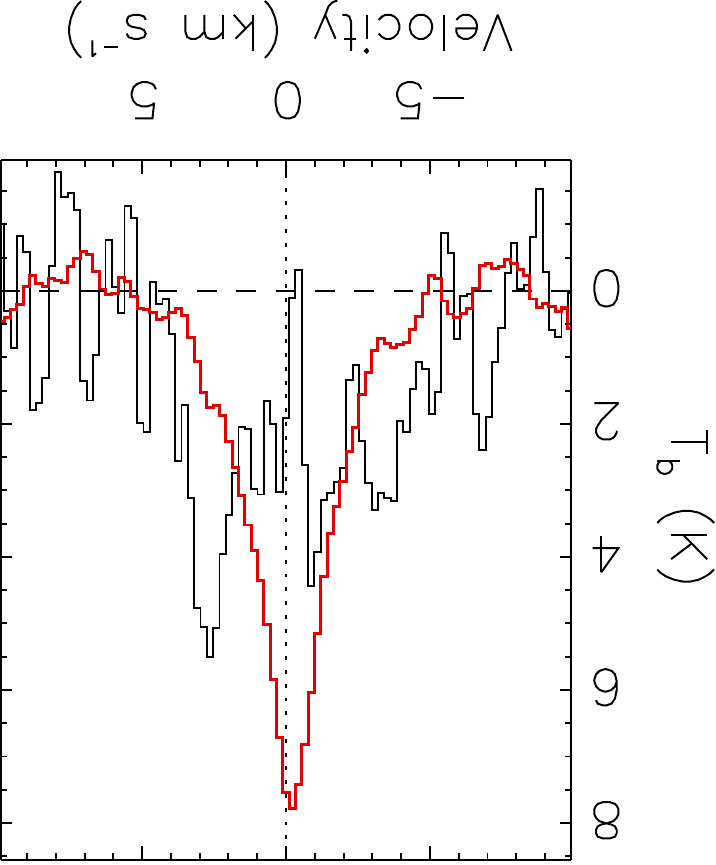}{0.3\textwidth}{MYLup, $^{13}$CO+CN+C$^{18}$O}} 
\caption{Original (black histograms) and velocity aligned (red histograms) spectra for the sources in the Lupus star-forming region}
\label{fig:appendix_lupus}           
\end{figure}

\begin{figure}[htb!]
\centering
\gridline{ \rotatefig{180}{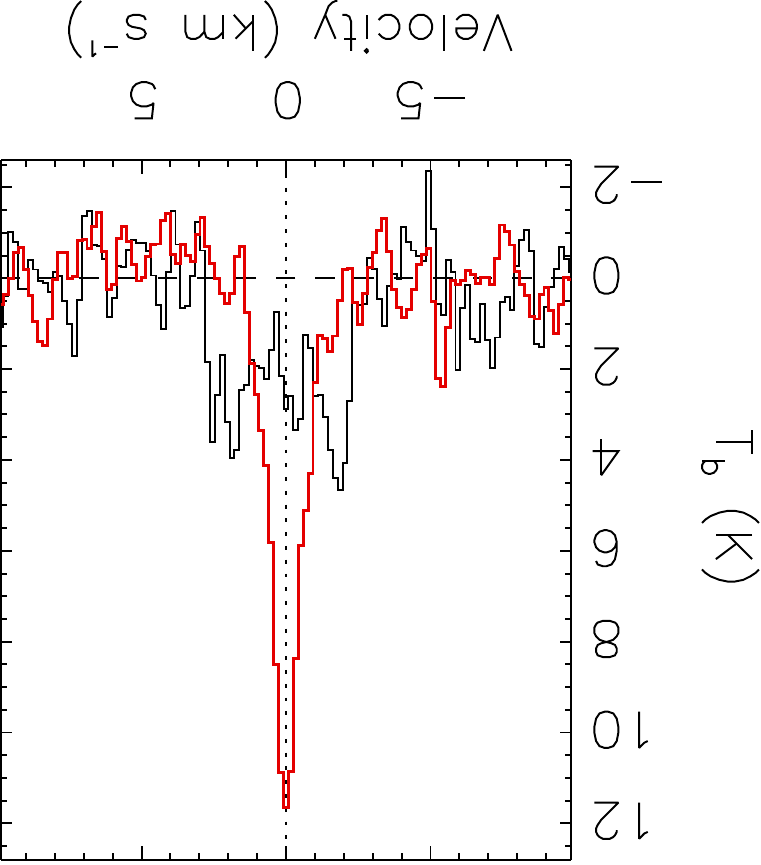}{0.3\textwidth}{CIDA9, $^{13}$CO+C$^{18}$O}
           \rotatefig{180}{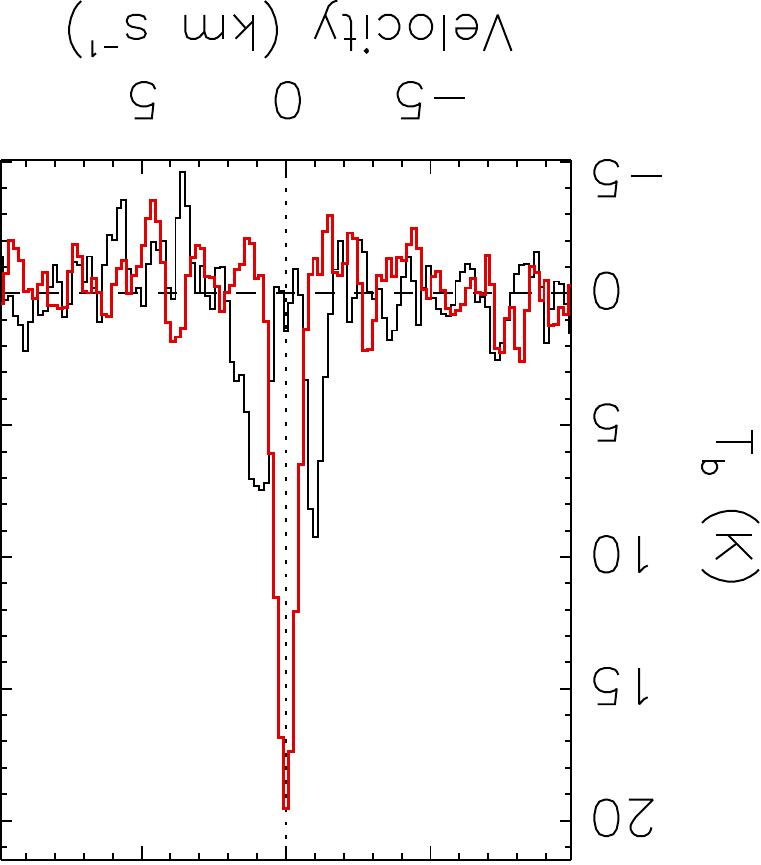}{0.3\textwidth}{DOTau, $^{13}$CO+C$^{18}$O}
           \rotatefig{180}{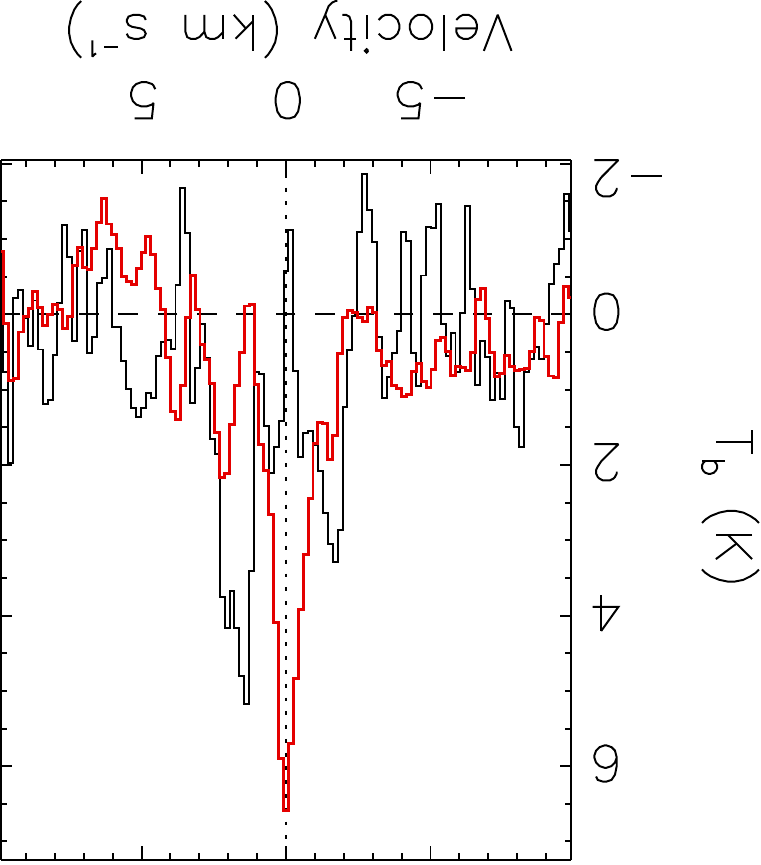}{0.3\textwidth}{DQTau, $^{13}$CO+C$^{18}$O}}
\gridline{ \rotatefig{180}{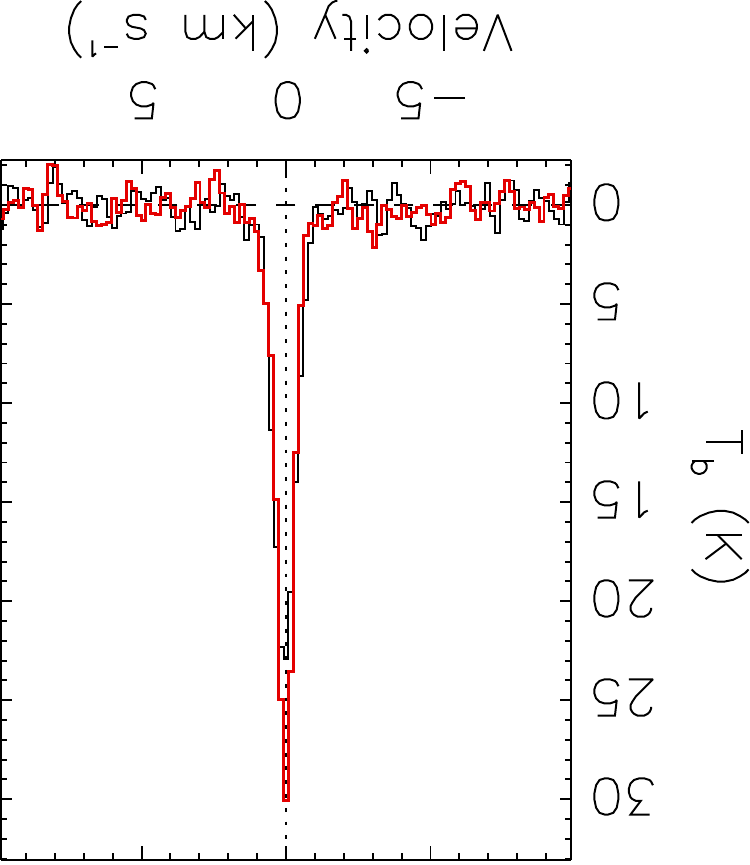}{0.3\textwidth}{DRTau, $^{13}$CO+C$^{18}$O}
           \rotatefig{180}{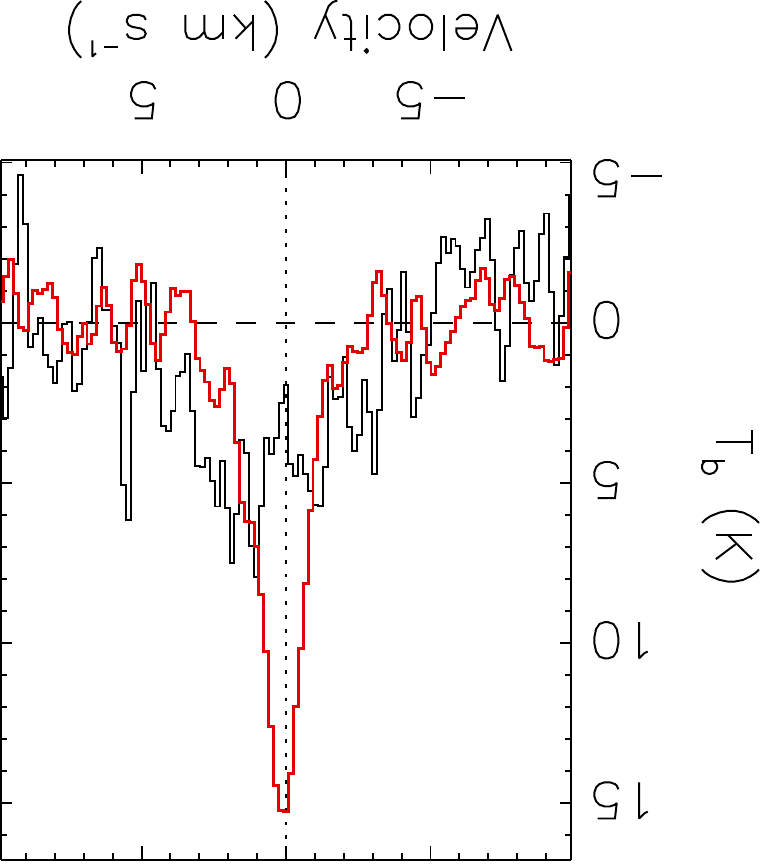}{0.3\textwidth}{DSTau, $^{13}$CO+C$^{18}$O}
           \rotatefig{180}{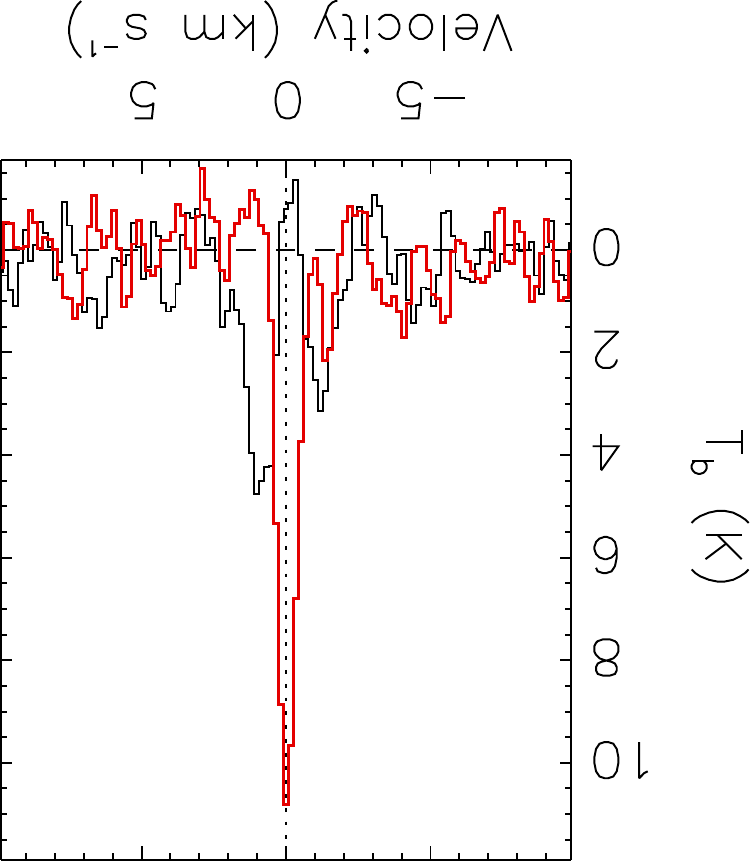}{0.3\textwidth}{FTTau, $^{13}$CO+C$^{18}$O}}
\caption{Original (black histograms) and velocity aligned  (red histograms) spectra for the sources in the Taurus star-forming region}
\label{fig:appendix_taurus}
\end{figure}

Figures \ref{fig:appendix_lupus} and \ref{fig:appendix_taurus} show the original, non-aligned (black) and the aligned (red) spectra of the sources in the Lupus and Taurus star-forming regions, respectively. These spectra are extracted from the combination of the molecular-line data which results in the highest S/N (Section \ref{sec:analysis}), and these are the spectra adopted to obtain the final measurements.

\begin{figure}
\centering
\gridline{ \rotatefig{180}{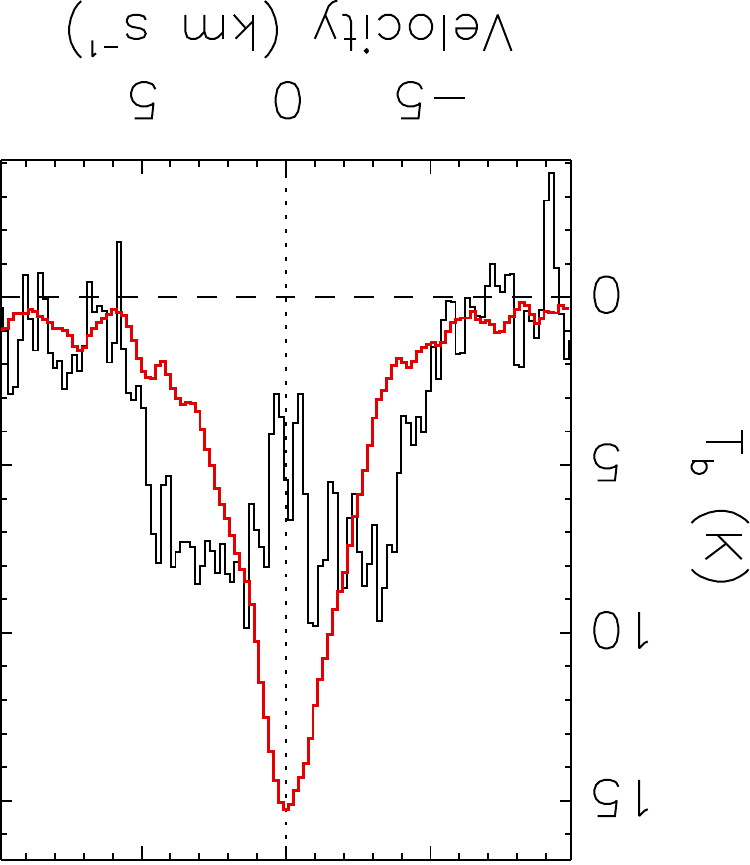}{0.3\textwidth}{RYLup, $^{13}$CO+C$^{18}$O}
           \rotatefig{180}{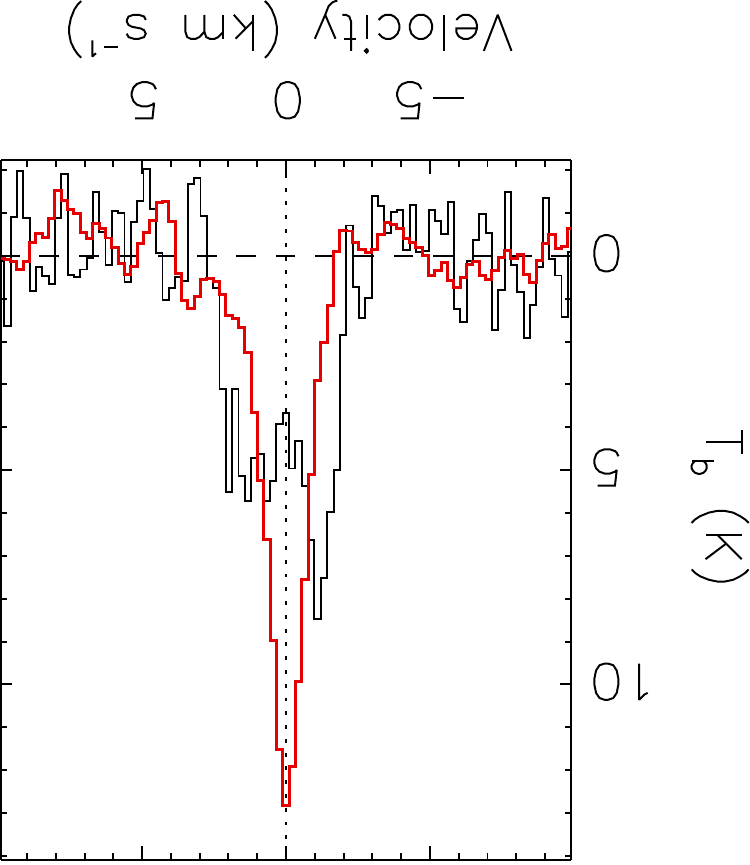}{0.3\textwidth}{Sz100, $^{13}$CO+CN}
           \rotatefig{180}{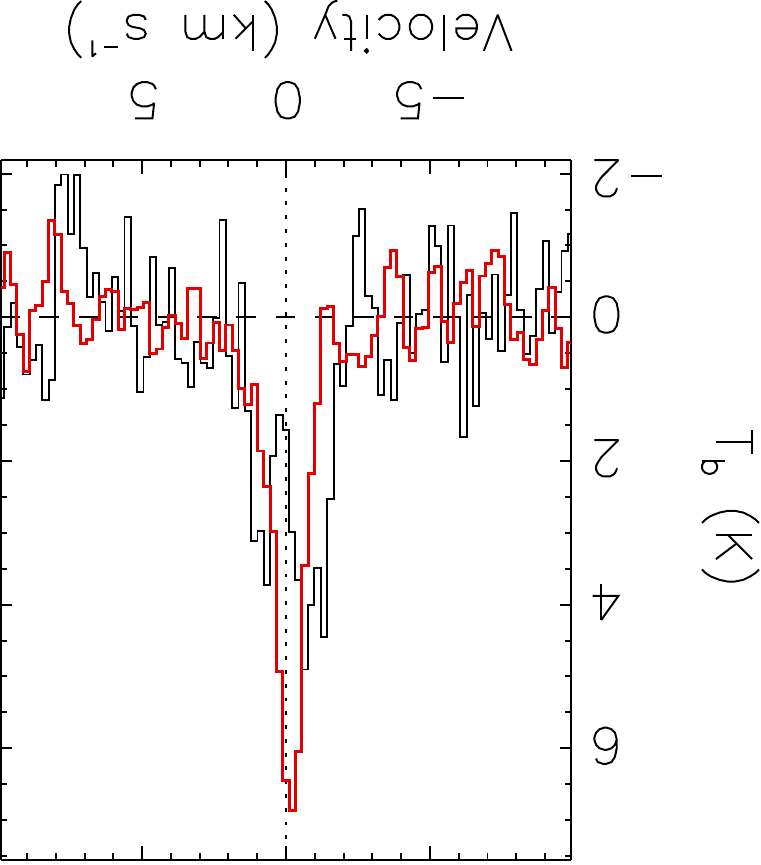}{0.3\textwidth}{Sz108B, $^{13}$CO+CN+C$^{18}$O}}
\gridline{ \rotatefig{180}{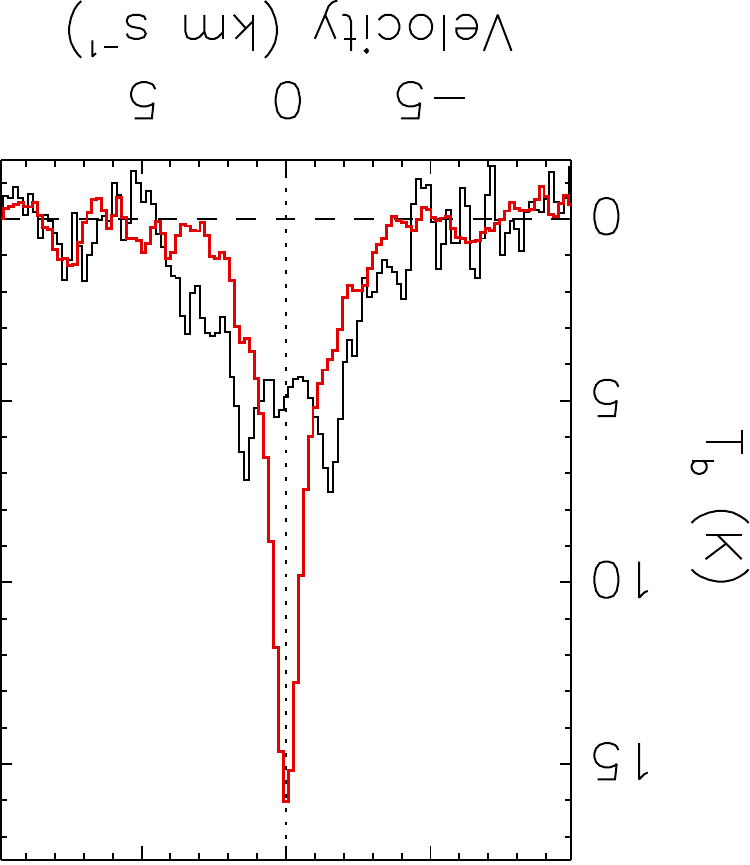}{0.3\textwidth}{Sz111, $^{13}$CO+C$^{18}$O}
           \rotatefig{180}{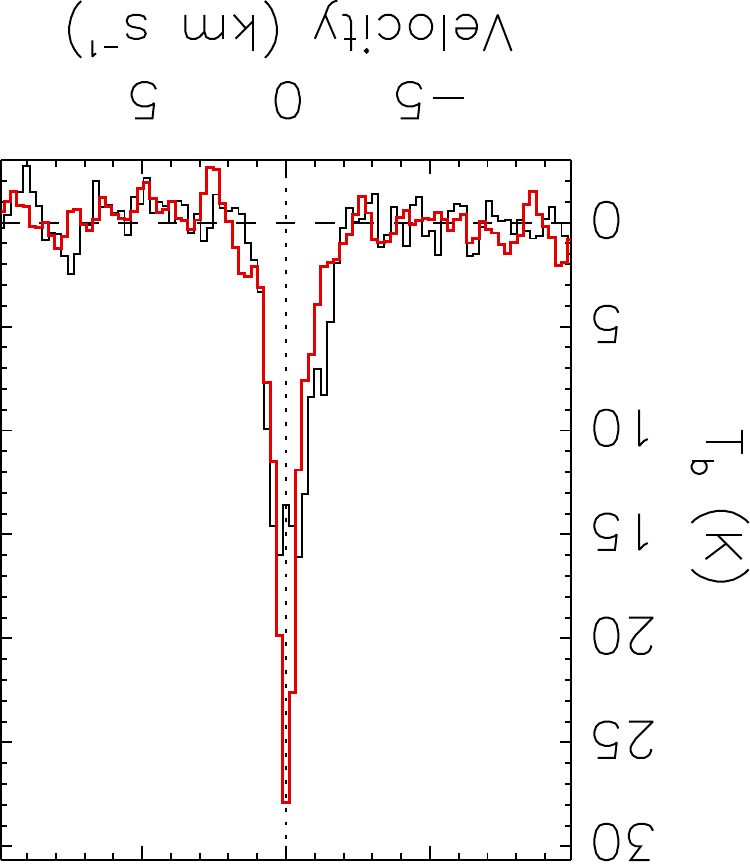}{0.3\textwidth}{Sz114, $^{13}$CO+CN+C$^{18}$O}
           \rotatefig{180}{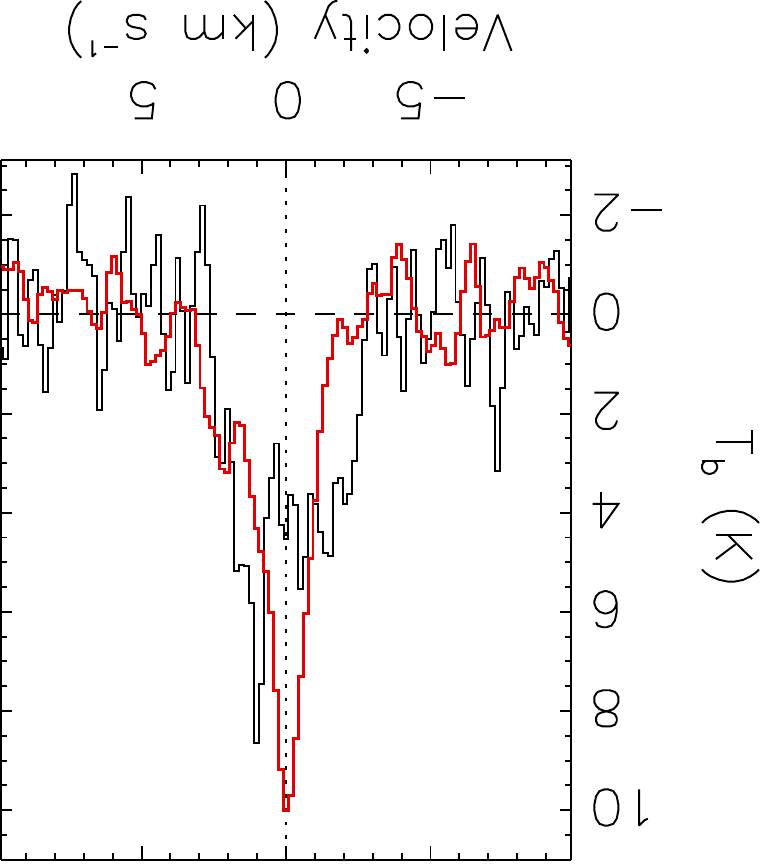}{0.3\textwidth}{Sz123A, $^{13}$CO+C$^{18}$O}}   
\gridline{ \rotatefig{180}{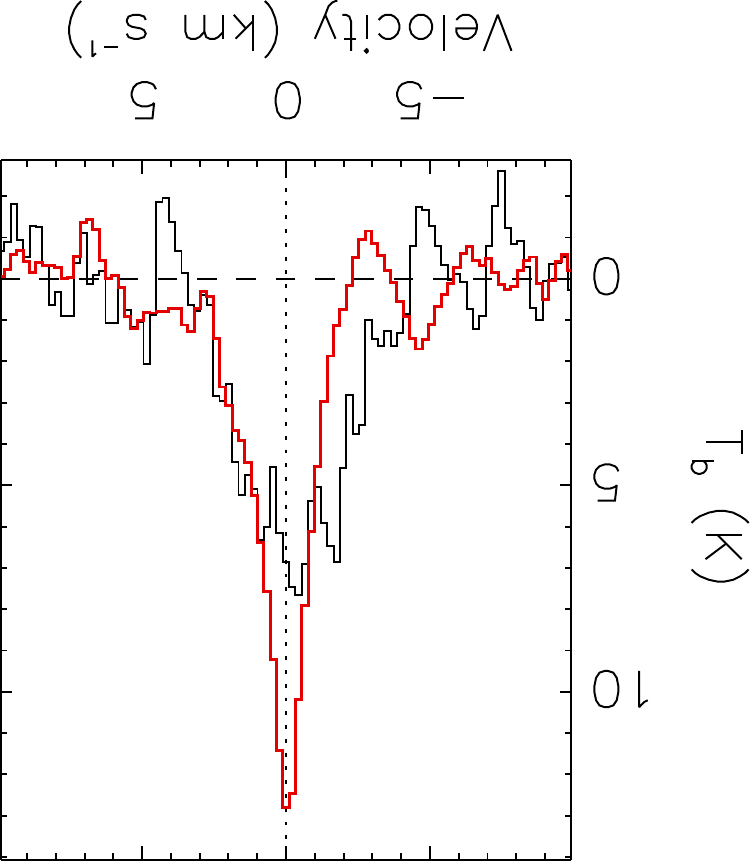}{0.3\textwidth}{Sz129, $^{13}$CO+CN+C$^{18}$O}
           \rotatefig{180}{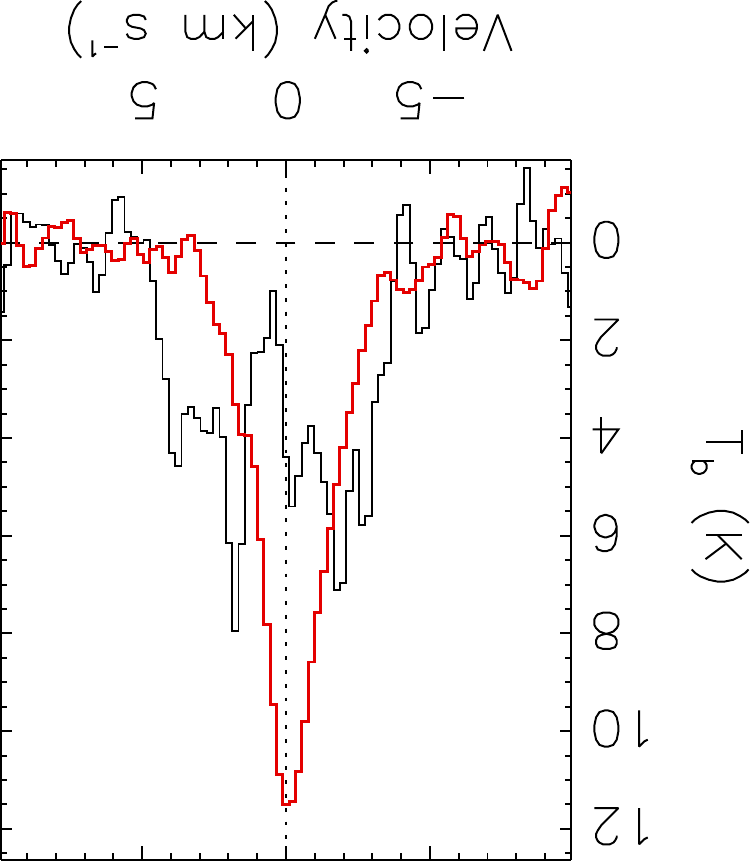}{0.3\textwidth}{Sz133, $^{13}$CO+CN+C$^{18}$O}
           \rotatefig{180}{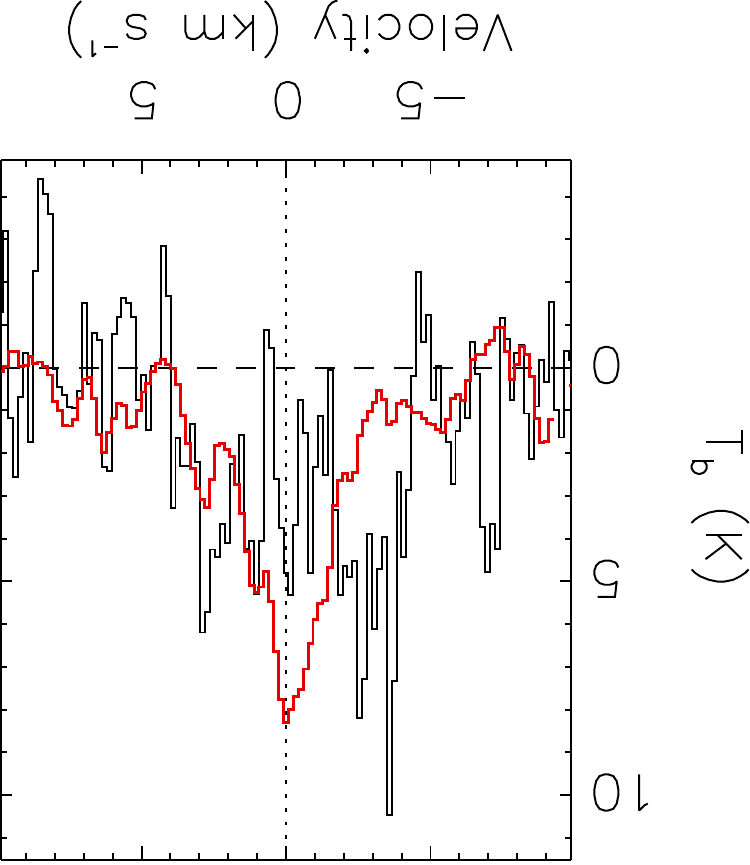}{0.3\textwidth}{Sz65, $^{13}$CO}}
\par \emph{Figure \ref{fig:appendix_lupus} continued}
\end{figure}

\begin{figure}[htb!]
\centering
\gridline{ \rotatefig{180}{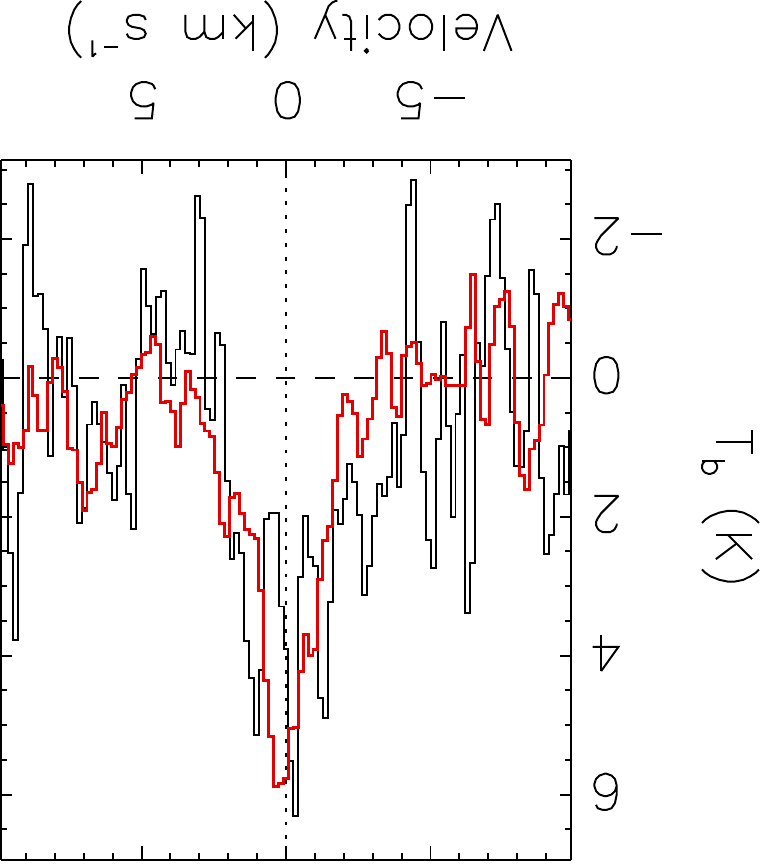}{0.3\textwidth}{Sz69, $^{13}$CO}
           \rotatefig{180}{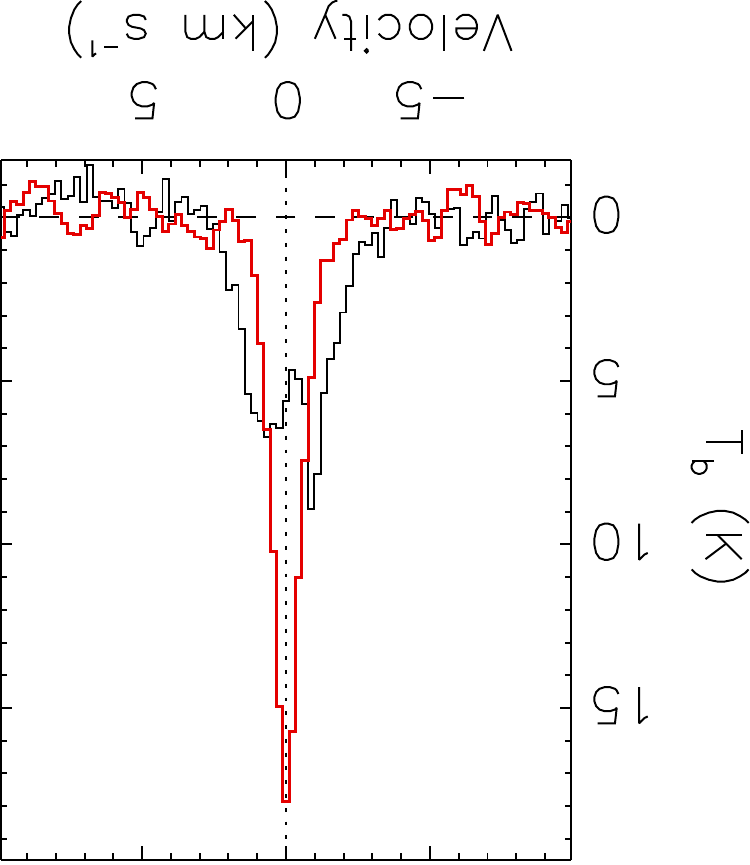}{0.3\textwidth}{Sz71, $^{13}$CO+CN+C$^{18}$O}
           \rotatefig{180}{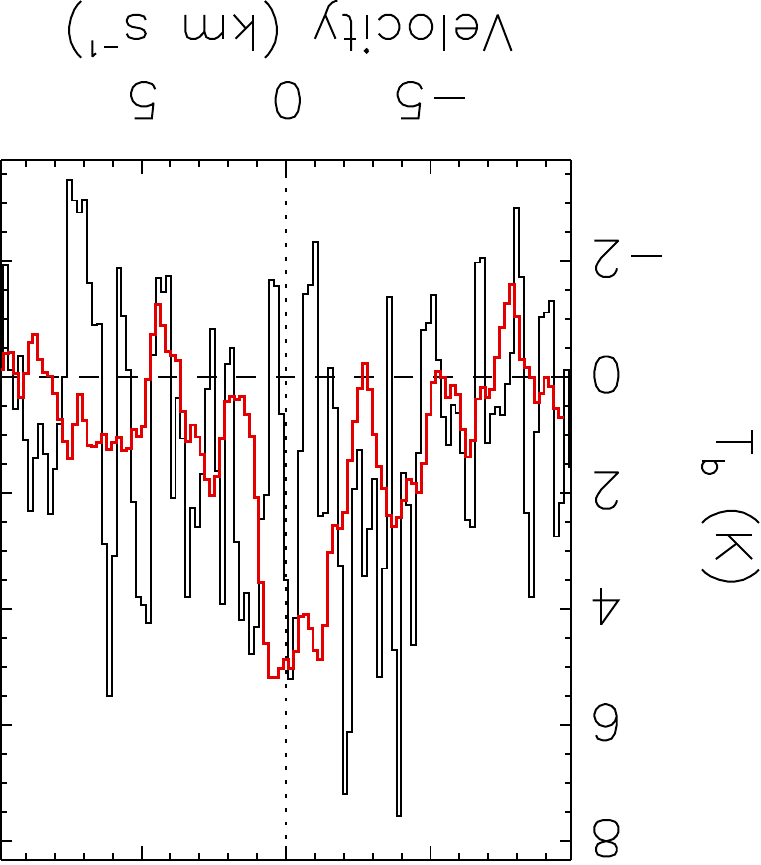}{0.3\textwidth}{Sz73, $^{13}$CO}}
\gridline{ \rotatefig{180}{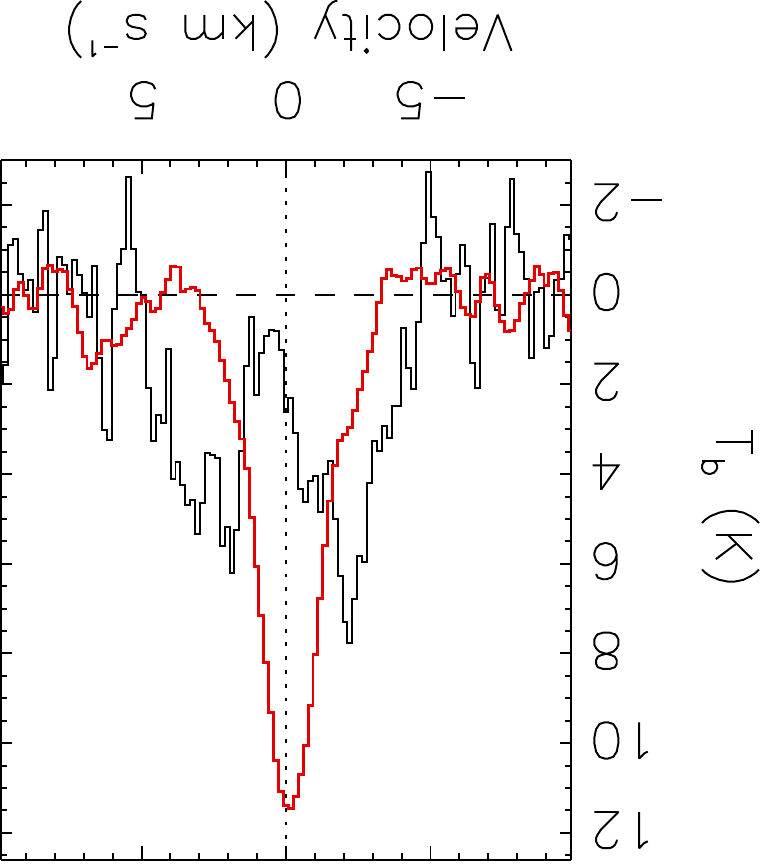}{0.3\textwidth}{Sz75, $^{13}$CO}
           \rotatefig{180}{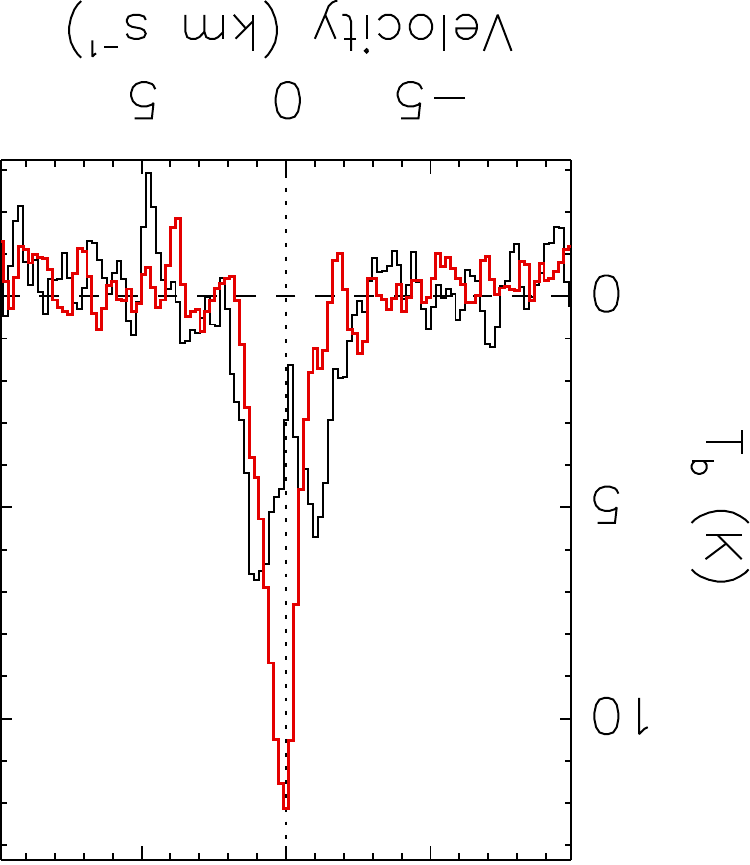}{0.3\textwidth}{Sz76, $^{13}$CO+C$^{18}$O}
           \rotatefig{180}{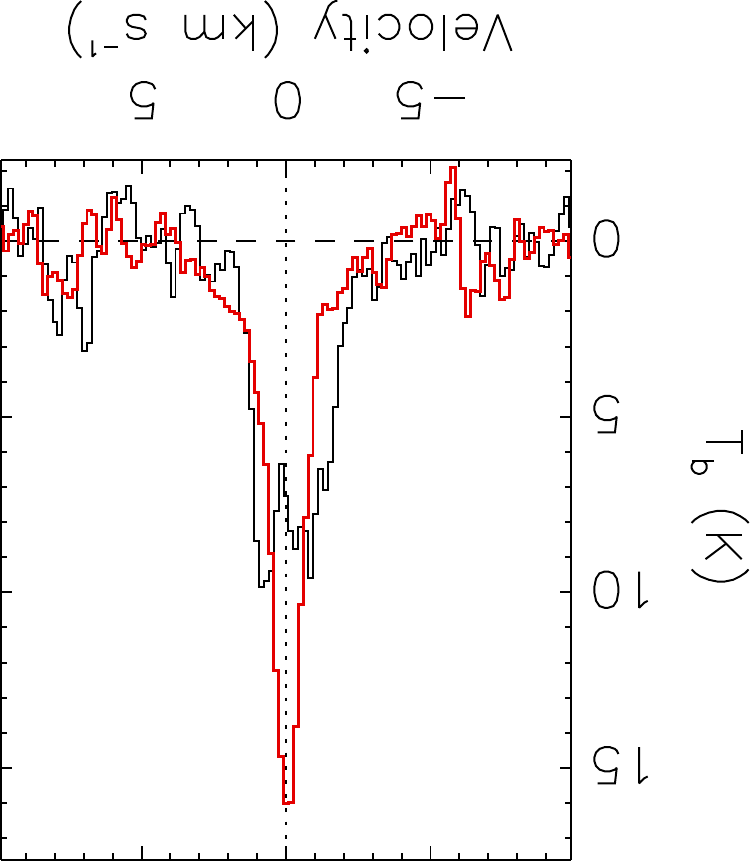}{0.3\textwidth}{Sz83, $^{13}$CO}}
\gridline{ \rotatefig{180}{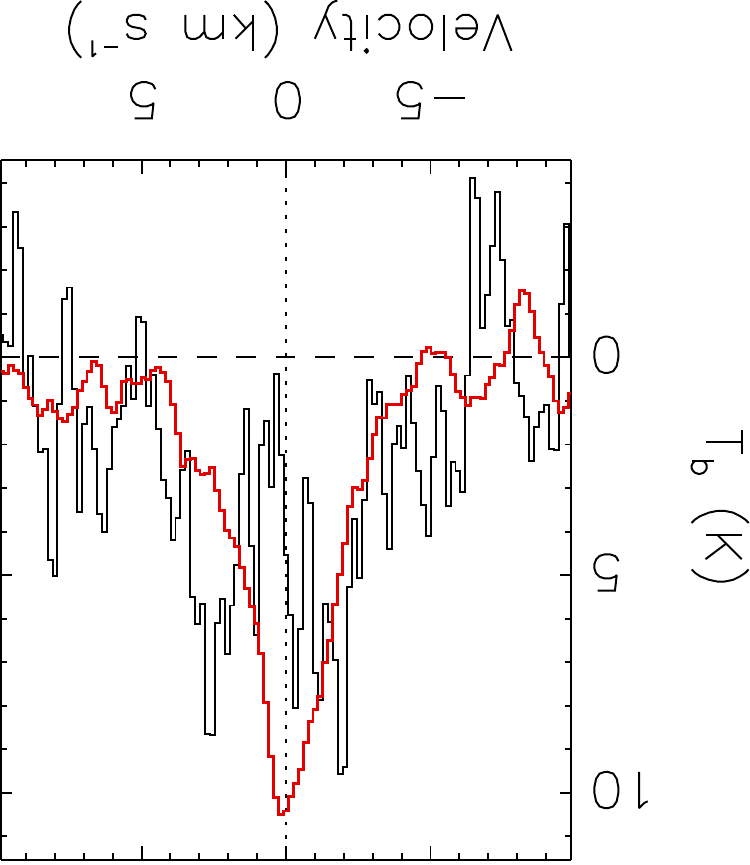}{0.3\textwidth}{Sz84, $^{13}$CO+C$^{18}$O}
            \rotatefig{180}{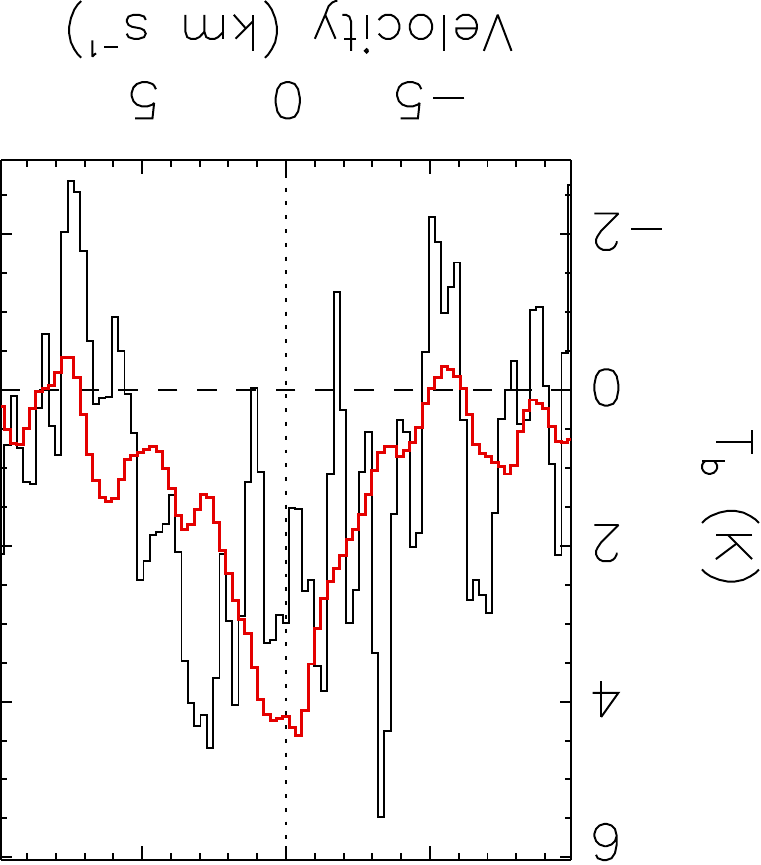}{0.3\textwidth}{Sz90, $^{13}$CO+CN}
           \rotatefig{180}{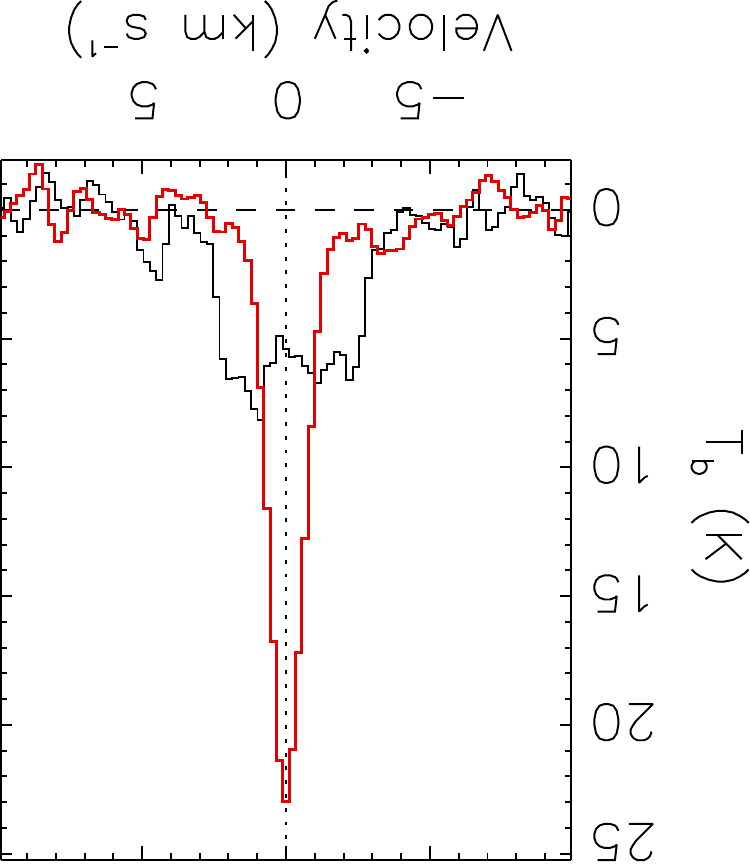}{0.3\textwidth}{Sz98, $^{13}$CO+CN}}
\par \emph{Figure \ref{fig:appendix_lupus} continued}
\end{figure}

\begin{figure}[htb!]
\centering
\gridline{ \rotatefig{180}{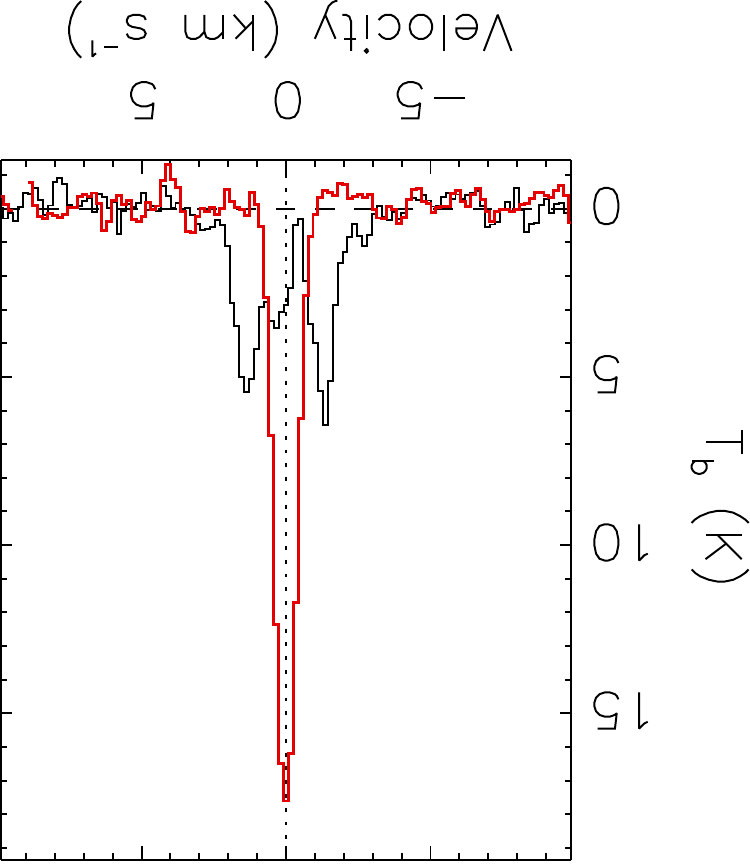}{0.3\textwidth}{SSTc2J160836.2-392302, $^{13}$CO}}
\par \emph{Figure \ref{fig:appendix_lupus} continued}
\end{figure}

\begin{figure}[htb!]{}
\centering
\gridline{ \rotatefig{180}{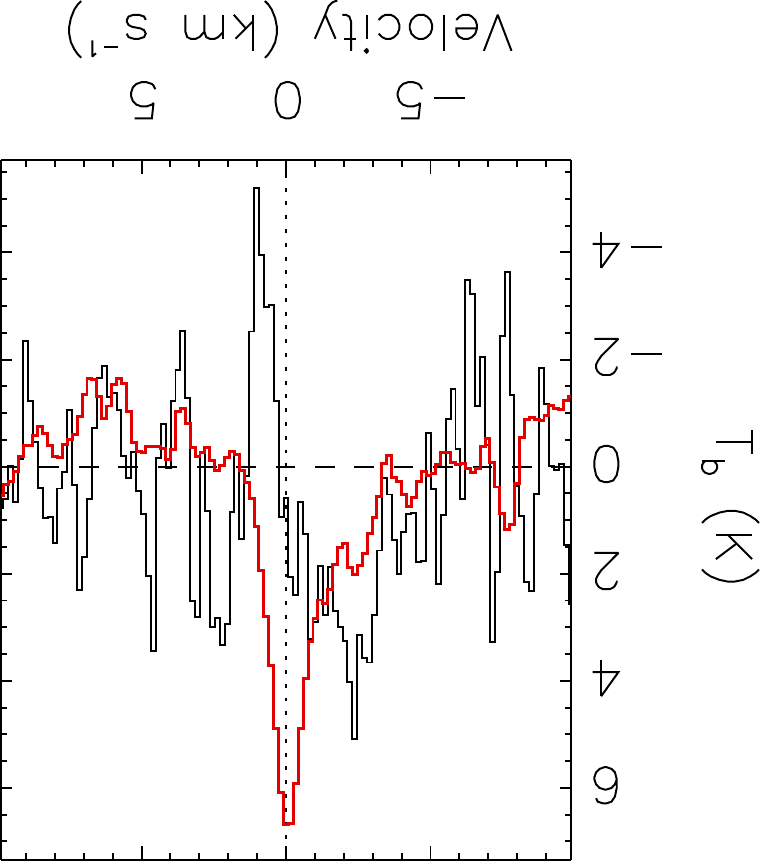}{0.3\textwidth}{HKTauA, $^{13}$CO+C$^{18}$O}
           \rotatefig{180}{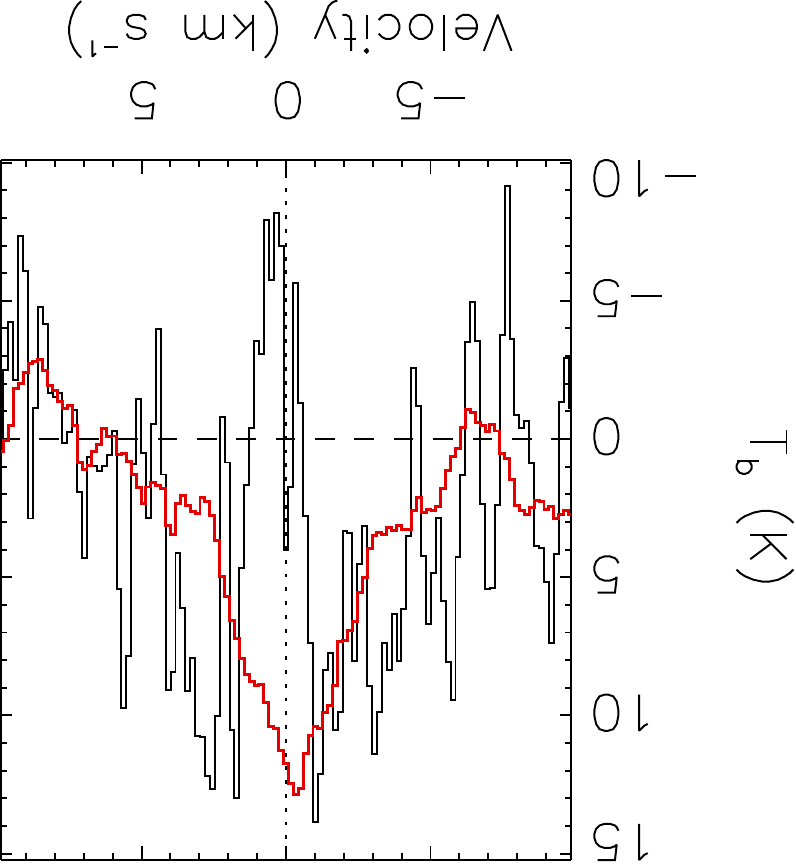}{0.3\textwidth}{HKTauB, $^{13}$CO+C$^{18}$O}
           \rotatefig{180}{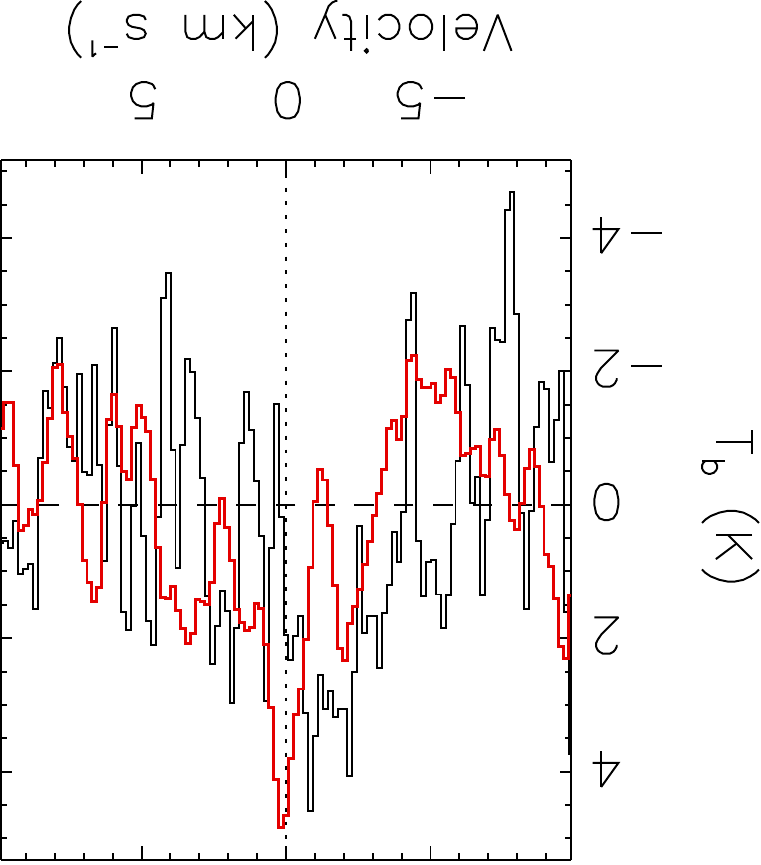}{0.3\textwidth}{HOTau, $^{13}$CO+C$^{18}$O}}
\gridline{ \rotatefig{180}{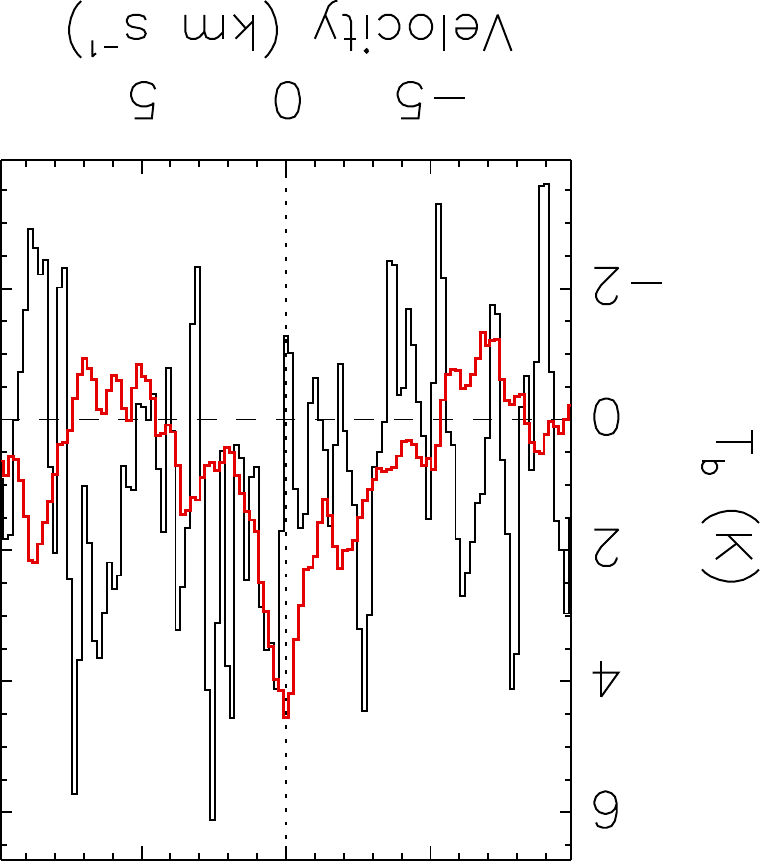}{0.3\textwidth}{IPTau, $^{13}$CO+C$^{18}$O}
           \rotatefig{180}{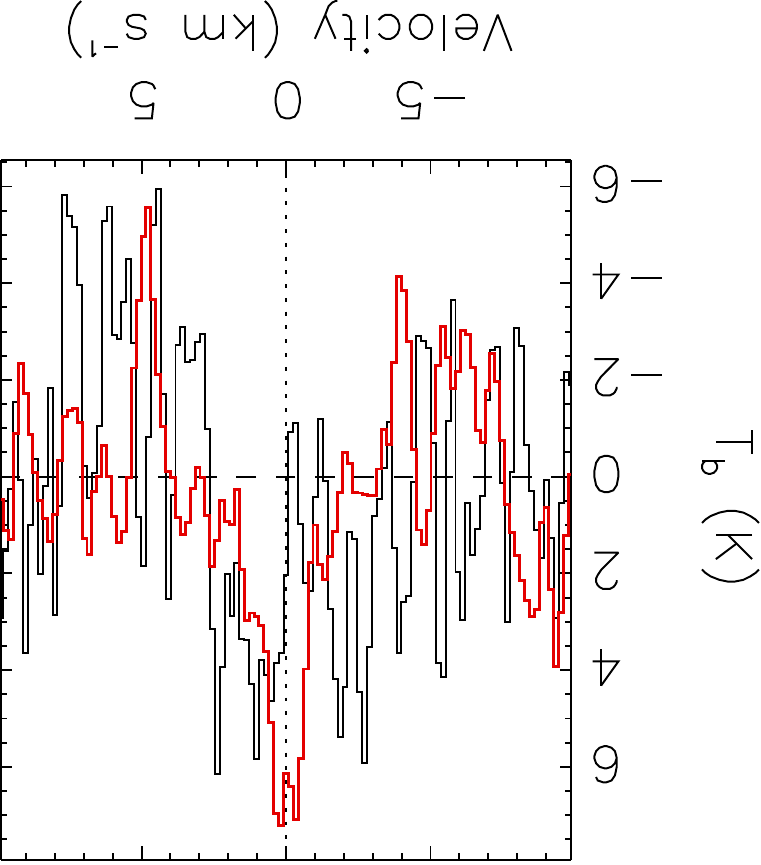}{0.3\textwidth}{IQTau, $^{13}$CO $J$ = 2--1}
           \rotatefig{180}{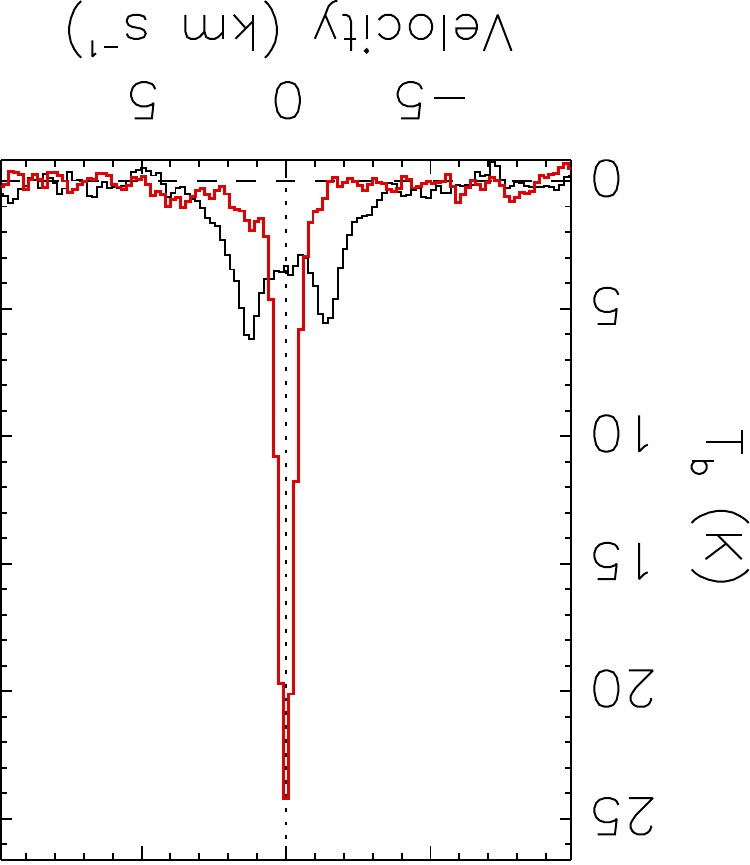}{0.3\textwidth}{MWC480, $^{13}$CO $J$ = 2--1}}
\gridline{ \rotatefig{180}{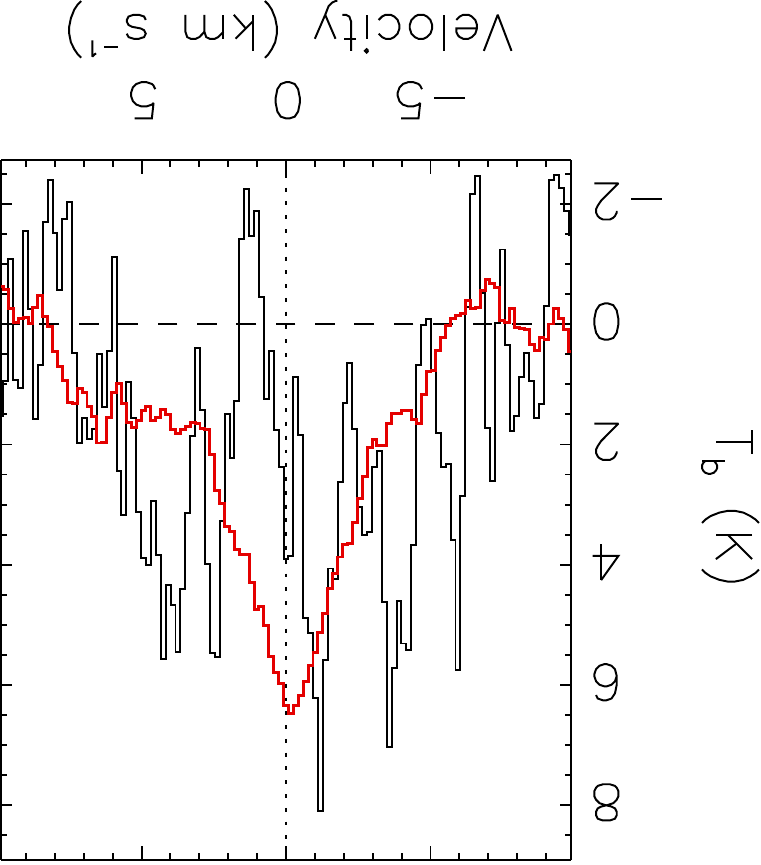}{0.3\textwidth}{RWAurA, $^{13}$CO+C$^{18}$O}
           \rotatefig{180}{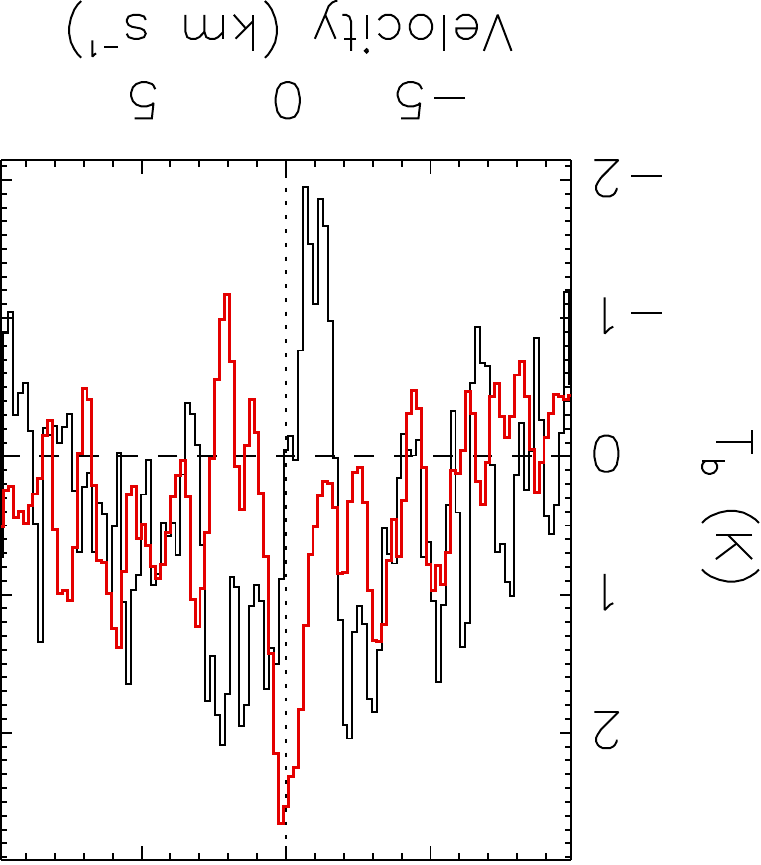}{0.3\textwidth}{TTauN, C$^{18}$O $J$ = 2--1}
           \rotatefig{180}{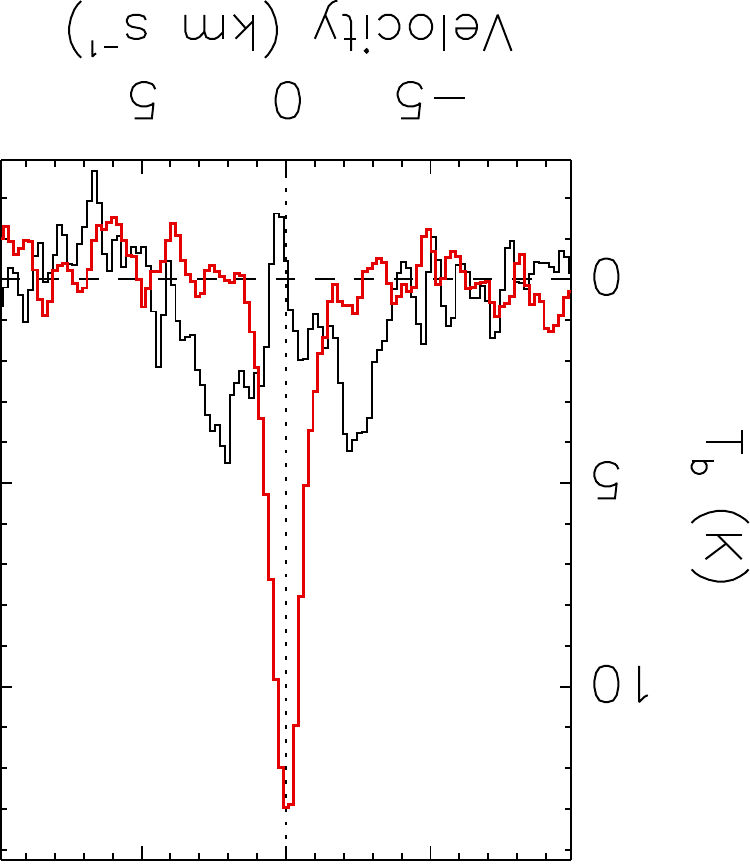}{0.3\textwidth}{UZTauE, $^{13}$CO $J$ = 2--1}}
\par \emph{Figure \ref{fig:appendix_taurus} continued}
\end{figure}

\begin{figure}[htb!]{}
\centering
\gridline{ \rotatefig{180}{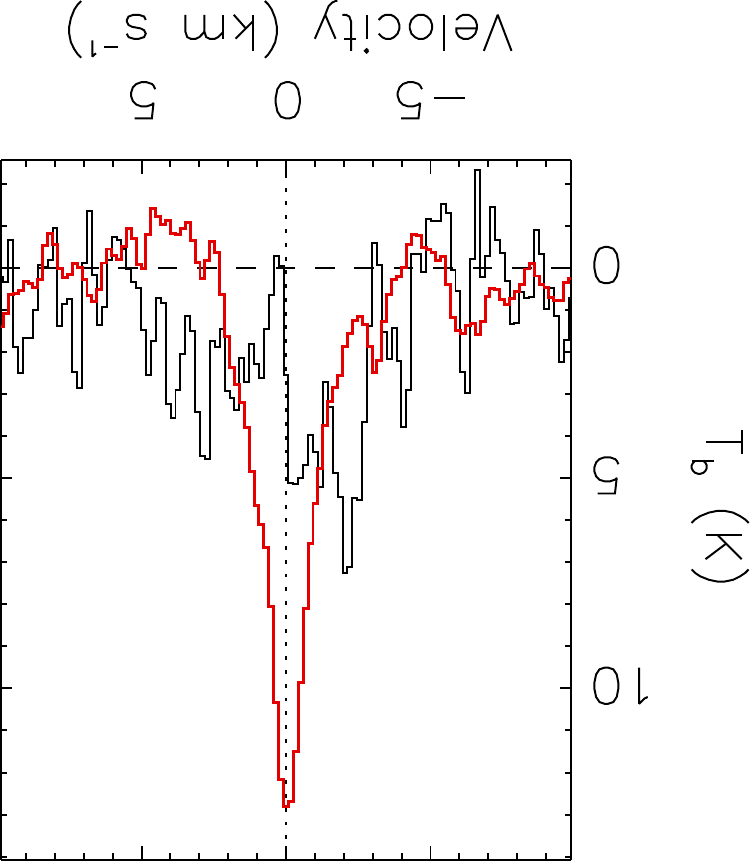}{0.3\textwidth}{V710Tau, $^{13}$CO+C$^{18}$O}
           \rotatefig{180}{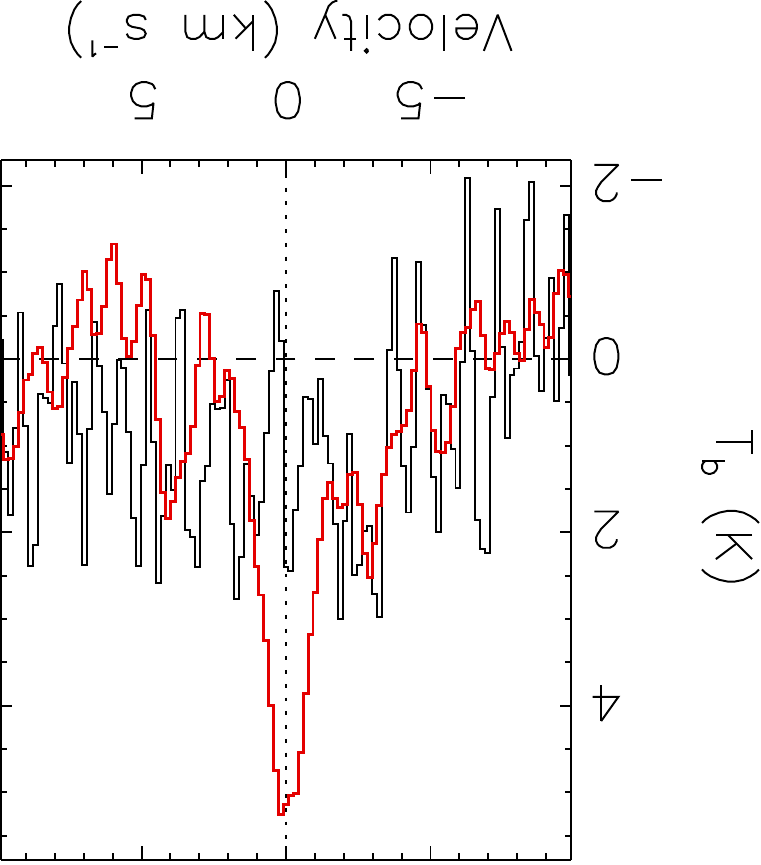}{0.3\textwidth}{V836Tau, $^{13}$CO+C$^{18}$O}}
\par \emph{Figure \ref{fig:appendix_taurus} continued}
\end{figure}

\FloatBarrier

\section{Testing the analysis on synthetic data}
\label{sec:appendix_synthetic}

\begin{figure}[htb!]
\centering
\gridline{ \rotatefig{180}{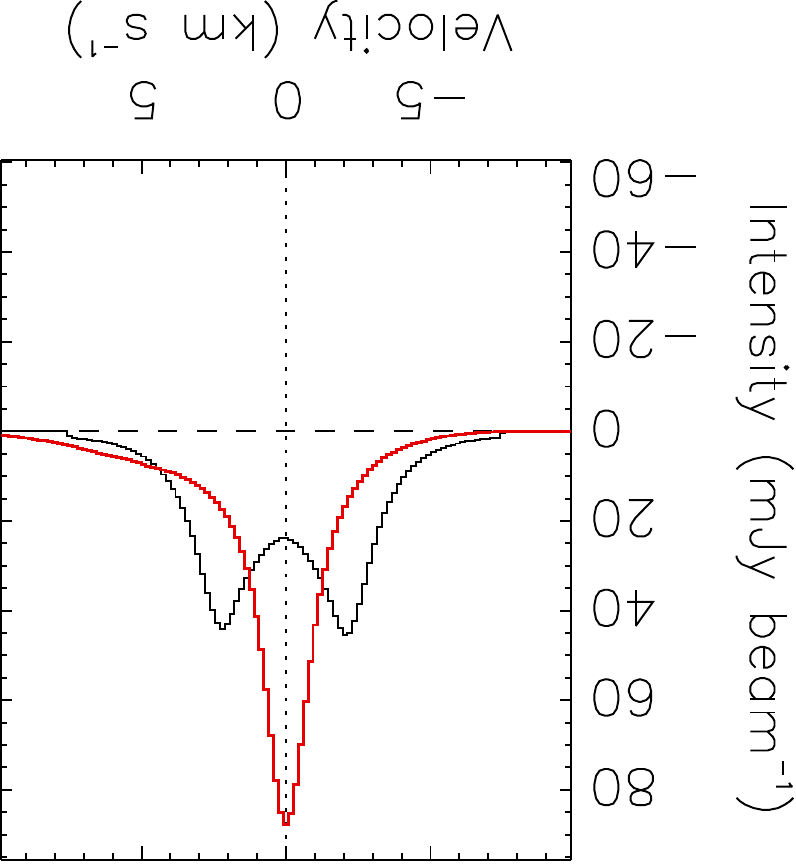}{0.4\textwidth}{(a) without noise}
           \rotatefig{180}{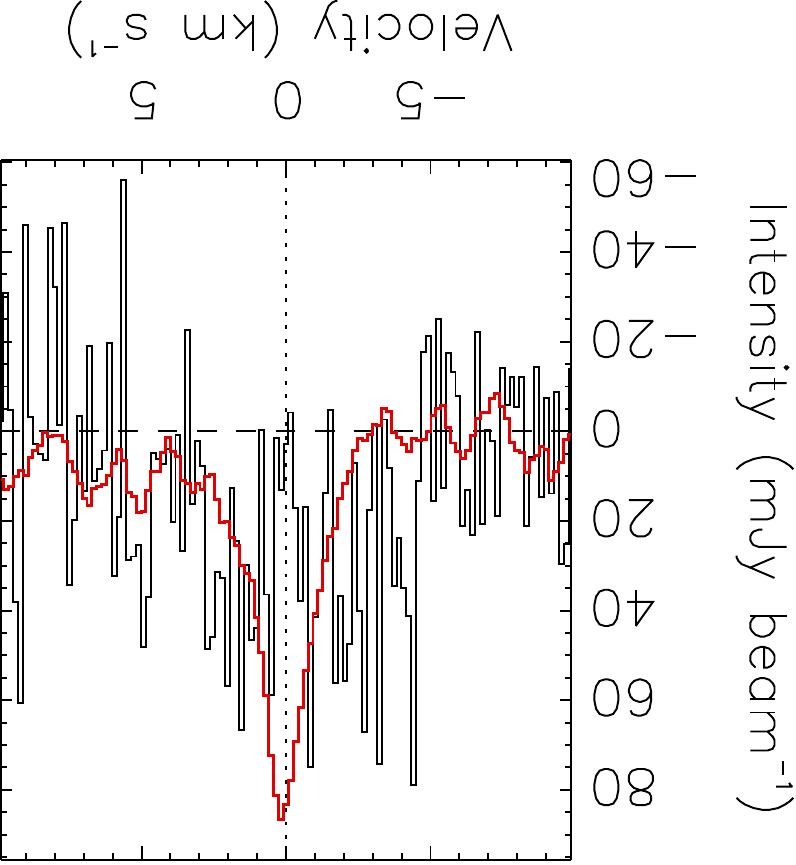}{0.4\textwidth}{(b) with random noise}}
\caption{Original (black histogram) and velocity aligned (red histograms) spectra for synthesised data.}
\label{fig:appendix_synthetic_data}
\end{figure}

Figure \ref{fig:appendix_synthetic_data} presents an example of the tests on our analysis using the synthetic $^{13}$CO (3--2) data. 
In this example, the synthetic data were generated from our model disk with an inclination angle 45$^{\circ}$, a position angle of 90$^{\circ}$, a scale height of $h/R = 0.2$, a disk mass of 0.001~$M_{\odot}$, and a stellar mass of $1.1~M_{\odot}$ using the CASA simulator.
The uv-coverage in the imaging simulations was adopted to be similar to that of the Band 7 data of the Lupus YSOs.
Figure \ref{fig:appendix_synthetic_data}a shows the synthetic spectra without noise.
The black histograms show the original spectrum directly extracted from the synthetic data cube, 
and the spectrum exhibits double peaks due to the Doppler shifted emission in the disk. 
The red histograms show the spectrum after the velocity alignment with the model parameters described above. 
After the velocity alignment, the spectrum becomes single-peaked because the Doppler shift due to the disk rotation is all corrected to zero velocity. 
Figure \ref{fig:appendix_synthetic_data}b shows the same synthetic data but with the noise included in the imaging simulations.
The noise level was adopted to be similar to that in our real data. 
The double peak is less visible because of the relatively low S/N. 
Then we applied our analysis described in Section \ref{sec:analysis} to measure the stellar mass from the synthetic data. 
After applying our analysis, the spectrum (red histograms) is properly aligned to have a clear single peak. 
The stellar mass is measured to be $1.1^{+0.1}_{-0.05}~M_{\odot}$ consistent with the model input within 1$\sigma$. 
This example demonstrates that our method indeed gives the expected result.

\section{Spectroscopic Masses}
\label{sec:appendix_Mspec}

All values for the spectroscopic masses calculated by using the evolutionary models of \citet{palla1999star}, \citet{siess2000}, \citet{bressan2012parsec}, \citet{chen2014improvingPARSEC}, \citet{baraffe2015} and \citet{feiden2016magnetic} are given in Table \ref{tab:Mspec_lupus} and Table \ref{tab:Mspec_taurus} for the sources in the Lupus and Taurus star-forming region, respectively.

\movetabledown=30mm
\begin{rotatetable}
\begin{deluxetable}{lcccccccccccc}
\label{tab:Mspec_lupus}
\tablecaption{Spectroscopic masses of different evolutionary models for the sources in the Lupus star-forming region}
\tablehead{\colhead{Name} & \colhead{\citetalias{palla1999star}} & \colhead{\citetalias{siess2000}} & \colhead{\citetalias{bressan2012parsec}} & \colhead{\citetalias{chen2014improvingPARSEC}} & \colhead{\citetalias{baraffe2015}} & \colhead{\citetalias{feiden2016magnetic}} & \colhead{magnetic \citetalias{feiden2016magnetic}} \\
\colhead{} & \colhead{[$M_{\odot}$]} & \colhead{[$M_{\odot}$]} & \colhead{[$M_{\odot}$]} & \colhead{[$M_{\odot}$]} & \colhead{[$M_{\odot}$]} & \colhead{[$M_{\odot}$]} & \colhead{[$M_{\odot}$]} &
}
\startdata
Sz65  & 0.80 $\pm$ 0.12 & 0.75 $\pm$ 0.11 &  0.57 $\pm$ 0.15 & 0.70 $\pm$ 0.12 & 0.71 $\pm$ 0.17 & 0.64 $\pm$ 0.16 & 1.14 $\pm$ 0.13 \\
J15450887-3417333 & 0.108 $\pm$ 0.016  & 0.127 $\pm$ 0.019   &  0.101 $\pm$ 0.010 & 0.32 $\pm$ 0.05 & 0.14 $\pm$ 0.03 & 0.114 $\pm$ 0.025 & 0.19 $\pm$ 0.04 \\
Sz68 & 2.1 $\pm$ 0.3 & 2.2 $\pm$ 0.3   &  1.6 $\pm$ 0.4 & 1.6 $\pm$ 0.4 & 1.40 $\pm$ 0.03 & 1.6 $\pm$ 0.4 & 1.68 $\pm$ 0.22 \\
Sz69 & 0.162 $\pm$ 0.024  & 0.192 $\pm$ 0.029    &  0.137 $\pm$ 0.025 & 0.40 $\pm$ 0.06 & 0.20 $\pm$ 0.04 & 0.18 $\pm$ 0.04 & 0.26 $\pm$ 0.06 \\
Sz71 & 0.41 $\pm$ 0.06 & 0.43 $\pm$ 0.06   &  0.35 $\pm$ 0.09 & 0.58 $\pm$ 0.13 & 0.42 $\pm$ 0.11 & 0.42 $\pm$ 0.11 & 0.68 $\pm$ 0.15 \\
Sz73 & 0.80 $\pm$ 0.12 & 0.80 $\pm$ 0.12   &  0.64 $\pm$ 0.1 & 0.73 $\pm$ 0.09 & 0.77 $\pm$ 0.16 & 0.75 $\pm$ 0.16 & 1.01 $\pm$ 0.10 \\
Sz75 & 0.97 $\pm$ 0.15  & 0.92 $\pm$ 0.14   &  0.66 $\pm$ 0.18 & 0.72 $\pm$ 0.15 & 0.84 $\pm$ 0.18 & 0.70 $\pm$ 0.18 & 1.31 $\pm$ 0.17 \\
Sz76 & 0.192 $\pm$ 0.029  & 0.25 $\pm$ 0.04    &  0.19 $\pm$ 0.03 & 0.38 $\pm$ 0.08 & 0.23 $\pm$ 0.04 & 0.23 $\pm$ 0.04 & 0.34 $\pm$ 0.07 \\
RXJ1556.1-3655 & 0.46 $\pm$ 0.07   & 0.47 $\pm$ 0.07   &  0.40 $\pm$ 0.11 & 0.69 $\pm$ 0.10 & 0.49 $\pm$ 0.13 & 0.49 $\pm$ 0.13 & 0.74 $\pm$ 0.14 \\
Sz82 & 1.19 $\pm$ 0.18  & 1.13 $\pm$ 0.17 &  0.77 $\pm$ 0.21 & 0.77 $\pm$ 0.20 & 1.06 $\pm$ 0.18 & 0.79 $\pm$ 0.21 & 1.4 $\pm$ 0.3 \\
Sz83 & 0.80 $\pm$ 0.12 & 0.74 $\pm$ 0.11     &  0.54 $\pm$ 0.14 & 0.62 $\pm$ 0.13 & 0.74 $\pm$ 0.15 & 0.58 $\pm$ 0.14 & 1.21 $\pm$ 0.23 \\
Sz84 & 0.140 $\pm$ 0.021   & 0.188 $\pm$ 0.028    &  0.130 $\pm$ 0.022 & 0.30 $\pm$ 0.08 & 0.170 $\pm$ 0.028 & 0.15 $\pm$ 0.03 & 0.22 $\pm$ 0.07 \\
Sz129& 0.78 $\pm$ 0.12 & 0.82 $\pm$ 0.12 &  0.64 $\pm$ 0.16 & 0.73 $\pm$ 0.09 & 0.78 $\pm$ 0.15 & 0.76 $\pm$ 0.15 & 1.00 $\pm$ 0.10 \\
RYLup& 1.42 $\pm$ 0.21 & 1.53 $\pm$ 0.23   &  1.40 $\pm$ 0.18 & 1.40 $\pm$ 0.18 & 1.40 $\pm$ 0.08 & 1.41 $\pm$ 0.17 & 1.51 $\pm$ 0.13 \\
J16000236-4222145&0.192 $\pm$ 0.029  & 0.25 $\pm$ 0.04    &  0.19 $\pm$ 0.03 & 0.38 $\pm$ 0.08 & 0.23 $\pm$ 0.04 & 0.23 $\pm$ 0.04 & 0.33 $\pm$ 0.07 \\
MYLup & 1.03 $\pm$ 0.15 & 1.06 $\pm$ 0.16 &  1.11 $\pm$ 0.11 & 1.11 $\pm$ 0.10 & 1.09 $\pm$ 0.11 & 1.11 $\pm$ 0.11 & 1.02 $\pm$ 0.10 \\
EXLup & 0.57 $\pm$ 0.08 & 0.56 $\pm$ 0.08    &  0.44 $\pm$ 0.11 & 0.57 $\pm$ 0.13 & 0.54 $\pm$ 0.13 & 0.50 $\pm$ 0.13 & 0.97 $\pm$ 0.16 \\
Sz90 & 0.78 $\pm$ 0.12 & 0.82 $\pm$ 0.12 &  0.65 $\pm$ 0.15 & 0.73 $\pm$ 0.08 & 0.78 $\pm$ 0.15 & 0.77 $\pm$ 0.15 & 0.99 $\pm$ 0.09 \\
Sz91 & 0.49 $\pm$ 0.07  & 0.46 $\pm$ 0.07    &  0.40 $\pm$ 0.11 & 0.70 $\pm$ 0.08 & 0.51 $\pm$ 0.13 & 0.52 $\pm$ 0.13 & 0.73 $\pm$ 0.12 \\
Sz98 & 0.80 $\pm$ 0.12 & 0.74 $\pm$ 0.11     &  0.54 $\pm$ 0.14 & 0.62 $\pm$ 0.13 & 0.74 $\pm$ 0.15 & 0.58 $\pm$ 0.14 & 1.21 $\pm$ 0.23 \\
Sz100& 0.105 $\pm$ 0.016 & 0.140 $\pm$ 0.021     &  0.105 $\pm$ 0.013 & 0.31 $\pm$ 0.06 & 0.142 $\pm$ 0.028 & 0.118 $\pm$ 0.026 & 0.18 $\pm$ 0.04 \\
J16083070-3828268 & 1.41 $\pm$ 0.21  & 1.53 $\pm$ 0.23    &  1.40 $\pm$ 0.18 & 1.40 $\pm$ 0.18 & 1.40 $\pm$ 0.08 & 1.40 $\pm$ 0.17 & 1.50 $\pm$ 0.13 \\
SSTc2dJ160836.2-392302  & 0.93 $\pm$ 0.14  & 0.92 $\pm$ 0.14   &  0.68 $\pm$ 0.18 & 0.73 $\pm$ 0.15 & 0.84 $\pm$ 0.19 & 0.74 $\pm$ 0.18 & 1.27 $\pm$ 0.12 \\
Sz108B &  0.140 $\pm$ 0.021   & 0.178 $\pm$ 0.027 &  0.125 $\pm$ 0.021 & 0.33 $\pm$ 0.07 & 0.169 $\pm$ 0.030 & 0.15 $\pm$ 0.03 & 0.22 $\pm$ 0.06 \\
J16085324-3914401 & 0.28 $\pm$ 0.04 & 0.31 $\pm$ 0.05    &  0.25 $\pm$ 0.04 & 0.48 $\pm$ 0.09 & 0.29 $\pm$ 0.05 & 0.31 $\pm$ 0.05 & 0.46 $\pm$ 0.08 \\
Sz111 & 0.48 $\pm$ 0.07   & 0.46 $\pm$ 0.07    &  0.40 $\pm$ 0.11 & 0.70 $\pm$ 0.08 & 0.51 $\pm$ 0.13 & 0.51 $\pm$ 0.13 & 0.73 $\pm$ 0.12 \\
Sz114 & 0.150 $\pm$ 0.023  & 0.22 $\pm$ 0.03     &  0.161 $\pm$ 0.026 & 0.26 $\pm$ 0.07 & 0.194 $\pm$ 0.027 & 0.18 $\pm$ 0.03 & 0.28 $\pm$ 0.07 \\
J16102955-3922144 & 0.168 $\pm$ 0.025  & 0.21 $\pm$ 0.03 &  0.146 $\pm$ 0.026 & 0.39 $\pm$ 0.07 & 0.20 $\pm$ 0.04 & 0.18 $\pm$ 0.04 & 0.27 $\pm$ 0.06 \\
Sz123A & 0.47 $\pm$ 0.07 & 0.46 $\pm$ 0.07    &  0.38 $\pm$ 0.10 & 0.69 $\pm$ 0.05 & 0.55 $\pm$ 0.12 & 0.56 $\pm$ 0.12 & 0.67 $\pm$ 0.10 \\
J16124373-3815031 & 0.49 $\pm$ 0.07 & 0.47 $\pm$ 0.07   &  0.39 $\pm$ 0.10 & 0.60 $\pm$ 0.12 & 0.46 $\pm$ 0.12 & 0.46 $\pm$ 0.12 & 0.76 $\pm$ 0.15 \\
\enddata
\tablecomments{(\citetalias{palla1999star}) \citet{palla1999star}; (\citetalias{siess2000}) \citet{siess2000}; (\citetalias{bressan2012parsec}) \citet{bressan2012parsec}; (\citetalias{chen2014improvingPARSEC}) \citet{chen2014improvingPARSEC}; (\citetalias{baraffe2015}) \citet{baraffe2015}; (\citetalias{feiden2016magnetic}) \citet{feiden2016magnetic}}
\tablecomments{We were not able to obtain reliable spectroscopic masses for Sz133, because the location of this source in the HR diagram does not overlap with the evolutionary models of PMS stars. Thus, Sz133 was excluded from the comparison.}
\end{deluxetable}
\end{rotatetable}

\begin{rotatetable}
\begin{deluxetable}{lcccccccccccc}
\label{tab:Mspec_taurus}
\tablecaption{Spectroscopic masses of different evolutionary models for the sources in the Taurus star-forming region}
\tablehead{\colhead{Name} & \colhead{\citetalias{palla1999star}} & \colhead{\citetalias{siess2000}} & \colhead{\citetalias{bressan2012parsec}} & \colhead{\citetalias{chen2014improvingPARSEC}} & \colhead{\citetalias{baraffe2015}} & \colhead{\citetalias{feiden2016magnetic}} & \colhead{magnetic \citetalias{feiden2016magnetic}} \\
\colhead{} & \colhead{[$M_{\odot}$]} & \colhead{[$M_{\odot}$]} & \colhead{[$M_{\odot}$]} & \colhead{[$M_{\odot}$]} & \colhead{[$M_{\odot}$]} & \colhead{[$M_{\odot}$]} & \colhead{[$M_{\odot}$]} &
}
\startdata
CIDA9A   &  0.38$\pm$0.06 & 0.39$\pm$0.06 & 0.33$\pm$0.09  & 0.64$\pm$0.11 &  0.41$\pm$0.12  & 0.43$\pm$0.12  & 0.62$\pm$0.14 \\
DOTau    &  0.57$\pm$0.09 & 0.54$\pm$0.08 & 0.47$\pm$0.12  & 0.71$\pm$0.06 &  0.59$\pm$0.14  & 0.61$\pm$0.14  & 0.80$\pm$0.11 \\
DQTau    &  0.52$\pm$0.08 & 0.50$\pm$0.07 & 0.39$\pm$0.10  & 0.43$\pm$0.11 &  0.54$\pm$0.10  & 0.41$\pm$0.10  & 0.92$\pm$0.26 \\
DRTau    &  0.93$\pm$0.14 & 0.97$\pm$0.15 & 0.75$\pm$0.18  & 0.75$\pm$0.13 &  0.89$\pm$0.17  & 0.84$\pm$0.17  & 1.12$\pm$0.10 \\
DSTau    &  0.57$\pm$0.09 & 0.53$\pm$0.08 & 0.46$\pm$0.12  & 0.71$\pm$0.07 &  0.57$\pm$0.14  & 0.58$\pm$0.14  & 0.80$\pm$0.12 \\
FTTau    &  0.30$\pm$0.05 & 0.32$\pm$0.05 & 0.25$\pm$0.04  & 0.56$\pm$0.08 &  0.32$\pm$0.06  & 0.33$\pm$0.06  & 0.48$\pm$0.08 \\
HKTauA   &  0.41$\pm$0.06 & 0.42$\pm$0.06 & 0.36$\pm$0.09  & 0.62$\pm$0.12 &  0.43$\pm$0.12  & 0.44$\pm$0.12  & 0.67$\pm$0.15 \\
HOTau    &  0.27$\pm$0.04 & 0.29$\pm$0.04 & 0.22$\pm$0.04  & 0.52$\pm$0.08 &  0.29$\pm$0.05  & 0.30$\pm$0.06  & 0.43$\pm$0.08 \\
IQTau    &  0.47$\pm$0.07 & 0.45$\pm$0.07 & 0.39$\pm$0.10  & 0.70$\pm$0.09 &  0.49$\pm$0.13  & 0.50$\pm$0.13  & 0.72$\pm$0.13 \\
IPTau    &  0.52$\pm$0.08 & 0.50$\pm$0.07 & 0.43$\pm$0.11  & 0.68$\pm$0.10 &  0.52$\pm$0.14  & 0.52$\pm$0.14  & 0.81$\pm$0.14 \\
MWC480   &  1.88$\pm$0.28 & 2.1$\pm$0.3  & 1.96$\pm$0.10  & 1.95$\pm$0.10 &  \nodata        & 1.97$\pm$0.10  & 1.689$\pm$0.023 \\
RWAurA   &  1.04$\pm$0.16 & 1.07$\pm$0.16 & 1.13$\pm$0.11  & 1.13$\pm$0.11 &  1.11$\pm$0.11  & 1.13$\pm$0.12  & 1.04$\pm$0.10 \\
TTauN    &  2.0$\pm$0.3  & 2.2 $\pm$0.3  & 2.14$\pm$0.26  & 2.14$\pm$0.26 &  \nodata & 2.16$\pm$0.27  & 1.70$\pm$0.07 \\
UZTauE   &  0.38$\pm$0.06 & 0.39$\pm$0.06 & 0.32$\pm$0.08  & 0.51$\pm$0.13 &  0.38$\pm$0.10  & 0.39$\pm$0.10  & 0.63$\pm$0.15 \\
V710Tau  &  0.39$\pm$0.06 & 0.40$\pm$0.06 & 0.34$\pm$0.09  & 0.60$\pm$0.12 &  0.40$\pm$0.12  & 0.42$\pm$0.12  & 0.65$\pm$0.15 \\
V836Tau  &  0.48$\pm$0.07 & 0.48$\pm$0.07 & 0.40$\pm$0.10  & 0.60$\pm$0.12 &  0.48$\pm$0.13  & 0.47$\pm$0.12  & 0.80$\pm$0.15 \\
\enddata
\tablecomments{(\citetalias{palla1999star}) \citet{palla1999star}; (\citetalias{siess2000}) \citet{siess2000}; (\citetalias{bressan2012parsec}) \citet{bressan2012parsec}; (\citetalias{chen2014improvingPARSEC}) \citet{chen2014improvingPARSEC}; (\citetalias{baraffe2015}) \citet{baraffe2015}; (\citetalias{feiden2016magnetic}) \citet{feiden2016magnetic}}
\end{deluxetable}
\end{rotatetable}

\end{document}